\documentclass[12pt]{article}
\usepackage{amsmath}
\usepackage{amssymb}

\def\be{\begin{eqnarray}}
\def\ee{\end{eqnarray}}

\begin{document}

\centerline{\large \bf Scaling and nonscaling finite-size effects
in the Gaussian and} \centerline{\large \bf the mean spherical
model with free boundary conditions}

\vspace*{0.2cm} \centerline{X.S. Chen$^{1,2}$ and V. Dohm$^{2}$}
\centerline{$^1$ Institute of Theoretical Physics, Chinese Academy
of Sciences,} \centerline{P.O. Box 2735, Beijing 100080, China}
\centerline{$^2$ Institut f\"{u}r Theoretische Physik, Technische
Hochschule Aachen,} \centerline{D-52056 Aachen, Germany}



We calculate finite-size effects of the Gaussian model in a $L
\times \widetilde L^{d-1}$ box geometry with free boundary
conditions in one direction and periodic boundary conditions in
$d-1$ directions for $2 < d < 4$. We also consider film geometry
$(\widetilde L \to \infty)$. Finite-size scaling is found to be
valid for $d < 3$ and $d > 3$ but logarithmic deviations from
finite-size scaling are found for the free energy and energy
density at the Gaussian upper borderline dimension $d^* =3$. The
logarithms are related to the vanishing critical exponent $1 -
\alpha - \nu = (d-3)/2$ of the Gaussian surface energy density.
The latter has a cusp-like singularity in $d > 3$ dimensions. We
show that these properties are the origin of nonscaling
finite-size effects in the mean spherical model with free boundary
conditions in $d \geq 3$ dimensions. At bulk $T_c$ in $d = 3$
dimensions we find an unexpected {\it non-logarithmic} violation
of finite-size scaling for the susceptibility $\chi \sim L^3$ of
the mean spherical model in film geometry whereas only a
logarithmic deviation $\chi \sim L^2 \ln L$ exists for box
geometry. The result for film geometry is explained by the
existence of the lower borderline dimension $d_l = 3$, as implied
by the Mermin-Wagner theorem, that coincides with the Gaussian
upper borderline dimension $d^* = 3$. For $3 < d < 4$ we find a
power-law violation of scaling $\chi \sim L^{d-1}$ at bulk $T_c$
for box geometry and a nonscaling temperature dependence
$\chi_{surface} \sim \xi^d$ of the surface susceptibility above
$T_c$. For $2 < d < 3$ dimensions we show the validity of
universal finite-size scaling for the susceptibility of the mean
spherical model with free boundary conditions for both box and
film geometry and calculate the corresponding universal scaling
functions for $T \geq T_c$.



\newpage

\section* {I. Introduction and summary}

Finite-size effects near phase transitions and the concept of finite-size
scaling near critical points have been the subject of many studies over the
past decades \cite{fisher,barber,finite,privman}. Consider, for example, the
susceptibility $\chi (t, L)$ of a ferromagnetic system at the reduced
temperature $t = (T - T_c)/T_c \geq 0$ near the bulk critical temperature $T_c$
in a cubic geometry with a linear size $L$ below the upper critical
dimension $d=4$. The property of finite-size
scaling means that, for sufficiently large $L$ and small $t$,  $\chi$ has the
asymptotic form
\be
\label{gleichung0} \chi (t, L) \; = \; \chi (t, \infty) f_\chi (L/\xi)
\ee
where $\chi (t, \infty) = A_\chi t^{- \gamma}$ is the bulk
susceptibilty and $\xi = \xi_0 t^{- \nu}$ is the bulk correlation length. An
appealing feature of finite-size scaling is universality which
means that all nonuniversal parameters of the confined system can
be absorbed entirely in the bulk amplitude $A_\chi$ and
in the {\it bulk} correlation length $\xi$,
thus finite-size scaling functions such as $f_\chi (x)$ are
expected to be independent of nonuniversal details (such as the
lattice structure, the lattice spacing and the magnitude of
coupling constants). This implies that the
amplitude $B_\chi$ of the small-$x$ behavior $f (x) = B_\chi
x^{\gamma / \nu}$ for $T \to T_c$ at fixed $L$ is also universal.
The specific shape and the amplitude $B_\chi$ of
such scaling functions do, of course, depend on the geometry and
on the kind of boundary conditions. A central prediction of
finite-size scaling is the size dependence at the bulk critical
temperature $T_c$
\be
\label{gleichung-2} \chi (0, L) \; = \; A_\chi \; \xi_0^{- \gamma/\nu}
\; B_\chi \; L^{\gamma / \nu}
\ee
with the {\it bulk} critical exponent $\gamma / \nu$ and the {\it
universal} amplitude $B_\chi$. For purely periodic boundary
conditions and short-range interactions, universal finite-size
scaling in the sense of Eqs. (\ref{gleichung0}) and
(\ref{gleichung-2}) has been largely confirmed, except for the
{\it nonuniversal} exponential behavior in the region $L >> \xi$
which has recently been shown
\cite{chen-dohm-1999,chen-dohm-99,chen-dohm-2000,chen-dohm-cond}
to depend on the lattice structure for lattice models and on the
cutoff procedure for continuum models.

Of particular interest are nonperiodic boundary conditions which
are relevant for real systems. For example, for the superfluid
transition of $^4$He, Dirichlet boundary conditions of field
theories are believed to be fairly realistic \cite{dohm-1993}. For
this system, however, accurate experiments have detected
nonscaling finite-size \cite{dohm-1993,gasparini,lipa,fig-2} and
surface \cite{kuehn} effects that are as yet unexplained.
Furthermore there exist unexplained finite-size effects in the XY
model with nonperiodic boundary conditions as detected by Monte
Carlo simulations \cite{schultka}.

On the theoretical side, the true conditions for the validity of
universal finite-size scaling for systems with nonperiodic
boundary conditions are not established. This includes the
important case of free boundary conditions for lattice models
which are believed to be asymptotically equivalent to Dirichlet
boundary conditions of continuum models. It is known that
universal finite-size scaling in the sense of Eq. (1) fails for
the mean spherical model in film geometry with free boundary
conditions in $d = 3$ and $d = 4$ dimensions
\cite{barber,barber-1973,barber-1974,dantchev1}, and similarly for
the ideal Bose gas with Dirichlet boundary conditions for cubic
and film geometry \cite{barber,barber-2,barber-3,singh-1985}. In
these models the bulk correlation length could not be used as the
only reference length and nonscaling finite-size effects were
incorporated in nonuniversal shifts of the temperature variable
\cite{barber-1973,barber-1974,dantchev1,barber-2,barber-3,singh-1985}.
Logarithmic nonscaling finite-size effects that depend on the
lattice spacing exist also in Gaussian interface models
\cite{gelfand} as well as in other models \cite{privman}.

On the other hand, universal amplitude ratios have been found for
critical systems contained in parallel plates with nonperiodic
boundary conditions \cite{cardy-1990}. Furthermore,
field-theoretic renormalization-group calculations have apparently
confirmed the validity of universal finite-size scaling within the
$\varphi^4$ field theory with Dirichlet boundary conditions:
Universal finite-size amplitude ratios \cite{eisenriegler} and
universal finite-size contributions to the free energy density and
to the critical Casimir force were calculated both in the Gaussian
(one-loop) approximation \cite{symanzik,krech-1992,sutter} as well
as in two-loop order \cite{krech-1992}. Universal finite-size
scaling functions have also been predicted for the specific heat
and the superfluid density in the presence of Dirichlet boundary
conditions \cite{dohm-1993,schmolke}. Related field-theoretic
predictions have also been presented for surface quantities
\cite{frank-dohm,mohr-dohm-2000}. In these papers \cite
{eisenriegler,symanzik,krech-1992,sutter,schmolke,frank-dohm,mohr-dohm-2000},
however, the method of dimensional regularization was employed
which neglects lattice and cutoff effects.

Recent work on finite-size effects
\cite{chen-dohm-1999,chen-dohm-99,chen-dohm-2000,chen-dohm-cond,
chen-dohm-1998,chen-dohm-1998-a,chen-dohm-1998-b,chen-dohm-1998-c,
chen-dohm-2001} has demonstrated that general
renormalization-group arguments are not sufficient to prove the
validity of universal finite-size scaling and that cutoff and
lattice effects are nonnegligible for confined systems with
periodic boundary conditions. Clearly these investigations need to
be extended to the case of nonperiodic boundary conditions.

The corresponding analytic calculations, at finite cutoff and at
finite lattice spacing, become quite difficult within the
$\varphi^4$ theory beyond the lowest order. Before embarking on
such an ambitious project it is of course necessary to first
examine the lowest-order case under the simplest nontrivial
conditions, i.e., with free (or Dirichlet) boundary conditions in
only one direction. Therefore, as a first step, we consider the
exactly solvable Gaussian model with short-range interaction on a
simple-cubic lattice with a lattice constant $\tilde a$ for a
finite rectangular $L \times \widetilde L^{d-1}$ box geometry with
free boundary conditions in one direction and periodic boundary
conditions in $d-1$ directions. Even at the Gaussian level, the
analytic calculations at finite lattice spacing in the range $2 <
d < 4$ turn out to be nontrivial.

For the specific heat and the susceptibility of the Gaussian model
we find full agreement with universal finite-size scaling. With
regard to the singular part of the free energy we find that the
finite-size scaling form is indeed valid for $d < 3$ and $d
> 3$ but logarithmic deviations from finite-size scaling occur at
$d = 3$ where the critical exponent $1 - \alpha -\nu = (d-3)/2$ of
the surface energy density vanishes. In order to describe the
logarithmic $d = 3$ behavior it is necessary to keep the lattice
spacing finite. We find that the same logarithmic deviations from
finite-size scaling exist in the continuum version of the Gaussian
model with Dirichlet boundary conditions provided that a finite
cutoff is used. This implies that the method of dimensional
regularization at infinite cutoff is not capable of correctly
describing the $d = 3$ behavior of the singular part of the free
energy density and of the energy density since it yields
unphysical divergences of these quantities in the form of a pole
term $\sim (d-3)^{-1}$ \cite{dohm-1989}. As discussed in Sect.
III. H, the dimension $d = 3$ can be considered as an upper
borderline dimension $d^*$ of the Gaussian model with free
boundary conditions above which lattice and cutoff effects become
nonnegligible for the surface energy density.

For $d > 3$ we find that the surface energy density
$U_{surface}(t)$ of the Gaussian model with free boundary
conditions has a cusp-like singularity at bulk $T_c$ as $T_c$ is
approached from above. For the lattice model at finite lattice
spacing $\tilde a$ the height of the cusp is
\be
\label{gleichung-3)} \lim_{t \to 0 +} U_{surface} (t) \; = \;
U_{surface} (0) \; = \; T_c \; \xi_0^{-2} \; \tilde a^{3-d} \;
\widetilde B_d
\ee
with
\be
\label{gleichung-3} \widetilde B_d = \frac{1}{8}
\int\limits_0^\infty dy \left\{ \left[1 + e^{-4y} - 2 \; e^{-2y} \; I_0
(2y) \right]  \left[ e^{-2y} \; I_0 (2y) \right]^{d-1} \right\}
> 0
\ee
where $I_0 (z)$ is the Bessel function of order zero. The
temperature dependent part of $U_{surface}(t)$ has a universal
scaling form $\sim \xi^{3-d}$ but it vanishes at $T_c$ and is
subleading compared to the nonuniversal finite regular part, Eq.
(\ref{gleichung-3)}), at $T_c$. The latter part yields a leading
nonscaling contribution $2 \; U_{surface} (0) / L$ to the total
energy density. These results remain valid also for the Gaussian
model in film geometry ($ \widetilde L \rightarrow \infty$) with
free boundary conditions.

In a second step we analyze the exactly solvable mean spherical
model with the same boundary conditions. Previously this model has
been studied for film geometry at integer dimensions $d=3,4,5,...$
\cite{barber-1973,barber-1974}. Here we extend this analysis to
{\it continuous} dimensions in the range $2 < d < 4$ and consider
both film and box geometry. This reveals $d = 3$ as a borderline
dimension between a universal scaling $(d < 3)$ and a nonuniversal
nonscaling $(d \geq 3)$ regime. In this paper we calculate the
nonscaling effects for $3 \leq d < 4$ as well as the analytic form
of the universal finite-size scaling function $f_\chi (L/\xi)$,
Eq. (\ref{gleichung0}), of the susceptibility for $2 < d < 3$
including the amplitude $B(s)$ of the scaling result, Eq.
(\ref{gleichung-2}) with $\gamma/\nu = 2 \;$,
\be
\label{gleichung-4} \chi (0, L) \; = \; B(s) L^2 \; \; , \; \; d <
3
\ee
at arbitrary shape factor $s = L/\widetilde L \geq 0$. The
amplitude $B (s)$ is shown to diverge for $d \to 3$.

The mean spherical model can be considered as a Gaussian model
with a constraint where the constraint can be expressed in terms
of the Gaussian energy density. Our results for the latter
quantity explain the origin of logarithmic nonscaling terms in
thermodynamic quantities of the mean spherical model at $d = 3$
and of power-law violations of finite-size scaling for $3 < d <
4$. While previous work suggested the existence of only {\it
logarithmic} deviations from finite-size scaling in $d = 3$
dimensions
\cite{barber,barber-1973,barber-1974,dantchev1,barber-2,barber-3,singh-1985}
we find, quite unexpectedly, a {\it non-logarithmic} violation of
the scaling prediction, Eq. (\ref{gleichung-4}), for the size
dependence of the susceptibility at bulk $T_c$ in $d = 3$
dimensions
\be
\label{gleichung-1a} \chi (0, L) \; = \; B_{film} \; \tilde a^{-1}
L^3
\ee
for film geometry whereas for box geometry, at fixed finite shape factor
$s = L/\widetilde L > 0$, we find the expected logarithmic deviation
from scaling
\be
\label{gleichung4a} \chi (0, L) = B_{box} (s) L^2 \ln
(L/\widetilde a) \; .
\ee
As will be shown in detail in Sect. IV. C, the special result of
Eq. (\ref{gleichung-1a}) for film geometry at $d = 3$ is due to
the simultaneous appearance of two logarithmic effects at $d = 3$
where two borderline dimensions coincide : it is a combined effect
of the logarithmic {\it surface} term of the Gaussian model at the
(upper) borderline dimension $d^* = 3$ where the exponent $1 -
\alpha - \nu$ vanishes and of a logarithmic {\it finite-size} term
arising from the mode continuum of the film system just at the
(lower) borderline dimension $d_l = 3$ at which the film critical
temperature vanishes in accordance with the Mermin-Wagner theorem
\cite{mermin}. Most striking is the {\it discontinuous} change of
the exponent 2 of the power law $\chi_{film} = B (0) L^2$ for $d <
3$, Eq.(\ref{gleichung-4}), to 3 of the power law $\chi_{film} =
B_{film} \; \tilde a^{-1} L^3$ for $d=3$, Eq.
(\ref{gleichung-1a}).

The result of Eq. (\ref{gleichung-1a}) is not contained in the
work of Barber and Fisher \cite{barber-1973} who calculated $\chi$
for film geometry in $d = 3$ dimensions only for $T \geq
\widetilde T (L)$ where $\widetilde T (L) > T_c$ is some
temperature that they called "quasicritical". Our $d = 3$ result
for $\chi (t, L)$ covers the entire critical region $T \geq T_c$
including the regime $T \geq \widetilde T (L)$. In the latter
regime, the explicit form of our result is at variance with the
simpler form of Barber and Fisher.

For box geometry in $3 < d < 4$ dimensions we find a power-law
violation of scaling at $T_c$
\be
\label{gleichung5} \chi (0, L) \; = \; B_{box} (s,d) \tilde
a^{3-d} \; L^{d-1}
\ee
where the amplitude $B_{box} (s,d)$ is proportional to the
amplitude $\widetilde B_d$, Eq. (\ref{gleichung-3}), of the cusp
of the Gaussian surface energy density. A nonscaling form is also
found for the temperature dependence of the surface susceptibility
for $3 < d < 4$ above $T_c$,
\be
\label{gleichung-7} \chi_{surface}  \; = \; \widetilde A_{surface}
\tilde a^{3-d} \; \xi^d \sim t^{- d/(d-2)}
\ee
with $\xi \sim t^{- \nu} \; , \nu = (d - 2)^{-1}\; $, whereas the
scaling form, Eq. (\ref{gleichung0}), would imply $\chi_{surface}
\sim O (\chi_b \; \xi) \sim t^{-3 / (d-2)}$ for the mean spherical
model. Again the amplitude $\widetilde A_{surface}$ in Eq.
(\ref{gleichung-7}) is proportional to $\widetilde B_d$.

For film geometry in $d > 3$ dimensions we find an anomalous enhancement
of the film critical temperature $T_{c,d} (L)$ {\it above} the bulk
critical temperature $T_{c,d} (\infty)$. A corresponding shift was first
found for $d \geq 4$ by Barber and Fisher \cite{barber-1973}.
This enhancement is most naturally expressed in terms of the
dimensionless parameter $2J \beta_{c,d} (L) = 2J [k_B T_{c,d} (L)]^{-1}$
where $J$ is the nearest-neighbor coupling. The result is for $d >
3$
\be
\label{gleichung7} 2 J \; [\beta_{c,d} \; (\infty) \; - \; \beta_{c,d} \; (L) ]
\; = \; 4 \widetilde B_d \;
\tilde a/L \; - \; \widetilde C_d \; (\tilde a/L)^{d-2}
\; + \; O(\tilde a^{d/2} \; L^{- d/2})
\ee
with the nonuniversal amplitude $\widetilde B_d$, Eq.
(\ref{gleichung-3}), and with a universal amplitude $\widetilde
C_d > 0$. Eq. (\ref{gleichung7}) implies $T_{c,d} (L) > T_{c,d}
(\infty)$ for large $L \gg \tilde a$. The leading term $\sim
L^{-1}$ in Eq. (\ref{gleichung7}) has a nonscaling $L$ dependence
whereas the subleading universal term has the scaling $L$
dependence $\sim L^{1/\nu}$.

In summary we see that both the anomalous nonscaling enhancement
of $T_{c,d} (L)$, Eq. (\ref{gleichung7}), and the power-law
violations, Eqs. (\ref{gleichung5}) and (\ref{gleichung-7}), for
$d > 3$ can be traced back to the same amplitude $\widetilde B_d$,
Eq. (\ref{gleichung-3}), of the nonscaling cusp of the Gaussian
model. Thus the analysis of the Gaussian model provides a better
understanding of the origin of the power-law nonscaling
finite-size effects in the mean spherical model for $d > 3$, and,
for box geometry, of the logarithmic deviations at the Gaussian
upper borderline dimension $d^* = 3$. For film geometry, however,
the Gaussian logarithmic effect at $d^* = 3$ is enhanced by a
second logarithmic effect due to the lower borderline dimension
$d_l = 3$ (where the film critical temperature vanishes), which
then yields the power law Eq. (\ref{gleichung-1a}).

We point out that all nonscaling effects are tied to the finite
lattice constant $\tilde a > 0$, as seen explicitly in Eqs.
(\ref{gleichung-3)}) and (\ref{gleichung-1a}) -
(\ref{gleichung7}). We expect that similar effects exist in the
ideal Bose gas with Dirichlet boundary conditions
\cite{barber-2,barber-3,singh-1985} with a finite cutoff (even if
a smooth cutoff is used) . These effects are not captured by the
standard method of dimensional regularization. It remains to be
seen whether the mechanism for nonscaling finite-size effects in
the mean spherical model and the ideal Bose gas is an artifact
restricted to these models or whether some of these features are
of more general significance. This question is of particular
interest below $T_c$ where an explanation of the pronounced
nonscaling finite-size effects in $^4$He remain to be a challenge
for future research.

In Section II we summarize the predictions implied by the finite-size
scaling hypothesis. Section III contains the detailed results for the
finite-size effects in the Gaussian lattice model with free boundary
conditions and in the Gaussian continuum model with Dirichlet boundary
conditions. In Sect. IV we analyze the consequences of our results for
the mean spherical model with free boundary conditions. The
derivation of our results is presented in several Appendices.

\newpage

\section*{II. Finite-size scaling predictions}

In the subsequent sections we shall present exact results for the
finite-size effects on the free energy density, energy density,
specific heat and susceptibility of lattice models in a
rectangular $L \times \widetilde L^{d-1}$ box geometry with free
boundary conditions in the direction of size $L$ and periodic
boundary conditions in the $d - 1$ directions of size $\widetilde
L$. For the sake of clarity we first summarize the predictions
implied by the finite-size scaling hypothesis which, for this
geometry and these boundary conditions, have not yet been
formulated explicitly in the literature. We denote the critical
temperature of the $d$-dimensional bulk ($L \rightarrow \infty,
\widetilde L \to \infty$) system by $T_{c,d}$. In the limit
$\widetilde L \to \infty$ at fixed $L$, the box becomes a film of
thickness $L$ which may have its own critical temperature $T_{c,
d} (L) \neq T_{c, d} \equiv T_{c,d} (\infty)$. In general one
expects $T_{c,d} (L) < T_{c,d}$ but it turns out (see Sect. IV,
see also Ref. \cite{barber-1973}) that for the mean spherical
model with $d
> 3$ the film critical temperature $T_{c,d} (L)$ exceeds the bulk
critical temperature $T_{c,d}$. For simplicity, in this Section,
we assume a $d$ dimensional box with a $\it finite$ shape factor
$L / \widetilde L > 0$ and confine ourselves to $T \geq T_{c,d}$.

First we consider the free energy density $f (t, L, \widetilde L)$
(in units of $k_B T$) at the reduced temperature $t = (T-T_{c,d}) / T_{c,d} \geq 0$
and at vanishing external field. It is expected that, for small $t$,
$f$ can be decomposed into a singular and a "nonsingular" part
\cite{privman-fisher,privman1}
\be
\label{gleichung1} f (t, L, \widetilde L) = f_s (t, L, \widetilde L) \; + \; f_{ns}
(t, L, \widetilde L)
\ee
where $f_{ns} (t, L, \widetilde L)$ has a regular $t$ dependence.
In the bulk limit the corresponding decomposition is
\be
\label{gleichung1a} f_b (t) \equiv f (t, \infty, \infty) \; = \; f_{bs}
(t) \; + \; f_0 (t)
\ee
where the regular part $f_0 (t) \equiv f_{ns} (t, \infty, \infty)$ can be
identified unambiguously. For systems with short-range interactions below
the upper critical dimension $d = 4$ and for large $L, \widetilde
L$ and $\xi$ it is expected that the singular part $f_s (t, L, \widetilde L)$
has the finite-size scaling form \cite{fisher,privman-fisher}
\be
\label{gleichung2} f_s (t, L, \widetilde L) \; = \; L^{-d} \; {\cal F} \;
(L/\xi, L/\widetilde L)
\ee
where $\xi (t) = \xi_0 t^{-\nu}$ is the (second-moment) correlation
length of the $d$-dimensional bulk system. For a given shape factor
$s = L/\widetilde L$, the scaling function ${\cal F} (x, s)$ is expected
to be universal. More specifically, the singular and nonsingular parts of
the free energy density are expected to have the asymptotic (small $t$,
large $L,$ large $\widetilde L$) form \cite
{fisher,privman,privman-fisher,privman1}
\be
\label{gleichung2a} f_s (t, L, \widetilde L) = R^+_\xi \; \xi^{-d} + 2 \; A_{surface}^+
\; \xi^{1-d} \; L^{-1} \; + \; L^{-d} \; {\cal G} (L/\xi, L/\widetilde L)
\ee
and
\be
\label{gleichung3} f_{ns} (t, L, \widetilde L) \; = \; f_0 (t) \; + \; 2 \Psi_1
(t) / L
\ee
with a universal bulk amplitude $R_\xi^+$ and a universal surface
amplitude $A^+_{surface}$, and with a universal finite-size
part ${\cal G} (L/\xi, L/\widetilde L)$ of the scaling function
\be
\label{gleichung6a} {\cal F}(L/\xi, L/\widetilde L) = R^+_\xi (L/\xi)^d \; + \; 2
\; A^+_{surface} (L/\xi)^{d-1} \; + \; {\cal G} (L/\xi, L/\widetilde L) \; .
\ee
Eqs. (\ref{gleichung2a}) - (\ref{gleichung6a}) imply that there
exists the surface free energy
\be
\label{gleichung7a} f_{surface} (t)
= \lim_{L \rightarrow \infty} \left\{\left[f (t, L,
\widetilde L) \; - \; f_b
(t)\right] \frac{L}{2} \right\} \;
= A_{surface}^+ \; \xi^{1-d}
\; + \; \Psi_1 (t)
\ee
with a universal amplitude $A^+_{surface}$ of the singular part.
The nonsingular surface contribution $\Psi_1 (t)$ is a {\it
regular} function of $t$. Nonasymptotic Wegner \cite{wegner}
corrections to scaling are neglected in Eqs. (\ref{gleichung2}),
(\ref{gleichung2a}) and (\ref{gleichung6a}). The phenomenological
finite-size scaling theory does not make specific predictions
about the dependence on $L$ and $\widetilde L$ of higher-order
terms in Eq. (\ref{gleichung3}).

Eqs. (\ref{gleichung1}) - (\ref{gleichung7a}) are expected to hold
also for continuum models with Dirichlet boundary conditions in
one direction, with the same universal quantities as for free
boundary conditions of lattice systems. As noted in the
Introduction, however, there exist {\it nonuniversal} exponential
terms in the regime $L \gg \xi, \; \widetilde L \gg \xi$, where
the lattice-dependent and cutoff-dependent exponential correlation
length \cite{chen-dohm-2000,chen-dohm-cond,fisher-burford} becomes
the appropriate reference length.

Although the energy density (internal energy per unit volume)
divided by $k_B$
\be
\label{gleichung6} U (t, L, \widetilde L) = \; - \; T^2 \; \frac{\partial f (t,
L, \widetilde L)}{\partial T} \; .
\ee
is completely determined by the free energy density $f (t, L, \widetilde L)$ it
turns out that a separate discussion of the energy density is warranted
because of its important role played in the mean spherical model
in Sect. IV. From Eqs. (\ref{gleichung1}) - (\ref{gleichung3}) one
obtains the prediction
\be
\label{gleichung8} U (t, L, \widetilde L) = U_s (t, L, \widetilde L)
\; + U_{ns} (t, L, \widetilde L)
\ee
where the singular part
\be
\label{gleichung9} U_s (t, L, \widetilde L) = T_c \; \xi_0^{- 1 / \nu} \; L^{-
(1-\alpha) / \nu}
\; {\cal U} (L/\xi, L/\widetilde L)
\ee
has the universal scaling function
\be
\label{gleichung10} {\cal U} (x,s) = - \nu x^{1 - 1/\nu} \; \partial
{\cal F} (x,s) / \partial x \qquad \qquad \qquad \qquad \qquad
\qquad \qquad
\ee
\be
\label{gleichung11a} = - d \nu \; R_\xi^+ \; x^{d - 1/\nu} - 2 (d-1) \nu
\; A_{surface}^+ \; x^{d-1-1/\nu} - \nu x^{1-1/\nu} \; \partial {\cal{G}} (x,s) /
\partial x
\ee
and where the leading nonsingular part
\be
\label{gleichung11} U_{ns} (t, L, \widetilde L) \; = \; U_0 (t) \; + \; 2 U_1 (t)
/ L
\ee
has a regular $t$ dependence with $U_0 (t) = - T^2 \partial f_0
(t) / \partial T$ and
\be
\label{gleichung12} U_1 (t) = - T^2 \partial \Psi_1 (t) /
\partial T.
\ee
For the surface energy density, Eq. (\ref{gleichung2a}) implies asymptotically
\be
\label{gleichung16}
U_{surface} (t) = - T^2 \; \partial f_{surface} (t) / \partial T
\ee
\be
\label{gleichung13} & = & U_1 (t) \; - \; T_c \; \xi_0^{- 1 / \nu}
\; (d-1) \nu \; A_{surface}^+ \; \xi^{- (1 - \alpha - \nu) / \nu}
 \; .
\ee
In Eqs. (\ref{gleichung9}) and (\ref{gleichung13}) we have used
the hyperscaling relation
\be
\label{gleichung14} d \nu = 2 - \alpha \; .
\ee
These scaling predictions have been confirmed by several
field-theoretic re\-nor\-ma\-li\-za\-tion-group (RG) calculations of
$f_s (t, L, \infty)$ \cite{krech-1992} and of $U_{surface}$
\cite{eisenriegler,dohm-1989,goldschmidt,mikheev}
based on the $\varphi^4$ continuum Hamiltonian
with Dirichlet boundary conditions for the field $\varphi
({\bf x})$. All calculations, however, were carried out within the
dimensional regularization scheme which neglects cutoff effects.
As pointed out by Dohm \cite{dohm-1993,dohm-1989}, an unresolved feature
of the dimensionally regularized perturbative results for $U_{surface}$
\cite{eisenriegler,krech-1992,dohm-1989,goldschmidt,mikheev} is a
pole term $\sim (d - 3)^{-1}$ that diverges in three dimensions.

We note that the critical exponent of $U_{surface}(t)$
\be
\label{gleichung15} 1 - \alpha - \nu = (d-1) \nu - 1
\ee
is positive for ordinary critical points of the O$(n)$ universality
class with $d > 2$ which implies a
{\it finite} critical value $U_{surface} (0) = U_1 (0)$. By
contrast, for the Gaussian model, $1 - \alpha - \nu = (d - 3) / 2$
is positive only for $d > 3$, thus $U_{surface} (t)$ diverges
for $t \to 0$ in $d \leq 3$ dimensions (see Sect. III).

We shall also consider the specific heat (divided by $k_B$)
\be
\label{gleichung16a} C (t, L, \widetilde L) &=& \frac{\partial U (t, L, \widetilde L)}
{\partial T} \; = \;  C_s (t, L, \widetilde L) \; + \; C_{ns} (t, L, \widetilde L)
\; .
\ee
From Eqs. (\ref{gleichung8}) - (\ref{gleichung11}) we obtain the
predictions
\be
\label{gleichung17a} C_s (t, L, \widetilde L) &=& \xi_0^{- 2/\nu} \; L^{\alpha /
\nu} \; {\cal{C}} (L / \xi, L/\widetilde L)
\ee
and
\be
\label{gleichung20a} C_{ns} (t, L, \widetilde L) &=& \partial U_0 (t) / \partial
t \; + \; 2 L^{-1} \partial U_1 (t) / \partial t
\ee
with the universal scaling function
\be
\label{gleichung17b} {\cal{C}} (x,s) &=& \nu \; x^{1 - 1/\nu} \;
\partial {\cal{U}} (x,s) / \partial x \; .
\ee
The scaling structure implies that the surface specific heat
$C_{surface} (t) \; = \;  \partial U_{surface} (t) / \partial T$ has a
divergent singular part,
\be
\label{gleichung21a} C_{surface} (t) = \xi_0^{1-d} A_{C,surface}^+\;
t^{- \alpha_s } \; + \; \partial U_1 (t) /
\partial T
\ee
with the surface scaling exponent
\be
\label{gleichung21b}\alpha_s = \alpha \; + \; \nu
\ee
and with a universal amplitude
\be
\label{gleichung22b} A_{C,surface}^+ \; = \; - \; (1 - \alpha - \nu)
(d - 1) \; \nu \; A_{surface}^+ \; .
\ee
Finally we recall the prediction for the asymptotic scaling form
of the susceptibility
\be
\label{gleichung27-a} \chi (t, L, \widetilde L) = \chi_b (t) f_\chi
(L/\xi, L/\widetilde L)
\ee
according to Eq. (1) where $\chi_b (t) = \chi (t, \infty) = A_\chi \; t^{- \gamma}$ is
the bulk susceptibility. For $L \gg \xi, \; \widetilde L \gg \xi$, the
scaling function is expected to have the expansion
\be
\label{gleichung22c}f_\chi (L/\xi, L/\widetilde L) = 1 +  c_\chi \xi/L
+ O (\xi^2/L^2, e^{- \widetilde L/\xi})
\ee
with the universal coefficient $c_\chi$. For $t > 0$ this implies
\be
\label{gleichung22c0}
\chi_{surface}(t) =
\lim_{L \rightarrow \infty} \left\{\left[\chi (t, L, \widetilde L) \; - \; \chi_b
(t)\right] \frac{L}{2} \right\} \;
= \; A^+_{\chi,surface} t^{-\gamma_s}
\ee
with the surface scaling exponent
\be
\label{gleichung22d}
\gamma_s \; = \; \gamma \; + \; \nu
\ee
and with the surface amplitude
\be
\label{gleichung22e}
A^+_{\chi,surface} \; = \; \frac{1}{2} \; A_\chi \; \xi_0 \; c_\chi \; .
\ee
For $T \to T_{c,d}$ the small $L/\xi$ behavior of the scaling function
is expected to be
\be
\label{gleichung31a} f_\chi (L / \xi, L/\widetilde L) \sim B_\chi
(L / \widetilde L) \; (L / \xi)^{\gamma/\nu}
\ee
with a finite universal amplitude $B_\chi (L / \widetilde L) > 0$
which implies
\be
\label{gleichung32a} \chi (0, L, \widetilde L) = A_\chi \; \xi_0^{-
\gamma / \nu} B_\chi (L / \widetilde L) \; L^{\gamma / \nu} \;
.
\ee
In the following we examine the range of validity of these
predictions for the exactly solvable Gaussian and mean spherical
models in $2 < d < 4$ dimensions.

\newpage

\section*{III. Gaussian lattice model with free boundary conditions}

\section*{A. Lattice Hamiltonian}

We consider $N$ continuous scalar variables $\varphi_j, - \infty \leq
\varphi_j \leq \infty$, on the lattice points ${\bf x}_j$ of a
simple-cubic lattice with a lattice spacing $\tilde a$ in a finite
rectangular $L \times \widetilde L^{d-1}$ box of volume $V = L \widetilde
L^{d-1} = N \tilde a^d$. We assume a Gaussian statistical weight $\sim \exp (-H)$ with
the lattice Hamiltonian
\be
\label{gleichung18} H = \tilde a^d \left[\sum_i \frac{r_0}{2}
\varphi_j^2 \; + \; \sum_{< i j >} \; \frac{J}{2 \tilde a^2} \;
(\varphi_i - \varphi_j)^2 \right]
\ee
with a nearest-neighbor coupling $J > 0$. The factor $(k_BT)^{-1}$ is
absorbed in $H$. The dimensionless partition function is
\be
\label{gleichung17} Z = \left[\prod_j \int\limits^\infty_{- \infty} \;
\frac{d \varphi_j}{\tilde a^{1-d/2}}\right] \; \exp (- H) \; .
\ee
In the bulk limit $\widetilde L \rightarrow \infty, L \rightarrow \infty$,
this model has a critical point at $r_0=0$ for arbitrary $d>0$. We assume
that the temperature $T$ enters only through
\be
\label{gleichung18a} r_0 \; = \; a_0 \; \frac{T - T_c}{T_c} \qquad ,
a_0 > 0 \; .
\ee
A serious shortcoming
of this model is the fact that it has no low temperature phase,
i.e., no bulk limit exists for $r_0 < 0$. Nevertheless there exist
nontrivial finite-size effects for $r_0 \geq 0$, as we shall see.

We assume free boundary conditions in the $d$-th
("vertical") direction and periodic boundary conditions in the $d-1$
("horizontal") directions. The $d-1$ "horizontal" coordinates and the
"vertical" coordinate of the lattice points ${\bf x}_j = {(\bf y}_j, z_j)$
are denoted by ${\bf y}_j$ and $z_j$, respectively. The "bottom" and "top"
surfaces perpendicular to the vertical direction have the coordinates
$z_j = \tilde a$ and $z_j = L$, respectively, thus we have $L/\tilde a$
layers of fluctuating variables. The variables in the bottom
and top surfaces have only one neighboring layer. This is equivalent to
assuming Dirichlet boundary conditions $(\varphi_j = 0)$ in the (fictitious)
layers $z_j = 0$ below the bottom surface and $z_j = L + \tilde a$
above the top surface.
The variables $\varphi_j$ can be represented as
\be
\label{gleichung18b} \varphi_j = \widetilde L^{-(d-1)} \; (L + \tilde a)^{-1} \;
\sum_{{\bf k},p} \; \hat \varphi_{{\bf k}, p} \; \; \exp (i\;{\bf k \cdot y}_j)
\; \; \sqrt{2} \; \sin (p\;z_j)\;
\ee
with the Fourier amplitudes
\be
\label{gleichung18c} \hat \varphi_{{\bf k}, p} = \tilde a^d \sum_j
\varphi_j \; \exp (- i {\bf k}\cdot {\bf y}_j) \sin (p z_j) \; .
\ee
The sum $\sum_{{\bf k}, p}$ runs over ($d-$1) dimensional
{\bf k} vectors with components $k_i = 2 \pi m_i / \widetilde L, \; i = 1, ..., \;
d-1$ with integers $m_i = 0, \; \pm 1, \; \pm 2, ..., $ in the range $- \pi /\tilde
a \leq k_i < \pi / \tilde a$ and over wave numbers $p = \pi n
/ (L+\tilde a), \; n = 1,2,..., L/\tilde a$ in the range $0 < p < \pi / \tilde
a$. We see that, for $L/\tilde a$ layers with free boundary conditions, the natural
unit wave number in $p$ space is $\pi/(L+ \tilde a)$ rather than $\pi/L$.
For each given $p$, there are $(\widetilde L/\tilde a)^{d-1}$ variables
$\hat \varphi_{{\bf k}, p}$. Eq. (\ref{gleichung18b}) implies $\varphi_j = 0$
at $z_j = 0$ and $\varphi_j = 0$ at $z_j = L+\tilde a$ for arbitrary ${\bf y}_i$,
thus we have a total number of $N = (L/\tilde a) (\widetilde L / \tilde a)^{d-1}$
variables $\hat \varphi_{{\bf k}, p}$. Substituting Eq. (\ref{gleichung18b})
into Eq. (\ref{gleichung18}) yields the diagonalized Hamiltonian
\be
\label{gleichung18c} H = \frac{1}{2} \;\; \widetilde L^{-(d-1)} \; (L +
\tilde a)^{-1} \; \sum_{{\bf k},p} \;
\left(r_0 + J_{{\bf k}, d-1} + J_p \right)\; \hat \varphi_{{\bf k}, p} \; \; \hat
\varphi_{-{\bf k}, p}
\ee
with
\be
\label{gleichung21} J_{{\bf k}, d-1} = \frac{4 J}{\tilde a^2}
\; \sum^{d-1}_{i = 1} \; \left[1 - \cos \; (k_i \tilde a)\right] \; ,
\ee
\be
\label{gleichung22} J_p \; = \; \frac{4J}{\tilde a^2} \; \left[1
- \cos (p \tilde a) \right] \;.
\ee
The Jacobian of the linear transformation $\varphi_j \rightarrow
\hat \varphi_{{\bf k},p}$ of Eq. (\ref{gleichung18b}) is
\be
\label{gleichung25a} \left| \frac{\partial \varphi_j}{\partial \hat
\varphi_{{\bf k},p}} \right| \; = \; \left(\widetilde L^{d-1} \; L
\right)^{-N} \; N^{N/2} \; .
\ee
Using Eqs. (\ref{gleichung17}), (\ref{gleichung18b}),
(\ref{gleichung18c}) and (\ref{gleichung25a}) we obtain the free
energy density divided by $k_B T$
\be
\label{gleichung24} f (t, L, \widetilde L)&=& - V^{-1} \; \ln Z \nonumber\\
                    &=& - \frac{1}{2} \; \tilde a^{-d} \left[\ln
                      \pi\; + \; (\widetilde L/\tilde a)^{1-d}
                      \; \ln 2
                      \right] \nonumber\\
                    &+& \frac{1}{2} \; \widetilde L^{- (d-1)} \;
                      L^{-1} \sum_{{\bf k},p} \; \ln
                      \left[\left(r_0 + J_{{\bf k}, d-1} + J_p
                      \right) \tilde a^2 \right]
                      \; .
\ee
In all calculations of this Section we shall keep the lattice spacing
$\tilde a$ finite.

In the following we shall also consider film geometry (bulk limit in the
$d-1$ horizontal directions). In Eq. (\ref{gleichung24}), this
corresponds to the replacement $\widetilde L^{-(d-1)} \sum_{{\bf k},p}
\rightarrow \sum_p \int_{\bf k}$ where $\int_{\bf k}
\equiv (2 \pi)^{1-d} \int d^{d-1} k$ with $|k_i | \leq \pi / \tilde a,
i = 1,2,..., d - 1$, hence
\be
\label{gleichung25b} f (t, L, \infty) = - \frac {1} {2} \; \tilde a^{-d}
\; {\rm ln} \;  \pi \; + \; \frac{1}{2} \; L^{-1} \; \sum_p
\int\limits_{\bf k} {\rm ln} \left[(r_0 + J_{{\bf k}, d-1} + J_p)
\tilde a^2 \right] \; .
\ee
A simplifying (but unrealistic) feature of the Gaussian model is that the
critical point of the film system of finite thickness $L$ is also determined
by $r_0=0$, i.e., it remains unshifted compared to the bulk critical point
for all $d$. This differs from the case of the spherical model to be
discussed in Sect.IV.

\section*{B. Bulk properties}

First we briefly summarize some of the known bulk properties.
The square of the second-moment bulk correlation length $\xi$ above
$T_c$ is defined by
\be
\label{gleichung28a} \xi^2 \; = \; \lim_{L \rightarrow \infty} \;
\lim_{\widetilde L \rightarrow \infty} \; \frac{1}{2 d} \; \; \;
\frac{\sum_{i,j} ({\bf x}_i - {\bf x}_j)^2 < \varphi_i \; \varphi_j
> } {\sum_{i,j} < \varphi_i \; \varphi_j > } \; .
\ee
It is given by $\xi^2 = J_0 \; r_0^{-1}$ or
\be
\label{gleichung26} \xi \; = \; \xi_0 \; t^{- \nu}\; , \; \nu = 1/2
\ee
with
\be
\label{gleichung26a} \xi_0 = (J_0 / a_0)^{1/2} \qquad , \; J_0 = 2
J \; .
\ee
From Eq. (\ref{gleichung24}) we have the bulk free energy
density for $r_0 = a_0 t \geq 0$
\be
\label{gleichung25} f_b = - \; \frac{1}{2} \; \tilde a^{-d} \; \ln \pi
\; + \; \frac{1}{2} {\int\limits_{\bf q}}^{(d)} \; \ln \left[(r_0 + J_{{\bf q},
d}) \; \tilde a^2 \right]
\ee
where $\int_{\bf q}^{(d)} = (2 \pi)^{-d} \int d^d q$ with $|q_i|
\leq\pi / \tilde a, i = 1, ..., d$. Eqs. (\ref{gleichung24}) -
(\ref{gleichung25})
are defined for all
integer dimensions $d=1,2,...$ They can be extended to continuous $d$, as
usual, by means of analytic continuation via Euler's Gamma function $\Gamma$.
From Eq. (\ref{gleichung25})
one obtains the singular part of $f_b$ for $0 < d < 2$ and $2 < d < 4$
\be
\label{gleichung27} f_{bs} = R_\xi^+ \xi^{-d}
\ee
with the universal bulk amplitude
\be
\label{gleichung28} R_\xi^+ \; = \; - \; \frac{A_d}{d (4-d)} \; ,
\ee
\be
\label{gleichung29}A_d = \; \frac{\Gamma (3-d/2)}{2^{d-2}
\pi^{d/2} (d-2)} \; .
\ee
The regular part of $f_b$ reads for $0 < d < 2$ and $2 < d < 4$
\be
\label{gleichung30} f_0 \; = \; \tilde a^{-d} \left[\tilde c_1 +
r_0 \; \tilde a^2 \; \tilde c_2 + r_0^2 \; \tilde a^4 \; \tilde c_3 +
O (r_0^3 \; \tilde a^6) \right]
\ee
with $d$-dependent constants $\tilde c_i$ \;. The constants $
\tilde c_1, \; \tilde c_2$ and $\tilde c_3$ diverge for $d \to 0,
\; d \to 2$ and $d \rightarrow 4$, respectively, where $f_b$
attains a logarithmic dependence on $r_0 \tilde a^2$. The bulk
susceptibility is simply $\chi_b = r_0^{-1} = J_0^{-1} \xi^2$
which implies $A_\chi = a_0^{-1}$. The critical exponents are
\be
\label{gleichung18aa} \eta = 0, \qquad \gamma = 2 \nu = 1 \qquad \mbox {for all} \; d>0
\ee
and
\be
\label{gleichung19a}
\alpha = (4 - d) / 2 \qquad {\rm for} \; 0 < d \leq 4
\ee
above $T_c$, in agreement with the hyperscaling relation Eq.
(\ref{gleichung14}) for $d \leq 4$. The prefactor in Eq.
(\ref{gleichung-2}) is simply $A_\chi \xi_0^{- \gamma/\nu} =
J_0^{-1}$.

The second-moment bulk correlation length $\xi$ must be distinguished
from the "exponential" bulk correlation length $\xi_{\bf e}$ in the
direction of the unit vector ${\bf e} = ({\bf x}_i - {\bf x}_j) /
| {\bf x}_i - {\bf x}_j |$ which is defined via the large-distance behavior of
anisotropic bulk correlation function $G ({\bf x}_i - {\bf x}_j) =
< \varphi_i \; \varphi_j > \; $ \cite{fisher-burford}. For the
special case where ${\bf x} = (x, 0, 0, ...)$ is directed along
one of the cubic axes the correlation function decays exponentially
as \cite{chen-dohm-2000}
\be
\label{gleichung31b} G ({\bf x}) = \frac{\tilde a^{2-d}}{4J} \;
\left(\frac{\tilde a}{2 \pi | x |} \right)^{(d-1)/2} \left[\sinh
\left(\frac{\tilde a}{\xi_1}\right) \right]^{(d-3)/2} \nonumber\\
\times \; e^{-|x|/\xi_1} \; \left[1 + O (|x|^{-1}) \right]
\ee
with the exponential correlation length
\be
\label{gleichung31c} \xi_1 = \left[\frac{2}{\tilde a} \rm{arcsinh}
\left(\frac{\tilde {\it a}}{2 \xi} \right) \right]^{-1} \; .
\ee
We shall see that it is $\xi_1$ rather than $\xi$ that determines
the exponential part of the finite-size effects above $T_c$ not
only for periodic boundary conditions \cite{chen-dohm-2000} but
also for free boundary conditions.

\section*{C. Free energy density}

In Appendix A we derive from Eq. (\ref{gleichung24})
the size dependent free energy density for box geometry for large
$L/\tilde a$ at fixed $L/\xi \geq 0$ and at fixed $L/\widetilde L$
for $d > 1$
\be
\label{gleichung31} f(t,L, \widetilde L) &=& f_b + 2 f_{surface} (t)
\; L^{-1} + {\cal G}(L/\xi, L/\widetilde L) L^{-d} - \frac{1}{2}
\tilde a^{-1} \widetilde L^{1-d} \ln 2
\nonumber\\
&+& O (\tilde a \; L^{-d-1},
\tilde a^{4-d} \; L^{-4})
\ee
where
\be
\label{gleichung32} f_{surface} (t) \; = \; \frac{\tilde a^{1-d}}{8}
\int\limits^\infty_0 dy \Big\{y^{-1} \left[1 + e^{- 4 y} - 2 e^{-
2 y} I_0 (2 y) \right] \nonumber\\
\times \left[e^{- 2 y} I_0 (2 y) \right]^{d-1} \; \exp (- y \;
r_0 \; \tilde a^2 J_0^{-1})\Big\}
\ee
with the Bessel function of order zero
\be
\label{gleichung33} I_0 (z) \; = \; \frac{1}{\pi}
\int\limits^\pi_0 d \theta \; \exp (z \; \cos \theta) \; .
\ee
Eq. (\ref{gleichung31}) contains the universal finite-size
part
\be
\label{gleichung58}
{\cal G} (x, s) = \frac{1}{2}
\int\limits_0^\infty dy \; y^{-1} \Bigg\{ \Big(\frac{\pi}{y}\Big)^{d/2} &-&
\frac{1}{2} \Big[s K (s^2 y)\Big]^{d-1} \Big[K \Big(\frac{y}{4}\Big) -
1 \Big] \nonumber\\
&-& \frac{1}{2} \Big(\frac{\pi}{y} \Big)^{(d-1)/ 2} \Bigg\} e^{-
y x^2 / 4 \pi^2}
\ee
with
\be
\label{gleichung35} K (z) \; = \; \sum^\infty_ {m = - \infty} \exp
(- m^2 z) \; .
\ee
We note that $f_{surface}$ depends on the lattice constant $\tilde
a$, unlike the finite-size part ${\cal G} (x, s)$.
Using $K (z) \sim (\pi / z)^{1/2}$ for $z \to 0$ we obtain
for film geometry ($\widetilde L \rightarrow \infty$)
\be
\label{gleichung34} {\cal G} (x, 0) = \frac{1}{2} \int\limits^\infty_0 dy \;
y^{-1} \left[\left(\frac{\pi}{y}\right)^{1/2} - \frac{1}{2}
\; K\Big(\frac{y}{4}\Big) \right] \left(\frac{\pi}{y}
\right)^{(d-1) / 2} e^{- y x^2 / 4 \pi^2} \; .
\ee
The surface part remains of course identical with Eq. (\ref{gleichung32}).
Eqs. (\ref{gleichung31}) - (\ref{gleichung34}) are applicable to
$T = T_c$ and to $T > T_c$ at fixed $L/\xi$. The correct exponential large
$L/\xi$ behavior of ${\cal{G}} (L/\xi, L/\widetilde L)$ at fixed $T > T_c$
is not yet included in Eqs. (\ref{gleichung58}) and (\ref{gleichung34}) as
it involves the "exponential" correlation length $\xi_1$, Eq.
(\ref{gleichung31c}). For large $L \gg \xi$ at fixed $T > T_c$, Eq.
(\ref{gleichung58}) must be replaced by
\be \label{gleichung42a} {\cal G} (L/\xi_1, L/\widetilde L) = -
2^{-d} [L / (\pi \xi_1)]^{(d-1) / 2} \; e^{- 2 L / \xi_1} \left[1
+ O (\xi_1^2 / L^2) \right] \nonumber\\
- (d-1) (L / \widetilde L)^{(d+1) / 2} \left[L/(2 \pi
\xi_1)\right]^{(d-1)/2} \; e^{- \widetilde L / \xi_1} \left[1 + O
(\xi_1^{1/2} \widetilde L^{- 1/2}) \right] \; . \ee
Correspondingly, Eq. (\ref{gleichung6a}) must be modified for large
$L \gg \xi$. The nonuniversal last term $\sim \widetilde L^{1-d}$
in Eq. (\ref{gleichung31}) contributes to the regular part $f_{ns}
(t, L, \widetilde L)$ of $f$, thus Eq. (\ref{gleichung3}) should
be complemented accordingly.
In order to clarify to what extent the $\tilde a$ dependent term
$f_{surface} (t)$ contains universal contributions we need to
distinguish the cases $1 < d < 3$, $d = 3$, and $3 < d < 5$. For
this purpose it will be useful to express the regular part linear
in $r_0$ in terms of generalized Watson functions defined by
\cite{barber-1973}
\be
\label{gleichung73} W_d (z) &=& \frac{1}{(2 \pi)^d} \int\limits^{2
\pi}_0 \ldots \int\limits^{2 \pi}_0 \frac{d \theta_1 \ldots d
\theta_d} {z + 2 \sum^d_{j = 1} (1 - \cos \theta_j)}
\ee
\be
\label{gleichung-74} &=& \int\limits_0^\infty dy \; e^{- z y}
[e^{- 2 y} I_0 (2 y) ]^d \; . \qquad \quad
\ee

\section*{$\bf{1 < d < 3}$}

For $0 \leq r_0 \; \tilde a^2 \ll 1$ and $1 < d < 3$ we obtain
from Eq. (\ref{gleichung32})
\be
\label{gleichung36} f_{surface} (t) \; = \; f_{surface} (0) & + &
A_{surface}^+ \; \xi^{1-d} \nonumber\\
& - & \tilde b_d \; r_0 \; \tilde a^{3-d} J_0^{-1}
\; + \; O (\tilde a / \xi^d)
\ee
with the universal amplitude
\be
\label{gleichung40} A^+_{surface} \; = \; - \;
\frac{\Gamma
\big((3-d) / 2 \big)}{2^{d+1} \; \pi^{(d-1)/2} \; (d-1)} \; < 0
\ee
and with the nonuniversal constant
\be
\label{gleichung36b} \tilde b_d \; = \; \frac{1}{8}
\int\limits_0^\infty\; dy \bigg\{\big[1 + e^{-4y} & - & 2
e^{-2y} I_0 (2 y) \big] \left[e^{-2y} I_0 (2y) \right]^{d-1}
\nonumber\\
& - & (4 \pi y)^{(1 - d) / 2} \bigg\}  \; .
\ee
This constant can be partially expressed in terms of generalized
Watson function as
\be
\label{gleichung-78} \tilde b_d & = &\frac{1}{8} \; \left[W_{d-1} (4) -
2 W_d (0) \right]
\nonumber\\
&+& \frac{1}{8} \int\limits_0^A dy \left[e^{- 2 y} I_0 (2 y)
\right]^{d-1} \nonumber\\
&+& \frac{1}{8} \int\limits_A^\infty \left\{ \left[e^{- 2 y} I_0 (2 y)
\right]^{d-1} \; - \; (4 \pi y)^{(1-d)/2} \right\} \nonumber\\
&+& 2^{-d-1} \pi^{(1-d)/2} (d-3)^{-1} \; A^{(3-d) / 2} \; .
\ee
(see Appendix A). We note that $\tilde b_d$ does not depend on the
arbitrary finite constant $A > 0$.

The second term in Eq. (\ref{gleichung36}) has the expected singular
scaling form $\sim \xi^{1-d}$. The first term $f_{surface} (0)$ and
the third term $\sim \tilde b_d$ contribute to the regular part
2$\Psi_1 (t) L^{-1}$ of $f_{ns} (t, L, \widetilde L)$. Thus the
surface contribution is in accord with the predicted universal scaling structure
of Eqs. (\ref{gleichung2}), (\ref{gleichung3}),
(\ref{gleichung7a}) and (\ref{gleichung6}).
We expect that $\tilde b_d$ depends on the
lattice structure (see also Subsect. G). We note that both
$A^+_{surface}$ and the coefficient $\tilde b_d$ diverge for
$d \rightarrow 3$ such that
\be
\label{gleichung-79} \lim_{d \to 3 -} \; \left[\tilde b_d \; - \;
A^+_{surface} \right] \; = \; \tilde b
\ee
has a finite limit $\tilde b$ (see Appendix A). The explicit
expression for the constant $\tilde b$ is given in Eqs.
(\ref{gleichung39}) and (\ref{gleichung-82}) below.

\section*{$\bf {d = 3}$}

For $0 \leq r_0 \; \tilde a^2 \ll 1 $ we obtain from Eq.
(\ref{gleichung32}) at $d = 3$
\be
\label{gleichung38} f_{surface} (t) = f_{surface} (0)
- (16 \pi)^{-1} \; \xi^{-2}
\; \ln (\xi / \tilde a) - \tilde b \; r_0 \; J_0^{-1} \; + \; O
(\tilde a / \xi^3) \; .
\ee
The analytic expression for the nonuniversal constant $\tilde b$ is
\be
\label{gleichung39} \tilde b & = & \frac{1}{8}
\int\limits^A_0 dy \left[1 + e^{-4y} - 2 e^{-2y} \; I_0 (2y)
\right] e^{-4y} \left[I_0 (2y) \right]^2 \nonumber\\
& + & \frac{1}{8} \int\limits^\infty_A dy \left\{\left[1 + e^{-4y}
- 2 e^{-2y} I_0 (2y) \right]
e^{-4y} \left[I_0 (2y) \right]^2 \; -
(4 \pi y)^{-1} \right\} \nonumber\\
& + & (32 \pi)^{-1} \; (1 - C_E - \ln A)
\ee
where $C_E $ is Euler's constant. This constant can be partially
expressed in terms of generalized Watson functions as
\be
\label{gleichung-82} \tilde b &=& \frac{1}{8} \; \left[W_2 (4) - 2
W_3 (0) \right]
\nonumber\\
&+& \frac{1}{8} \int\limits_0^A dy \left[e^{- 2 y} I_0 (2 y)
\right]^2 \nonumber\\
&+& \frac{1}{8} \int\limits_A^\infty \left\{ \left[e^{- 2 y} I_0 (2 y)
\right]^2 \; - \; (4 \pi y)^{-1} \right\} \nonumber\\
&+& (32 \pi)^{-1} (1 - C_E - \ln A) \; .
\ee
Note that $\tilde b$
does not depend on the arbitrary finite constant $A > 0$. We
expect that $\tilde b$ depends on the lattice structure (see also
Subsect. G).

While the first and third terms of Eq. (\ref{gleichung38})
contribute to the nonuniversal regular part 2 $\Psi_1 (t) L^{-1}$
of $f_{ns} (t, L, \widetilde L)$ the second (logarithmic) term is
clearly a singular contribution to the free energy density. This
implies that, for $d = 3$, Eqs. (\ref{gleichung2}) and
(\ref{gleichung6a}) must be replaced by
\be
\label{gleichung41} f_s (t, L, \widetilde L, \tilde a) \; = \; L^{-3} \;
\widetilde {\cal F} (L/\xi, L/\widetilde L, \tilde a / \xi)
\ee
where
\be
\label{gleichung42} \widetilde {\cal F} (L/\xi, L/\widetilde L,
\tilde a/\xi) \; = \; {\cal
F} (L/\xi, L/\widetilde L) \; - \; (8 \pi)^{-1} \; (L/\xi)^2 \; \ln
(\xi / \tilde a)
\ee
contains a logarithmic nonscaling term that depends on the
nonuniversal lattice constant $\tilde a$. We conjecture, however,
that the coefficient $- (8 \pi)^{-1}$ is universal, i.e.,
independent of the lattice structure (see also Subsect. G). The
scaling part reads
\be
\label{gleichung43} {\cal F} (L/\xi, L/\widetilde L) \; = \; (12 \pi)^{-1} \;
(L/\xi)^3 \; + \; {\cal G} (L/\xi, L/\widetilde L)
\ee
where ${\cal G} (L/\xi, L/\widetilde L)$ is given by Eq.
(\ref{gleichung58}) for box geometry and by Eq. (\ref{gleichung34})
for film geometry. Correspondingly the universal scaling prediction
of Eq. (\ref{gleichung2a}) must be replaced by
\be
\label{gleichung44} f_s (t, L, \widetilde L, \tilde a) &=& R_\xi^+ \; \xi^{-3} \; -
\; \left[(8 \pi)^{-1} \; \xi^{-2} \; \ln (\xi/\tilde a) \right]
L^{-1} \nonumber\\ &+& L^{-3} {\cal G} (L/\xi, L/\widetilde L) \; + \; O (L^{-4}) \;
,
\ee
with the universal bulk amplitude $R_\xi^+ = - (12 \pi)^{-1}$ but
{\it without a universal surface amplitude}. In Subsection D we
shall see that this logarithmic behavior is related to the vanishing of the
surface critical exponent of the Gaussian energy density at $d=3$.

\section*{${\bf 3 < d < 5}$}

For $0 \leq \tilde a / \xi \ll 1$ we obtain from Eq.
(\ref{gleichung32}) for $3 < d < 5$
\be
\label{gleichung45} f_{surface} (t)
&=& f_{surface} (0) \; - \; \widetilde B_d \; r_0 \; \tilde a^{3-d} J_0^{-1} \;
\nonumber\\
&+& A_{surface}^+ \; \xi^{1-d} \; + \; O (\tilde a / \xi^d)
\ee
with the universal amplitude
\be
\label{gleichung49b} A^+_{surface} \; = \;
\frac{\Gamma \big((5-d) / 2 \big)}{2^d \; \pi^{(d-1)/2} \; (d-1)
(d-3)} \; > 0
\ee
and with the nonuniversal constant
\be
\label{gleichung56} \widetilde B_d = \frac{1}{8}
\int\limits_0^\infty dy \left\{ \left[1 + e^{-4y} - 2 \; e^{-2y} \; I_0
(2y) \right]  \left[ e^{-2y} \; I_0 (2y) \right]^{d-1} \right\}
> 0 \;.
\ee
This constant can be expressed in terms of generalized Watson
functions as
\be
\label{gleichung90} \widetilde B_d = \frac{1}{8} \left[W_{d-1} (0)
+ W_{d-1} (4) - 2 W_d (0) \right] \; .
\ee
We note that $W_{d-1} (0)$ exhibits a pole $\sim (d-3)^{-1}$ near
$d = 3$ where it can be represented as
\be
\label{gleichung91} W_{d-1} (0) &=& 2^{2-d} \pi^{(1-d) / 2}
(d-3)^{-1} A^{(3-d) / 2}  \; + \; \int\limits^A_0 dy \left[e^{- 2
y} I_0 (2 y) \right]^{d-1} \nonumber\\
&+& \int\limits^\infty_A dy \left\{ \left[e^{- 2 y} I_0 (2 y)
\right]^{d-1} - (4 \pi y)^{(1-d)/2} \right\} \; .
\ee
The right-hand side of Eq. (\ref{gleichung91}) is independent of
the arbitrary constant $A > 0$.
The first two terms of Eq. (\ref{gleichung45}) contribute to the
regular part 2 $\Psi_1(t) L^{-1}$ of $f_{ns} (t, L)$ whereas the third term
has the expected singular scaling form $\sim \xi^{1-d}$. Thus the
surface contribution is in accord with the predicted universal scaling
structure of Eqs. (\ref{gleichung2}), (\ref{gleichung2a}), (\ref{gleichung6a})
and (\ref{gleichung7a}). We expect that $\widetilde B_d$ depends on
the lattice structure (see also Subsect. G). We note that Eqs.
(\ref{gleichung49b}) and (\ref{gleichung56}) are the analytic
continuations of Eqs. (\ref{gleichung40}) and
(\ref{gleichung36b}), respectively, from $d < 3$ to $d > 3$ and
that both $A^+_{surface}$ and the coefficient $\widetilde B_d$ diverge
for $d \rightarrow 3$ such that
\be
\label{gleichung92} \lim_{d \to 3 +} \left[\widetilde B_d -
A^+_{surface} \right] \; = \; \tilde b
\ee
has a finite limit $\tilde b$, Eq. (\ref{gleichung39})
(see Appendix A).

\section*{D. Energy density}

Eqs. (\ref{gleichung6}) and (\ref{gleichung24}) yield the
Gaussian energy density (divided by $k_B$)
\be
\label{gleichung26b} U (t, L, \widetilde L) \;
\equiv \; - \; \frac{T^2 a_0} {2 T_c} \; E (r_0, L, \widetilde L, \tilde a)
\ee
with
\be
\label{gleichung-d68} E (r_0, L, \widetilde L, \tilde a)
\; = \; \widetilde L^{1-d} \; L^{-1} \; \sum_{{\bf k}, p}\;
\left(r_0 + J_{{\bf k}, d-1} + J_p \right)^{-1} \; .
\ee
In the following we must distinguish the cases $T = T_c$ and $T >
T_c$. Using Eqs. (\ref{gleichung6}) and (\ref{gleichung31}) -
(\ref{gleichung35}) we obtain for $T > T_c$ and for large $L/\tilde a$
and large $\xi/\tilde a$ at fixed $L/\xi > 0$ and fixed $\widetilde L/L$
in $2 < d < 4$ dimensions
\be
\label{gleichung47} U (t, L, \widetilde L) = U_b (t) \; + \; 2 U_{surface} (t)
\; L^{-1} \; + \; T_c \xi_0^{-2} \;
{\cal E} (L/\xi, L/\widetilde L) L^{2-d} \; + \; O \left(\tilde
a \; L^{1-d}\right)
\ee
where $U_b (t) = - T^2 \partial f_b / \partial T$ is the bulk part
of $U (t, L, \widetilde L)$. Near $T_c$ the surface energy density is given by
\be
\label{gleichung-d70} U_{surface} (t) \; \equiv \; \frac{T_c
\tilde a^{3-d}} {8 \xi_0^2} \; E_{surface} (r_0 \; \tilde a^2 \;
J_0^{-1})
\ee
with
\be \label{gleichung-d71} E_{surface} (z) \;=\;
\int\limits^\infty_0 dy \Big\{\left[1 + e^{- 4 y} - 2 e^{- 2 y}
I_0 (2 y) \right] \left[e^{- 2 y} I_0 (2 y) \right]^{d-1} \;
e^{-yz} \Big\}
\ee
where $z \equiv r_0 \; \tilde a^2 \; J_0^{-1}$. Eq.
(\ref{gleichung-d71}) can be expressed completely in terms of
generalized Watson functions as
\be
\label{gleichung-98} E_{surface} (z) \; = \; W_{d-1} (z) \; +
W_{d-1} (z + 4) \; - \; 2 W_d (z) \; .
\ee
The universal function ${\cal{E}} (x,s) = - \frac{1}{2} \; x^{-1}
\partial{\cal{G}} (x,s) / \partial x$ of the finite-size part reads
\be \label{gleichung50a} {\cal E} (x,s) = \frac {1}{8 \pi^2}
\int\limits^\infty_0 dy \;
&\Bigg\{&\left(\frac{\pi}{y}\right)^{d/2} \; - \; \frac{1}{2} \;
[s K (s^2 y)]^{d-1} [K\left(\frac{y}{4}
\right)-1] \nonumber\\
&-& \frac{1}{2} \left(\frac{\pi}{y} \right)^{(d-1) / 2} \Bigg\}
e^{- y x^2 / 4 \pi^2}
\ee
for box geometry and
\be
\label{gleichung50} {\cal E} (x,0) =
\frac {1}{8 \pi^2} \int\limits^\infty_0 dy \;
\left[\left(\frac{\pi}{y}\right)^{1/2} \; - \; \frac{1}{2}
\; K\left(\frac{y}{4} \right) \right] \left(\frac{\pi}{y}
\right)^{(d-1) / 2} e^{- y x^2 / 4 \pi^2} \;
\ee
for film geometry. We note that $E_{surface}$ depends on the
lattice constant $\tilde a$, unlike the finite-size part ${\cal E}
(x, s)$. Eqs. (\ref{gleichung47}) - (\ref{gleichung50}) are not
applicable to $T = T_c$ for $d \leq 3$ where the functions $U_{surface} (0)$ and
${\cal E} (0,0)$ diverge. For $3 < d < 4$, Eqs. (\ref{gleichung47}) -
(\ref{gleichung50}) are applicable to both $T = T_c$ and $T > T_c$
at fixed $L/\xi \geq0$. The correct exponential large-$L$ behavior
in terms of $\xi_1$ at fixed $T > T_c$ is not yet included in Eq.
(\ref{gleichung50}). It can be derived from ${\cal G} (L/\xi_1, L/\widetilde
L)$, Eq. (\ref{gleichung42a})\; .

In order to see to what extent $E_{surface}$ contains universal
contributions we need to distinguish the cases $2 < d < 3, d = 3$
and $3 < d < 4$.

\section*{${\bf 2 < d < 3}$}

For $0 < r_0 \; \tilde a^2 \ll 1$ and $2 < d < 3$ we obtain from
Eqs. (\ref{gleichung16}) and (\ref{gleichung36})
\be
\label{gleichung51} U_{surface} (t) \; = \; \frac{1}{2} \; T_c
\xi_0^{-2} \left[-(d-1) A^+_{surface} \xi ^{3-d} \; + \; 2 \tilde
a^{3-d} \tilde b_d \right] + O \left(\tilde a / \xi^{d-2}\right)
\ee
where $\tilde b_d$ and $A^+_{surface} < 0$ are given by Eqs.
(\ref{gleichung36b}) and (\ref{gleichung40}). Eq. (\ref{gleichung51})
implies a divergent surface energy density $\sim t^{1-\alpha -
\nu}$ with a universal surface amplitude $(1-d) \nu A^+_{surface} > 0$
and with the critical exponent
\be
\label{gleichung66} 1 - \alpha - \nu \; = \; (d-3) / 2 < 0 \; ,
\ee
in accordance with the singular finite-size scaling part of Eq.
(\ref{gleichung13}). Thus, for $2 < d < 3, \; U_s (t, L, \widetilde L)$
satisfies the scaling prediction of Eqs. (\ref{gleichung8}) and
(\ref{gleichung9}) with the critical exponent $(1 - \alpha) / \nu
= d - 2$ and with the universal scaling function for $x > 0$
\be
\label{gleichung74} {\cal{U}} (x,s) &=& -  d  \nu  R_\xi^+ \;
x^{d-2} \; - \; 2 (d - 1)  \nu  A^+_{surface} \; x^{d-3} + 2 \nu
{\cal{E}} (x,s)
\ee
where the universal bulk amplitude $R_\xi^+$ is given by Eq.
(\ref{gleichung27}). The function ${\cal{E}} (x,s)$ diverges as
$\sim x^{d-3}$ for $x \to 0$. This divergence is cancelled by the
surface term which implies the finite limit
\be
\label{gleichung74a} {\cal{U}} (0,s) &=& \lim_{x \to 0} \;
{\cal{U}} (x,s) \; = \; {\cal{E}}_d (s)
\ee
where
\be \label{gleichung68a} {\cal E}_d (s) \; = \; \frac{1}{8 \pi^2}
\int\limits^\infty_0 dy \; \left\{\left(\frac{\pi}{y}\right)^{d/2}
\; - \; \frac{1}{2} \; [s K (s^2 y)]^{d-1} \left[K
\left(\frac{y}{4} \right) - 1 \right] \right\} \ee
for box geometry and
\be
\label{gleichung68} {\cal E}_d (0) =
\frac{1}{8 \pi^2}
\int\limits^\infty_0 dy \left(\frac{\pi}{y}\right)^{(d-1)/2} \;
\left\{\left(\frac{\pi}{y}\right)^{1/2} \; - \;
\frac{1}{2} \; \left[K \left(\frac{y}{4} \right) - 1 \right] \right\} \;
\ee
for film geometry. By a separate calculation at $T = T_c$ we find from
Eq. (\ref{gleichung26b})
\be
\label{gleichung67} U (0, L, \widetilde L) = U_b (0) &+& T_c
\xi_0^{-2} \left[ {\cal E}_d (s) L^{2-d}  +  2 \tilde b_d \tilde
a^{3-d} L^{-1} \right] \nonumber\\ &+&  O (\tilde a^{2-d/2} \;
L^{- d/2} \; , \; e^{- \widetilde L / \tilde a})
\ee
in agreement with the scaling parts, Eqs. (\ref{gleichung68a}) and
(\ref{gleichung68}). We note that $A^+_{surface}$, ${\cal E}_d$
and $\tilde b_d$ diverge for $d \rightarrow 3$.

\section*{${\bf d = 3}$}

For $0 < r_0 \;  / \tilde a^2 \ll 1$ and for $d = 3$ we obtain
from Eqs. (\ref{gleichung12}), (\ref{gleichung38}) and
(\ref{gleichung47}) - (\ref{gleichung50})
\be
\label{gleichung61} U (t, L, \widetilde L) = U_b (t) + \left[2 U_{surface} (t)
\; + \; T_c \; \xi_0^{-2} \; {\cal E} (L/\xi, L/\widetilde L) \right] L^{-1} +
O (\tilde a L^{-4})
\ee
with
\be
\label{gleichung63} U_{surface} (t) &=& \frac{1}{2} T_c \;
\xi_0^{-2} \; \left[(8 \pi)^{-1} {\rm ln} \; (\xi/\tilde a) + 2
\tilde b - (8 \pi)^{-1} \; + \; O (\tilde a / \xi) \right]
\ee
where $\tilde b$ is given by Eq. (\ref{gleichung39}). Thus, as a
special property of the Gaussian model, there exists a
logarithmically divergent surface energy density with an explicit
dependence on the lattice spacing $\tilde a$. This could have been
anticipated on general grounds because of $1 - \alpha - \nu = 0$
for $d = 3$. (This is parallel to logarithmic terms for systems
with periodic boundary conditions at the borderline dimension
where the specific heat exponent $\alpha$ vanishes
\cite{privman-rudnick}.) Thus, Eq. (\ref{gleichung13}) is not
applicable and the universal scaling prediction for the singular
part $U_s (t, L, \widetilde L)$, Eqs. (\ref{gleichung9}) -
(\ref{gleichung11a}), must be replaced by
\be
\label{gleichung77a} U_s (t, L, \widetilde L, \tilde a) &=& T_c \; \xi_0^{-2} \;
L^{-1} \; {\cal{{\widetilde U}}} (L/\xi, L/\widetilde L, \tilde a / \xi)
\ee
where
\be
\label{gleichung78b} {\cal{{\widetilde U}}} (L/\xi, L/\widetilde L,
\tilde a / \xi) &=& {\cal{U}} (L/\xi, L/\widetilde L) \; + \;
(8 \pi)^{-1} \ln (\xi / \tilde a)
\ee
with the scaling part
\be
\label{gleichung80} {\cal{U}} (x, s) &=& - \; d \; \nu \; R_\xi^+
x \; + \; {\cal{E}} (x,s) \; - \; (16 \pi)^{-1} \; .
\ee
We identify the nonsingular part as
\be
\label{gleichung81} U_{ns} (t, L, \widetilde L) &=& U_b (t) \; - \; U_{bs} (t)
\; + \; T_c \; \xi_0^{-2} \tilde b \; L^{-1} \; + \; O (\tilde a L^{-4}) \; .
\ee
The function ${\cal{E}} (x, s)$ diverges logarithmically for $x \to
0$. This divergence is cancelled by $U_{surface} (t)$ which implies
that Eq. (\ref{gleichung61}) has a finite limit for $t \to 0$ at fixed $L$
and $\widetilde L$
\be
\label{gleichung64} U (0, L, \widetilde L) = U_b (0) &+& T_c
\xi_0^{-2} \left\{- (8 \pi)^{-1} \; L^{-1} \; \ln (\tilde a / L) +
\left[\overline b (s) + 2 \tilde b \right] L^{-1} \right\} \nonumber\\
&+& O (\tilde a^{1/2} L^{- 3/2}\; , \; e^{- \widetilde L / \tilde
a})
\ee
with the universal constant
\be
\label{gleichung71} \bar b (s) &=& \frac{1}{8 \pi^2}
\int\limits^\infty_A dy \left\{(\pi / y)^{3/2} -
\frac{1}{2} \; [sK(s^2 y) ]^2 \left[K (y / 4) - 1 \right] \right\} \nonumber\\
&+& \frac{1}{8 \pi^2} \int\limits^A_0 dy \; \left\{(\pi / y)^{3/2}
- \frac{1}{2} [sK (s^2 y)]^2 \left[K (y / 4)-1\right] - \frac{\pi}{2y} \right\} \nonumber\\
&+& (16 \pi)^{-1} \; {\rm ln} \; A \; - \; (16 \pi)^{-1} \;
\left[1 - \; C_E + 2 \; {\rm ln} (2 \pi) \right] \; .
\ee
Note that $\bar b (s)$ is independent of the arbitrary constant A.
We have confirmed the validity of Eqs. (\ref{gleichung64}) and
(\ref{gleichung71}) for $d = 3$ by calculating $U (0, L,
\widetilde L)$ directly from Eq. (\ref{gleichung26b}) with $r_0 =
0$. Thus there exists a logarithmic nonscaling $L$ dependence of
the energy density at $T_c$, with an explicit dependence on
$\tilde a$. As noted above we conjecture, however, that the
coefficients $(8 \pi)^{-1}$ and $- (8 \pi)^{-1}$ in Eqs.
(\ref{gleichung63}) and (\ref{gleichung64}) are universal, i.e.,
independent of the lattice structure (see also Subsect. G). In
Sect. IV we shall see that the logarithms in Eqs.
(\ref{gleichung63}) and (\ref{gleichung64}) are the origin of the
logarithms appearing in the three-dimensional mean spherical model
with free boundary conditions.

\section*{${\bf 3 < d < 4}$}

For $0 < r_0 \; \tilde a^2 \ll 1$ and for $3 < d < 4$ we obtain
from Eqs. (\ref{gleichung16}) and (\ref{gleichung45})
\be
\label{gleichung53} U_{surface} (t)=
U_{surface} (0) - T_c \; \xi_0^{-2} (d-1) A^+_{surface} \;
\xi^{3-d} + O (\tilde a \xi^{2-d})
\ee
with the finite critical value
\be
\label{gleichung86} U_{surface} (0) \; = \; T_c \; \xi_0^{-2} \; \tilde
a^{3-d} \; \widetilde B_d > 0
\ee
where $A^+_{surface} > 0$ and $\widetilde B_d$ are given by Eqs.
(\ref{gleichung49b}) and (\ref{gleichung56}). The singular second
term $\sim \xi^{3-d}$ yields a divergent slope $\sim t^{(d-5)/2}$
for $t \rightarrow 0+$, thus $U_{surface} (t)$
has a nonuniversal finite cusp at $t=0$ for $3 <d < 5$, in contrast to the case
$d \leq 3$. As a consequence, only the temperature dependent
contributions $\sim \xi^{3-d}$ and $\sim {\cal{E}} (L/\xi, L/\widetilde L)$
to the energy density
\be
\label{gleichung83} U (t, L, \widetilde L) &=& U_{ns} (0, L) \; - \; T_c \;
\xi_0^{-2} (d-1) \; A_{surface}^+ \; \xi^{3-d} \; L^{-1}
\nonumber\\
&+& T_c \; \xi_0^{-2} \; {\cal{E}} (L/\xi, L/\widetilde L) \; L^{2-d} \; + \; O (\tilde
a L^{1-d})
\ee
have a universal scaling form. The nonuniversal critical value
$U_{surface} (0)$, Eq. (\ref{gleichung86}), increases for $d \rightarrow 3$.
It enters the finite energy density at $T_c$
\be
\label{gleichung81a} U_{ns} (0, L, \widetilde L) &=& U_b (0) \; + \; 2
U_{surface} (0) / L \; + \; O (\tilde a L^{1-d})
\ee
which belongs to the nonuniversal nonsingular part of $U (t, L,
\widetilde L)$ and which has a nonscaling $L$ dependence $\sim
L^{-1}$. This $L$-dependence is nonnegligible. This will have
significant consequences for the mean spherical model in $d > 3$
dimensions to be discussed in Sect. IV.

\section*{E. Specific heat}

Eqs. (\ref{gleichung16a}) and (\ref{gleichung26b}) yield
the specific heat (divided by $k_B$)
\be
\label{gleichung27a} C (t, L) & = & \frac{T^2 a_0^2} {2 T_c^2} \;
L^{-1} \; \widetilde L^{1-d}  \sum_{{\bf k}, p}
\left(r_0 \; + \; J_{{\bf k},d-1} \; + \; J_p \right)^{-2} \nonumber\\
& - & \frac{T a_0} {T_c} \;\; L^{-1} \; \widetilde L^{1-d}
\sum_{{\bf k},p} \left(r_0 \; + \; J_{{\bf k},d-1} \; + \; J_p
\right)^{-1} \; .
\ee
From the first term of Eq. (\ref{gleichung27a}) we find full agreement
with the finite-size scaling prediction, Eq. (\ref{gleichung17a}), in $2 <
d < 4$ dimensions with $2 / \nu = 4, \; \alpha/\nu = 4 - d$.
Specifically we find, for large $L/\tilde a$ and $\xi / \tilde a$
at fixed $L / \xi \geq 0$ and fixed $\widetilde{ L}/L$, the universal scaling
function for $x \geq 0$
\be
\label{gleichung87a} {\cal{C}} (x,s ) \; = \; \frac{1} {64 \pi^4}
\int\limits_0^\infty \; dy \; y \left[s K (s^2 y) \right]^{d-1} \;
\left[K \left(\frac{y} {4} \right) -1 \right] \; e^{- y x^2 / 4
\pi^2}
\ee
for box geometry and
\be
\label{gleichung87} {\cal{C}} (x,0) &=& \frac{1}{32 \pi^4}
\int\limits_0^\infty \; dy \; y \; \left(\frac{\pi}{y}
\right)^{(d-1)/2} e^{- x^2 y / 4 \pi^2} \sum_{n = 1}^\infty \;
e^{- n^2 y / 4} \;
\ee
for film geometry. The evaluation of the second term of Eq.
(\ref{gleichung27a}) can be taken directly from subsection D for
the energy density and yields only subleading corrections to the
first term of Eq. (\ref{gleichung27a}). For $T
> T_c$,  Eqs.  (\ref{gleichung87a}) and (\ref{gleichung87}) can be
decomposed as
\be \label{gleichung107} {\cal{C}} (x,s) = - \frac{1}{4} \; d \;
(d-2) \; R_\xi^+ \; x^{d-4} \; + \; A_{C, surface}^+ \; x^{d-5} \nonumber\\
- \; \frac{1} {32 \pi^4} \int\limits^\infty_0 dy \; y
\Bigg\{\left(\frac{\pi}{y}\right)^{d/2} \; - \; \frac{1}{2} \; [s
K (s^2 y)]^{d-1} [K\left(\frac{y}{4}
\right)-1] \nonumber\\
- \frac{1}{2} \left(\frac{\pi}{y} \right)^{(d-1) / 2} \Bigg\} e^{-
y x^2 / 4 \pi^2} \qquad \qquad \qquad
\ee
and
\be
\label{gleichung86a}
{\cal{C}} (x, 0) = - \frac{1}{4} \; d \; (d-2) \; R_\xi^+ \; x^{d-4}
\; + \; A_{C, surface}^+ \; x^{d-5} \nonumber\\
- \; \frac{1} {32 \pi^4} \int\limits^\infty_0 dy \; y (\pi /
y)^{(d-1) / 2} \; \left[ (\pi / y)^{1/2} \; - \; \frac{1}{2} \; K
\Big(\frac{y}{4}\Big) \right]\; e^{- y x^2 / 4 \pi^2}
\ee
where the bulk part (first term) contains the universal bulk
quantity $R_\xi^+$, Eq. (\ref{gleichung25a}), and where the surface
part (second term) has the universal surface amplitude
\be
\label{gleichung77b} A_{C, surface}^+ \; = \; - \; 2^{-d-1} \; \pi^{(1-d) /
2} \; \Gamma \big( (5-d) / 2 \big)
\ee
\be
\label{gleichung82} = \; \frac{1}{2} \; (d-1)
\; (3-d) \; A^+_{surface}
\ee
with $A^+_{surface}$ given by Eqs. (\ref{gleichung40}) or
(\ref{gleichung49b}), in agreement with the predicted structure of
the surface specific heat, Eqs. (\ref{gleichung21a}) and
(\ref{gleichung22b}). Eqs. (\ref{gleichung86a}) -
(\ref{gleichung82}) can be easily confirmed by calculating the
derivative $\partial U / \partial T$ from Eqs. (\ref{gleichung47})
- (\ref{gleichung50}).

Eqs. (\ref{gleichung107}) and (\ref{gleichung86a}) do not yet
include the correct exponential part of the large-$L$ behavior at
fixed $T
> T_c$ which involves the exponential correlation length $\xi_1$
\cite{chen-dohm-2000} and which can be derived from Eq.
(\ref{gleichung42a}).

\section*{F. Susceptibility}

The thermodynamic quantity of primary interest in the mean
spherical model in Sect. IV will be the susceptibility. Important
steps in the calculation of its finite-size properties can be
performed already on the level of the Gaussian model. For box
geometry the susceptibility is defined by
\be
\label{gleichung94} \chi (t,L, \widetilde L) \; = \;
\frac{\tilde a^{2d}} {\widetilde L^{d-1} \; L} \;
\sum_{i,j} \; < \varphi_i \varphi_j > \; .
\ee
Substituting the representation Eq. (\ref{gleichung18b}) into Eq.
(\ref{gleichung94}) we find
\be
\label{gleichung112} \sum_j \varphi_j \; = \; \frac{\tilde a^{1-d}} {(L
+ \tilde a) \sqrt{2}} \; \sum_p \; \hat \varphi_{{\bf 0}, p} \left[1
- (-1)^n \right] \; \frac{\sin (p \tilde a)} {1 - \cos (p \tilde a)}
\; ,
\ee
\be
\label{gleichung95} \chi (t, L, \widetilde L) \; = \;
\frac{\tilde a^2}{L(L
+ \tilde a)^2} \sum_p [1 - (-1)^n] \; \cot^2 (p
\tilde a/2) \; < \hat \varphi_{{\bf 0}, p}
\; \; \hat \varphi_{{\bf 0}, p} >
\ee
with $n \equiv p \; (L + \tilde a) / \pi = 1, 2, \ldots,
L/\tilde a$. From the Gaussian Hamiltonian Eq. (\ref{gleichung18c})
we have $< \hat \varphi_{{\bf 0},p} \; \hat \varphi_{{\bf 0},p} > \; = \;
(L + \tilde a) (r_0 + J_p)^{-1}$ which leads to
\be
\label{gleichung96}
\chi (t, L, \widetilde L) \; = \; \frac{\tilde a^2}{L (L + \tilde a)^2} \;
\sum_p \; [1 - (-1)^n] \; \; \frac{\cot^2 (p \tilde a / 2)}{r_0 +
J_p}\; .
\ee
We note that this expression is independent of the dimension $d$
and of $\widetilde L$ which is a special property of the Gaussian model.
In App. B we evaluate Eq. (\ref{gleichung96}) for large $L / \tilde a$.
For large $\xi / \tilde a$ at fixed ratio $L / \xi \geq 0$ we find the scaling form
\be
\label{gleichung115} \chi (t, L, \widetilde L) \; = \; \chi_b \; f (L/\xi)
\; = \; L^{\gamma/\nu} \; \Phi (L / \xi)
\ee
with $\gamma / \nu = 2$ and the universal scaling function
\be
\label{gleichung97} f_\chi (x) &=& \frac{4} {\pi^2}
\int\limits_0^\infty \; dy \; (1 - e^{-yx^2 / \pi^2}) \;
\left[K (y) - K (4 y) \right] ,
\ee
where $K (z)$ is given by Eq. (\ref{gleichung35}) and
\be
\label{gleichung98} \Phi (x) &=& J_0^{-1} \; x^{-2} \; f_\chi (x)
\; ,
\ee
with $x = L/\xi$. The leading terms of $f_\chi (x)$ for large $x$
are
\be
\label{gleichung99} f_\chi (x) &=& 1 - 2 x^{-1} \; + \; O (x^{-2})
\; .
\ee
For $x \to 0$ we find $\lim_{x \to 0} x^{-2} f_\chi (x) = 1/12$,
thus the leading $L$ dependence at $T = T_c$ is in accord with the
scaling prediction Eq. (\ref{gleichung32a}), with $A_\chi \xi_0^
{- \gamma / \nu} = J_0^{-1}$ and
\be
\label{gleichung100} B_\chi \; = \; \frac{1}{12} \; ,
\ee
independent of the shape factor $L / \widetilde L$. In the limit
$L \to \infty$ at fixed $T > T_c$, Eqs. (\ref{gleichung115}) -
(\ref{gleichung99}) are valid only up to a nonuniversal
exponential contribution $\sim e^{- L/\xi_1}$ in terms of $L /
\xi_1$ rather than $L / \xi$ \cite{chen-dohm-2000}.

\section* {G. Continuum approximation with Dirichlet boundary
conditions}

As a first step towards the $\varphi^4$ field theory with Dirichlet boundary
conditions at finite cutoff $\Lambda$ in a confined geometry we briefly
consider the continuum version of the Gaussian model. The comparison with
the lattice version will serve to distinguish universal from nonuniversal
contributions.

The Gaussian continuum Hamiltonian for the scalar field
$\varphi ({\bf x}) = \varphi ({\bf y}, z)$ reads
\be
\label{gleichung22a} H_{field} = \int d^d x \left[\frac{r_0}{2}
\; \varphi^2 + \frac{1}{2} \; (\nabla \varphi)^2 \right] \; .
\ee
This corresponds to the replacements $2 J \rightarrow 1$ and
$J_{{\bf k}, d-1} \rightarrow {\bf k}^2, J_p \rightarrow p^2$ in
Eqs. (\ref{gleichung18c}), (\ref{gleichung21}), (\ref{gleichung22}),
(\ref{gleichung24}), (\ref{gleichung26a}) and (\ref{gleichung25}).
The wave numbers of the Fourier components of the field
$\varphi ({\bf x})$ are assumed to have a finite sharp cutoff
$\Lambda$. Although there exists no real system with a sharp
cutoff the sharp-cutoff procedure is of significant {\it
conceptual} relevance as it may signal important physical effects
in real systems with subleading long-range interactions
\cite{chen-dohm-2000,chen-dohm-cond,dantchev-2001}.

As a consequence of the sharp-cutoff procedure, the bulk
correlation function $G ({\bf x}) = < \varphi ({\bf x})
\varphi (0) > $ has an oscillatory power-law decay above $T_c$.
For the anisotropic cutoff $|q_i| \leq \Lambda\; , i = 1, ..., d$
we find the anisotropic non-exponential large-distance behavior
\be
\label{gleichung76} G ({\bf x}) = 2^d \Lambda^{d-2} \; (d +
\xi^{-2} \; \Lambda^{-2})^{-1} \;  \prod^d_{i = 1} \; \frac{\sin
(\Lambda x_i)} {\Lambda x_i} + O \; (e^{- | {\bf x} | /
\xi} ) \; ,
\ee
in contrast to the exponential decay of the lattice correlation
function, Eq. (\ref{gleichung31b}). Thus the sharp cutoff induces
long-range correlations, as expected \cite
{chen-dohm-2000,chen-dohm-cond,dantchev-2001}. For an isotropic sharp
cutoff $|{\bf q}| \leq\Lambda$, see Ref. \cite{chen-dohm-cond}.

The bulk free energy density above $T_c$ is for $2 < d < 4$ and
for $0 < d < 2$
\be
\label{gleichung78a} f_{b, field} \; = \; R^+_\xi \; \xi^{-d} \; +
\; f_{0, field}
\ee
with the regular part
\be
\label{gleichung36a} f_{0, field} \; = \; \Lambda^d \left[\hat c_1
+ r_0 \; \Lambda^{-2} \; \hat c_2 + r_0^2 \; \Lambda^{-4} \; \hat c_3
+ ... \right] \; .
\ee
The constants $\hat c_1, \; \hat c_2$ and $\hat c_3$ diverge for $d \to 0,
\; d \rightarrow 2$ and $d \rightarrow 4$, respectively, where $f_{b,field}$
attains a logarithmic dependence on $r_0 \; \Lambda^{-2}$ \; .

In the following we assume a $L \times \infty^{d-1}$ film geometry, with Dirichlet
boundary conditions $\varphi ({\bf y}, 0) = \varphi ({\bf y}, L) = 0$ at the
top and bottom surfaces. The continuum version of Eq. (\ref{gleichung25b}) is
\be
\label{gleichung75} f_{field} (t, L)
                    = - \frac{1}{2} \; \Lambda^d \ln
                      \pi \; + \; \frac{1}{2} \;
                      L^{-1} \sum_p \; \int\limits_{\bf k} \ln \left[\left(r_0 + {\bf
                      k}^2 + p^2 \right) \Lambda^{-2} \right] \;.
\ee
The sum $\sum_p$ runs over wave numbers $p = \pi n
/ L, \; n = 1,2,...$ in the range $0 < p < \Lambda$, and the components
$k_i$ are restricted to $|k_i| \leq \Lambda, i = 1,2, \ldots,
d-1$. For large $L \Lambda$ at fixed $L/\xi \geq 0$ we find for $2 < d < 4$
\be
\label{gleichung39a} f_{field} (t, L) \; = \; f_{b,field} &+& 2 \;
\hat f_{surface} (t) \; L^{-1} \; + \; \hat f_2 (r_0
\Lambda^{-2}) \Lambda^{d-2} \; L^{-2} \nonumber\\
&+& {\cal G} (L/\xi) \; L^{-d} \; + \; O (\Lambda^{d-4} \; L^{-4})
\ee
where
\be
\label{gleichung39b} \hat f_{surface} (t) \; = \; \frac{\Lambda^{d-1}}{8}
\int\limits_0^\infty dy \; y^{-1} \;(1 - e^{-y}) \;S (y)^{d-1} \; \exp
(- y \; r_0 \; \Lambda^{-2})
\ee
and
\be
\label{gleichung43a} \hat f_2 (r_0 \Lambda^{-2}) \; = \; \frac{\pi}{12}
\int\limits_0^\infty dy \; e^{-y} \; S(y)^{d-1} \; \exp (- y \;
r_0 \Lambda^{-2})
\ee
with
\be
\label{gleichung39c} S (y) \; = \; \frac{1}{2 \pi}
\int\limits_{-1}^1 dq \; \exp (- q^2 y)\; .
\ee
For the universal function ${\cal G} (L/\xi)$ see Eq.
(\ref{gleichung34}).

In order to exhibit the universal and nonuniversal contributions
to $\hat f_{surface} (t)$ we need to
distinguish the cases $2 < d < 3, d = 3, 3 < d < 4$.
For $2 < d < 3$ we find from Eq. (\ref{gleichung39b})
\be
\label{gleichung78} \hat f_{surface} (t) & = & \hat f_{surface} (0)
\; + \; A^+_{surface} \; \xi^{1-d} \nonumber\\
&-& \hat b_d \; r_0 \; \Lambda^{d-3} \; + \; O (r_0^2 \;
\Lambda^{d-5})
\ee
with the nonuniversal constant
\be
\label{gleichung79} \hat b_d \; = \; \frac{1}{8}
\int\limits^\infty_0 dy \left[(1 - e^{-y}) \; S (y)^{d-1} \; - (2
\pi)^{1-d} \; (\pi / y)^{(d-1)/2} \right] \; .
\ee
The universal amplitude $A^+_{surface} < 0$ is given by Eq.
(\ref{gleichung40}). For $d = 3$ we find a logarithmic nonscaling
term similar to that of Eq. (\ref{gleichung38}),
\be
\label{gleichung48a} \hat f_{surface} (t) = \hat f_{surface} (0)
- (16 \pi)^{-1} \; \xi^{-2} \; \ln (\Lambda \xi)
\; - \; \hat b \; r_0 \; + \; O (\Lambda^{-2} \;
\xi^{-4})
\ee
with the universal prefactor $(16 \pi)^{-1}$ and with the
nonuniversal constant
\be
\label{gleichung49a} \hat b \; = \; \frac{1}{32 \pi} \; + \; \frac{1}{8}
\int\limits^\infty_0 dy \left\{(1 - e^{-y}) \left[S (y)^2 - (4 \pi y)^{-1}
\right] \right\} \; .
\ee
For $3 < d < 4$ we find
\be
\label{gleichung57} \hat f_{surface} (t) &=& \hat f_{surface} (0) \; -
\; \hat B_d \; r_0 \; \Lambda^{d-3} \nonumber\\ & + & A_{surface}^+ \; \xi^{1-d}
\; + \; O (\Lambda^{-2} \; \xi^{-4})
\ee
with the universal amplitude $A^+_{surface} > 0$, Eq.
(\ref{gleichung49b}), and with the nonuniversal constant
\be
\label{gleichung57a} \hat B_d = \frac{1}{8} \int\limits^\infty_0
dy \; S (y)^{d-1} \; (1 - e^{-y}) > 0 \; .
\ee
As expected, $f_{field} (t, L)$ and $\hat f_{surface} (t)$ contain
the same universal parts as the corresponding functions of the
lattice model. The nonuniversal constants $\hat b_d, \hat b$ and
$\hat B_d$, however, differ from the corresponding constants
$\tilde b_d, \tilde b$ and $\widetilde B_d$ of the lattice model.
For $d \rightarrow 3, \hat b_d$ and $\hat B_d$ are divergent.

As an additional effect of the sharp cutoff, there exists the
nonuniversal contribution $L^{-2}$ to the free energy in Eq.
(\ref{gleichung39a}). For $d > 2$ this term is nonnegligible
compared to the universal scaling term $\sim L^{-d}$ but it has a
regular dependence on $r_0 \sim t$ and therefore can be considered
to belong to the nonsingular part $f_{ns} (t, L)$ of the free
energy density. Nevertheless it yields a leading nonuniversal
contribution $\sim \Lambda^{d-2} \; L^{-2}$ to the Casimir force
at $T_c$, similar to that for periodic boundary conditions
discussed in Ref. \cite{chen-dohm-cond}.

We briefly summarize the results for the energy density $U_{field}
(t, L)$ as derived from Eqs. (\ref{gleichung39a}) -
(\ref{gleichung57a}). For $2 < d < 3$, we obtain
\be
\label{gleichung-c118} U_{field} (t, L) &=& U_{b,field} (t) +
T_c \; \xi_0^{-2} \left[(1-d) A^+_{surface} \; \xi^{3-d} +
\Lambda^{d-3} \; \hat b_d \right] L^{-1} \nonumber\\
&+& T_c \; \xi_0^{-2} \; {\cal{E}} (L/\xi) \; L^{2-d} \; + \; O \;
(\Lambda^{d-2} \; L^{-2}) \; .
\ee
The singular part is in full agreement with the finite-size
scaling structure. For $d = 3$, the energy density reads for large
$L \Lambda$ at fixed $t > 0$
\be
\label{gleichung-c119} U_{field} (t, L) = U_{b, field} (t) & + &
\left[2 \; \hat U_{surface} (t) + T_c \; \xi^{-2} \; {\cal{E}}
(L/\xi) \right] L^{-1} \nonumber\\
&+& O (\Lambda L^{-2})
\ee
with
\be
\label{gleichung-c120} \hat U_{surface} (t) = \frac{1}{2} T_c \;
\xi_0^{-2} \left[(8 \pi)^{-1} \; \ln (\Lambda \xi) + 2 \hat b - (8
\pi)^{-1} + O (\Lambda^{-1} \; \xi^{-1}) \right] \; .
\ee
In the limit $t \to 0$ at fixed $L$ we obtain
\be
\label{gleichung-c121} U_{field} (0, L) = U_{b, field} (0) &+& T_c
\; \xi_0^{-2} \; \big[(8 \pi)^{-1} L^{-1} \; \ln (\Lambda L) + (\bar b +
2 \hat b) L^{-1} \nonumber\\
&+& O (\Lambda^{-1/2} \; L^{-3/2} ) \big]
\ee
with the universal constant $\bar b$, Eq. (\ref{gleichung71}), and
the nonuniversal constant $\hat b$, Eq. (\ref{gleichung49a}). The
logarithmic nonscaling behavior in Eqs. (\ref{gleichung-c120}) and
(\ref{gleichung-c121}) is parallel to that in Eqs.
(\ref{gleichung63}) and (\ref{gleichung64}) of the lattice model
in Section III D. The prefactors $(8 \pi)^{-1}$ in Eqs.
(\ref{gleichung-c120}) and (\ref{gleichung-c121}) are the same as
in Eqs. (\ref{gleichung63}) and (\ref{gleichung64}) of the lattice
model and are expected to be universal.

For $3 < d < 4$, Eqs. (\ref{gleichung53}) - (\ref{gleichung81a})
remain valid if $\tilde a^{3-d} \widetilde B_d$ is replaced
by $\Lambda^{d-3} \hat B_d$. Thus the surface energy density is
\be
\label{gleichung-c122} \hat U_{surface} (t) = \hat
U_{surface} (0) -  T_c \; \xi^{-2}_0 (d-1) \;
A_{surface}^+ \; \xi^{3-d} + O (\Lambda^{-1} \; \xi^{2-d})
\ee
with a finite critical value
\be
\label{gleichung123} \hat U_{surface} (0) = T_c \; \xi^{-2}_0
\; \Lambda^{d-3} \; \hat B_d > 0 \;.
\ee
The singular part $\sim \xi^{3-d}$ is in agreement with finite-size
scaling but is subleading compared to the regular part, as expected
from the lattice model.

For completeness we note that the specific heat and the
susceptibility of the Gaussian continuum model with free boundary
conditions are in full agreement with finite-size scaling for $2 <
d < 4$, with the same scaling functions as those in Sects. III E
and F for the lattice model, as expected.

\section*{H. Dimensional regularization}

The method  of dimensional regularization has been employed in all
previous analytic calculations of finite-size and surface effects
within the $\varphi^4$ theory with Dirichlet boundary conditions.
This method neglects cutoff and lattice effects. This is justified
provided that the leading terms are universal. This is the case,
however, only for $d < d^*$ where $d^*$ is a certain {\it upper}
borderline dimension. In the present context of the Gaussian model
with Dirichlet boundary conditions there exist the following upper
borderline dimensions : $d^* = 0$ for the bulk free energy density
$f_b$, $d^* = 1$ for the surface free energy $f_{surface}$, $d^* =
2$ for the bulk energy density $U_b$, $d^* = 3$ for the surface
energy density $U_{surface}$, $d^* = 4$ for the bulk specific heat
$C_b$ and bulk susceptibility $\chi_b$, and $d^* = 5$ for the
surface specific heat $C_{surface}$ and surface susceptibility
$\chi_{surface}$. The method of dimensional regularization
correctly accounts for the leading universal parts only for $d <
d^*$ (where cutoff and lattice effects are negligible corrections)
and provides an analytic continuation to $d > d^*$. It does not
correctly describe, however, the cutoff and lattice dependent
terms for $d \geq d^*$. The upper borderline dimension $d^* = 3$
for the Gaussian surface energy density will play an important
role for the three-dimensional mean spherical model in Sect. IV.

We begin with the analytic expression for the bulk free energy
density of the Gaussian model in $d$ dimensions within the
dimensional regularization scheme \cite{gelfand}
\be
\label{gleichung-d123} f_{b,dim} (t) \; = \; - \; 2^{- d
- 1} \; \pi^{- d/2} \; \Gamma (- d/2) \; \xi^{-d} \; .
\ee
According to Eqs. (\ref{gleichung27}) - (\ref{gleichung29}),
$f_{b, dim}(t)$ indeed agrees with the singular part $f_{bs}(t)$
of the bulk free energy $f_b$ in $2 < d < 4$ and $0 < d < 2$
dimensions. The neglected terms are just those of the regular part
$f_0$. The latter is ultraviolet divergent for $d \geq 0$
dimensions according to Eqs. (\ref{gleichung30}) and
(\ref{gleichung36a}). Near $d = 2$ the right-hand side of Eq.
(\ref{gleichung-d123}) has a pole $\sim (d-2)^{-1}$ and therefore
does not capture the logarithmic temperature dependence of $f_{bs}
(t)$ in $d = 2$ dimensions.

Next we consider the size-dependent free energy density of the
Gaussian model $f_{dim} (t, L)$ for film geometry within the
dimensional regularization scheme. We find for general $d$
\be
\label{gleichung-d124} f_{dim} (t, L) &=& f_{b, dim} (t) \; + \; 2 \;
A_{surface}^+ \; \xi^{1-d} \; L^{-1} \; + \; {\cal{G}} (L / \xi) L^{-d}
\ee
where $A_{surface}^+$ and ${\cal{}}$ are given by Eqs.
(\ref{gleichung40}) and (\ref{gleichung34}). An alternative
representation is given in Eq. (6.3) of Ref. \cite{gelfand}. Eq.
(\ref{gleichung-d124}) indeed agrees with the singular part $f_s
(t, L)$ calculated in Sect. III for $2 < d < 3$ and for $3 < d <
4$. For $d = 3$, however, the singular part depends explicitly on
the cutoff or the lattice spacing according to Eqs.
(\ref{gleichung48a}) or (\ref{gleichung44}). This is reflected in
the dimensionally regularized result Eq. (\ref{gleichung-d124})
only as a pole term $\sim (d-3)^{-1}$ arising from $A^+_{surface}$.
Thus Eq. (\ref{gleichung-d124}) does not correctly describe the
leading singular temperature dependence $\sim t \ln t$ of the
surface free energy of the Gaussian continuum and lattice model in
three dimensions according to Eqs. (\ref{gleichung44}) and
(\ref{gleichung48a}).

Finally we consider the dimensionally regularized result for the
size-dependent energy density above $T_c$
\be
\label{gleichung-d125} U_{dim} (t, L) = U_{b, dim} (t) &+& T_c \;
\xi_0^{-2} \big[(1-d) A^+_{surface} \; \xi^{3-d} \; L^{-1}
\nonumber\\
&+& {\cal{E}} (L/\xi) L^{2-d} \big] \; .
\ee
For $T \to T_c$ this yields
\be
\label{gleichung-d126} U_{dim} (0, L) &=& T_c \; \xi_0^{-2} \;
{\cal{E}}_d \; L^{2-d}
\ee
as confirmed by a direct calculation at $T = T_c$.
We see that these expressions fail at $d = 3$ where
$A_{surface}^+$ and ${\cal{E}}_d$ do not exist because of pole
terms $\sim (d-3)^{-1}$ as noted already by Dohm \cite{dohm-1989}.
Eq. (\ref{gleichung-d125}) does not capture the logarithmic divergence
$\sim \ln t$ of the surface energy density for $T \to T_c$ at $d =
3$ according to Eqs. (\ref{gleichung63}) and (\ref{gleichung-c120}),
and in Eq. (\ref{gleichung-d126}) the leading size dependence $\sim L^{-1}
\ln L$ of $U (0, L)$ for $L \to \infty$ at $d = 3$ is lacking,
compare Eqs. (\ref{gleichung64}) and (\ref{gleichung-c121}).

Also for $d > 3$, Eqs. (\ref{gleichung-d125}) and
(\ref{gleichung-d126}) are not satisfactory since they contain
only terms that are subleading compared to the finite energy
density at $T_c$, Eq. (\ref{gleichung81a}). The latter term that
exhibits a {\it nonscaling} $L$ dependence is missing in Eqs.
(\ref{gleichung-d125}) and (\ref{gleichung-d126}).

As far as the $\varphi^4$ field theory is concerned, it is not
clear at present whether these shortcomings are only an artifact
restricted to the Gaussian approximation (corresponding to
one-loop order of the $\varphi^4$ theory) or whether there exist
further shortcomings at two-loop order. For this reason we do not
consider universal finite-size scaling to be firmly established
for nonperiodic boundary conditions since the earlier
field-theoretic results of Refs.
\cite{eisenriegler,symanzik,krech-1992,sutter,schmolke,
frank-dohm,mohr-dohm-2000} are based on a perturbation approach
using Gaussian propagators within the dimensional regularization
scheme. On the other hand we note that the singular parts of both
the specific heat and the susceptibility are correctly described
for the Gaussian model in $2 < d <4$ dimensions by means of
dimensional regularization.

\newpage

\section* {IV. Mean spherical model with free boundary conditions}

Again we consider $N$ scalar spin variables $S_i \;, - \infty \leq
S_i \leq \infty$, on the lattice points ${\bf x}_i$ of a simple-cubic
lattice with a lattice spacing $\tilde a$ in a finite rectangular
$L \times \widetilde L^{d-1}$ box of volume $V = L \; \widetilde
L^{d-1} = N \; \tilde a^d$. We assume the statistical weight $\propto\;
e^{- \beta H}$ with
\be
\label{gleichung127} H = \tilde a^d \left\{- \frac{J}{\tilde a^2}
\sum_{<ij>} \; S_i \; S_j \; + \; \mu \sum_i \; S^2_i \right\}
\ee
with a nearest-neighbor coupling $J > 0$. The "spherical field"
$\mu (T,L, \widetilde L, \tilde a)$ is determined as a function
of $\beta = (k_B T)^{-1}$ and of $L, \widetilde L, \tilde a$
through the constraint
\be
\label{gleichung128} \tilde a^{d-2} \sum_i \; <S^2_i > \;=
\; N \; = \; (L/\tilde a) (\widetilde L / \tilde a)^{d-1} \; .
\ee
For $\tilde a = 1$, Eqs. (\ref{gleichung127}) and
(\ref{gleichung128}) yield the standard formulation of the mean spherical model
\cite{barber-1973}. Keeping $\tilde a$ as an independent
nonuniversal parameter will facilitate the distinction between
nonuniversal and universal contributions.

In the following we assume the same boundary conditions as for the
Gaussian model of Sect. III. Thus the spin variables $S_j$ can be represented as
\be
\label{gleichung145} S_j = \widetilde L^{-(d-1)} \; (L + \tilde a)^{-1} \;
\sum_{{\bf k},p} \; \hat S_{{\bf k}, p} \; \; \exp (i\;{\bf k \cdot y}_j)
\; \; \sqrt{2} \; \sin (p\;z_j)\;,
\ee
and the diagonalized Hamiltonian reads
\be
\label{gleichung146} H = \frac{1}{2} \;\; \widetilde L^{-(d-1)} \; (L +
\tilde a)^{-1} \; \sum_{{\bf k},p} \;
\left(\tilde \mu + J_{{\bf k}, d-1} + J_p \right)\; \hat S_{{\bf k}, p} \; \; \hat
S_{-{\bf k}, p}
\ee
with the shifted spherical field
\be
\label{gleichung147} \tilde \mu \; = \;  2 \mu - 2 J_0 d \tilde
a^{-2}
\ee
with $J_0 = 2 J$.
For $J_{{\bf k},d-1}$ and $J_p$ see Eqs. (\ref{gleichung21}) and
(\ref{gleichung22}). The parameter $\tilde \mu (T, L, \widetilde L,
\tilde a)$ is determined implicitly as a function of $T$, $L,
\widetilde L$ and $\tilde a$ through the constraint equation
(\ref{gleichung128}) which now reads
\be
\label{gleichung148}
\tilde a^{d-2} \; \widetilde
L^{1-d} L^{-1} \; \sum_{{\bf k},p}
\left(\tilde \mu + J_{{\bf k}, d-1} \; + \; J_p \right)^{-1} =
\beta \;.
\ee
The susceptibility for $T \geq T_{c,d}$ is
\be
\label{gleichung150a} \chi (T, L, \widetilde L, \tilde a) &=&
\beta \; \frac{\tilde a^{2d}} {\widetilde L^{d-1} \; L} \;
\sum_{i,j} \; < S_i S_j >
\ee
\be
\label{gleichung150b}
& = & \; \beta \frac{\tilde a^2}{L(L
+ \tilde a)^2} \sum_p [1 - (-1)^n] \; \cot^2 (p
\tilde a/2) \; < \hat S_{{\bf 0}, p}
\; \; \hat S_{{\bf 0}, p} > \\
\label{gleichung151a} & = & \frac{\tilde a^2}{L (L + \tilde a)} \;
\sum_p \; [1 - (-1)^n] \; \; \frac{\cot^2 (p \tilde a / 2)}
{\tilde \mu (T, L, \widetilde L, \tilde a) + J_p}  \quad \quad
\ee
with $n \equiv p(L + \tilde a)/\pi = 1,2,\ldots, L/\tilde a$
which is parallel to Eqs. (\ref{gleichung94}) - (\ref{gleichung96})
for the Gaussian model. A significant difference, however, is the
dependence of $\chi$ on $d$, $L, \widetilde L$ and $\tilde a$ through
$\tilde \mu (T, L, \widetilde L, \tilde a)$.

\section*{A. Bulk properties}

First we recall some of the bulk properties. The bulk
susceptibility at finite wave vector $\bf q$ above $T_{c,d}$ is
defined by
\be
\label{gleichung132}
\chi_b ({\bf q}) = \lim_{V \rightarrow \infty} \; \beta \; \frac{\tilde
a^{2d}}{V} \; \sum_{i,j} \; <S_i \; S_j> \; e^{- i{\bf q} \cdot ({\bf
x}_i - {\bf x}_j)} \; .
\ee
The Gaussian structure of $H$, Eq. (\ref{gleichung146}), implies
\cite{chen-dohm-1998-a}
\be
\label{gleichung133} \chi_b ({\bf q})^{-1} \; = \; \tilde \mu_b \; + \;
J_{{\bf q}, d}
\ee
where
\be
\label{gleichung153} \tilde \mu_b\;  = \; \tilde \mu (T, \infty, \infty,
\tilde a) \; = \; \chi_b ({\bf 0})^{-1} \; \equiv \; \chi_b^{-1}
\ee
is determined implicitly by
\be
\label{gleichung134}
\tilde a^{d-2} \; \beta^{-1} \;
\int\limits_{\bf q} (\tilde \mu_b \; + \; J_{{\bf q},d})^{-1} \; =
\; 1 \; .
\ee
Eq. (\ref{gleichung134}) is the bulk version of the constraint
equation (\ref{gleichung148}). The square of the second-moment
bulk correlation length is determined by the susceptibility
according to \cite{chen-dohm-1998-a}
\be
\label{gleichung135} \xi^2 &=& \chi_b ({\bf 0}) \;\;
\frac{\partial}{\partial q^2} \; \left [\chi_b ({\bf q})^{-1}
\right ] _{{\bf q} = {\bf 0}} \; = \; J_0 \; \chi_b \; .
\ee
Setting $\chi^{-1}_b = \tilde \mu_b = 0$ yields the bulk critical
temperature $T_{c,d}$
\be
\label{gleichung136} \frac{1}{k_B T_{c,d}} \; = \; \beta_{c,d} \; = \; \tilde a^{d-2} \;
\int\limits_{\bf q} \; J_{{\bf q},d}^{-1} \;.
\ee
We note that $T_{c,d}$ is nonzero for $d > 2$ and
$\lim_{d\rightarrow 2 +} T_{c,d} = 0$. It is well known that the
bulk critical behavior of the mean spherical model belongs to the
universality class of the $n$-vector model in the large-$n$ limit
\cite{stanley-1968}, thus the vanishing of $T_{c,d}$ at $d = 2$ is
expected from the Mermin-Wagner theorem \cite{mermin}.

In order to elucidate the role played by the borderline dimension
$d=3$ for the confined system we extend our analysis to {\it
continuous} dimensions in the range $2 < d < 4$. Eqs.
(\ref{gleichung134}) and (\ref{gleichung136}) can be combined as
\be
\label{gleichung137} \tilde a^{2-d} \;
(\beta_{c,d} - \beta) = \chi^{-1}_b
\int\limits_{\bf q} \left[J_{{\bf q},d} \;(J_{{\bf q},d} \; + \; \chi^{-1}_b)
\right ]^{-1}.
\ee
This leads to the asymptotic critical behavior above $T_{c,d}$ for $2 < d < 4$
\be
\label{gleichung138} \chi_b = A_\chi \; t^{-\gamma}\;, \hspace{1.0cm} \xi =
\xi_0 \; t^{-\nu}\;, \hspace{1.0cm} t = \frac{T - T_{c,d}} {T_{c,d}} \; ,
\ee
where
\be
\label{gleichung139} \gamma = 2 / (d-2) \; , \hspace{1.0cm}
\nu = (d-2)^{-1} \;,
\ee
and
\be
\label{gleichung139a} \alpha = 2 - d \nu = (d-4)/(d-2)\;.
\ee
We note that these critical exponents can be considered as the
Fisher-renormalized \cite{fisher-1968} Gaussian critical exponents
\be
\label{gleichung140} \gamma \; = \; \frac{\gamma_{Gauss}} {1 -
\alpha_{Gauss}} \hspace{1.0cm}, \hspace{1.0cm} \nu \; = \;
\frac{\nu_{Gauss}} {1 - \alpha_{Gauss}} \; ,
\ee
and
\be
\label{gleichung140a} \alpha \; = \; \frac{\-\alpha_{Gauss}} {1 -
\alpha_{Gauss}} \; ,
\ee
as expected from the general theory of constrained systems
\cite{fisher-1968}, with the Gaussian exponents of Sect. III
\be
\label{gleichung141} \gamma_{Gauss} \; = \; 1 \; , \; \; \nu_{Gauss}
\; = \; 1/2 \; , \; \; \alpha_{Gauss} \; = \; (4-d) / 2 \; .
\ee
The amplitudes are $A_\chi \; = \; \xi^2_0 / J_0$ and
\be
\label{gleichung142} \xi_0 = \tilde a \;[A_d / (\beta_{c,d} \;J_0
\varepsilon) ]^{1/(d-2)}
\ee
with $\varepsilon = 4-d$ and the geometrical factor $A_d$, Eq.
(\ref{gleichung29}). The factor $A_\chi \xi_0^{- \gamma / \nu}$ in
Eqs. (\ref{gleichung-2}) and (\ref{gleichung32a}) becomes simply
$A_\chi \xi_0^{- \gamma / \nu} \; = \; J_0^{-1}$.

For completeness we note that at the lower critical dimension $d =
2$, the asymptotic behavior of $\chi_b$ and $\xi$ for $T
\rightarrow T_{c,2} = 0$ is exponential \cite{shapiro} and is
derived from Eqs. (\ref{gleichung134}) and (\ref{gleichung135}) as
\be
\label{gleichung143} \xi = \; c \; \tilde a \; \exp \; (2 \pi \beta J_0)\; ,
\ee
\be
\label{gleichung144} \chi_b = \; c^2 \; \tilde a^2 J_0^{-1} \;
\exp \; (4 \pi \beta J_0)
\ee
with
\be
\label{gleichung168a} c \; = \; 0.03125 \; .
\ee
The validity of universal finite-size scaling to be derived in
Subsection C below for $2 < d < 3$ is expected to hold also for $d
= 2$ above $T_{c,2} = 0$ in terms of the correlation length Eq.
(\ref{gleichung143}).

\section*{B. Film critical temperature}

For film geometry $(\widetilde L \to \infty)$ Eqs. (\ref{gleichung148})
and (\ref{gleichung151a}) are replaced by
\be
\label{gleichung154} \tilde a^{d-2} \; L^{-1} \; \sum_p \int\limits_{\bf k}
\left(\tilde \mu + J_{{\bf k}, d-1} \; + \; J_p \right)^{-1} =
\beta
\ee
and
\be
\label{gleichung155} \chi (T, L, \infty, \tilde a) & = &
\frac{\tilde a^2}{L (L + \tilde a)} \; \sum_p \; [1 - (-1)^n] \;
\; \frac{\cot^2 (p \tilde a / 2)} {\tilde \mu (T, L, \infty,
\tilde a) + J_p}  \; .\qquad
\ee
Unlike the box geometry, the film geometry
introduces a considerable complication in that for $d>3$ the film
system of thickness $L$ has its own sharp critical temperature
$T_{c,d}(L) > 0$ different from the critical temperature $T_{c,d}
\equiv T_{c,d}(\infty)$ of the $d$-dimensional bulk system. As shown
by Barber and Fisher \cite{barber-1973}, $T_{c, d} (L) > T_{c,d}
(\infty)$ for $d \geq 4$. Here we shall show that this
is true also for $3 < d \leq 4$.

The condition for criticality of the film system is $\chi (T, L,
\infty, \tilde a)^{-1} = 0$. This condition is satisfied at a
critical value $\tilde \mu = \tilde \mu_c < 0$ where $\tilde
\mu_c (L)$ is determined by the vanishing of the denominator of
the lowest-mode $(n = 1)$ term in the sum of Eq. (\ref{gleichung155}),
\be
\label{gleichungB.2} \tilde \mu_c (L) \; = \; - \; \frac{2 J_0}{\tilde a^2}
\left(1 - \cos \frac{\pi \tilde a} {L + \tilde a} \right) \; .
\ee
We note that $\tilde \mu_c (L)$ is independent of $d$. The leading
large $L$ behavior is
\be
\label{gleichung183} \tilde \mu_c (L) \; = \; - \; J_0 \pi^2
(L + \tilde a)^{-2} \; +
\; O [\tilde a^2 (L + \tilde a)^{-4}] \; .
\ee
According to Eq. (\ref{gleichung154}), the corresponding critical
value of $\beta_{c,d} (L) = \left[k_B T_{c,d} (L) \right]^{-1}$
is then given by
\be
\label{gleichungB.3} \beta_{c,d} \; (L) &=&
\frac{\tilde a^{d-2}} {L} \;
\sum_p \int\limits_{\bf k} (\tilde \mu_c \; + \; J_{{\bf k},d-1} \;
+ \; J_p)^{-1} \; .
\ee
Separating the lowest-mode $(n = 1)$ term we obtain for general
$d$
\be
\label{gleichungB.4} \beta_{c,d} (L) =
\frac{\tilde a^{d-2}} {L} \int\limits_{\bf k} J_{{\bf k},d-1}^{-1}
\; + \; \frac{\tilde a^{d-2}} {L} \sum_{n = 2}^{L/\tilde a}
\int\limits_{\bf k} \left(\tilde \mu_c + J_{{\bf k},d-1} + J_p ^{-1} \right) \; .
\ee
The first integral in Eq. (\ref{gleichungB.4}) is directly related to
the critical temperature of a $d-1$ dimensional bulk system
[compare Eq. (\ref{gleichung136})] and is infrared divergent for
$d \leq 3$, hence $\beta_{c,d} (L) = \infty$ or \cite{barber-1973}
\be
\label{gleichungB.5} T_{c,d} (L) &=& 0 \; \; \; \; {\rm for} \; \; d \leq 3
\ee
for any finite $L$, as expected from the Mermin-Wagner theorem
\cite{mermin}. We see that for the film system of finite thickness
the dimension $d = 3$ plays the role of a {\it lower critical
dimension} $d_l = 3$ up to which $T_{c,d} (L)$ vanishes. Thus, at
{\it finite} temperature and in $2 < d \leq 3$ dimensions, there
exists only one type of critical behavior for large $L$ near the
bulk critical temperature $T_{c,d} > 0$ for the $d$-dimensional
film system of finite thickness.

An analysis of Eq. (\ref{gleichungB.3}) for $d > 3$ is presented
in Appendix B. We find that $T_{c,d} (L)$ is enhanced above
$T_{c,d} (\infty)$ for $d > 3$ for sufficiently large $L \gg
\tilde a$.
This enhancement is most naturally
expressed in terms of the dimensionless parameter
\be
\label{gleichung186} \Delta \beta \; = \; J_0 \; [\beta_{c,d}
\; (\infty) \; - \; \beta_{c,d} \; (L) ] \; .
\ee
For large $L \gg \tilde a$ the result is
\be
\label{gleichung187} \Delta \beta \; = \; 4 \widetilde B_d \;
\tilde a/L \; - \; \widetilde C_d \; (\tilde a/L)^{d-2}
\; + \; O(\tilde a^{d/2} \; L^{-d/2})
\ee
with the nonuniversal amplitude $\widetilde B_d > 0$, Eq.
(\ref{gleichung56}), and the universal amplitude
\be
\label{gleichung188} \widetilde C_d = \frac{1}{8 \pi^2}
\int\limits_0^\infty dz \left\{1 - 2
\left(\frac{\pi}{z}\right)^{1/2} + e^{z/4} \left[K
\left(\frac{z}{4} \right) - 1 \right] \right\} \left(\frac{\pi}{z}
\right)^{(d-1)/2}
\ee
with $\widetilde C_d > 0$. Thus there are competing effects on
$T_{c,d} (L)$ from the scaling term $\sim L^{2-d} = L^{1 / \nu}$
and the nonscaling term $\sim L^{-1}$. The leading terms of the
fractional shift are
\be
\label{gleichung189} \frac{T_{c,d} (L)- T_{c,d} (\infty)}
{T_{c,d} (\infty)} \; = \;
a_d \tilde a/L \; - \; c_d (\tilde a/L)^{d-2} \; + \;
O (\tilde a^{d/2} \; L^{-d/2} )
\ee
with the positive amplitudes
\be
\label{gleichung189a} a_d \; = \; 4 \; \widetilde B_d \;
\left\{\int\limits_0^\infty dy \; \left[e^{-2y} I_0 (2y) \right]^d
\right\}^{-1} \; ,
\ee
\be
\label{gleichung190} c_d \; = \; \widetilde C_d \left\{
\int\limits_0^\infty dy \; [e^{-2y} I_0 (2y)]^d \right\}^{-1}
\ee
where $I_0$ is given by Eq. (\ref{gleichung33}). The amplitude
$a_d$ can be expressed in terms of generalized Watson functions,
see Eqs. (\ref{gleichung90}) and (\ref{gleichung-74}).
For $d = 4, a_4$
agrees with corresponding amplitude of Barber and Fisher
\cite{barber-1973}.

We see that the positive shift of $T_{c,d} (L)$ for $d > 3$ is
proportional to the same amplitude $\widetilde B_d > 0$ that
determines the finite cusp of the Gaussian surface energy density,
Eq. (\ref{gleichung71}). Thus the nonscaling Gaussian cusp and the
nonscaling enhancement of $T_{c,d} (L)$ for film geometry are
closely connected. In the next Subsection we shall find that the
Gaussian cusp is also responsible for nonscaling finite-size
effects on the susceptibility for both box and film geometry in
the mean spherical model for $d > 3$.

\section*{C. Constraint equation and susceptibility}

The crucial question is whether and for which dimension $d$ the
susceptibility, Eq. (\ref{gleichung151a}), attains the universal
scaling form of Eq. (\ref{gleichung27-a}) for large $L, \widetilde
L, \xi$ at fixed $\tilde a$. This requires to first analyze the
size dependence of $\tilde \mu$ implied by Eq.
(\ref{gleichung148}).
Up to a constant factor, the left-hand side of Eq. (\ref{gleichung148})
has the same form as the right-hand side of Eq. (\ref{gleichung-d68}) for the
Gaussian energy density, thus the constraint equation (\ref{gleichung148})
can be rewritten as
\be
\label{gleichung151} E (\tilde \mu, L, \widetilde L, \tilde a) &=& \beta
\; \tilde a^{2-d}
\ee
where $E (r_0, L, \widetilde L, \tilde a)$ is defined by Eq.
(\ref{gleichung-d68}). It is clear that any nonscaling $L$
dependence of the Gaussian function $E (\tilde \mu, L, \widetilde
L, \tilde a)$ will cause a nonscaling form of the $L$ dependence
of $\tilde \mu$ which, through Eq. (\ref{gleichung151a}), will in
turn cause a corresponding nonscaling $L$ dependence of the
susceptibility. This mechanism explains the existence of a
borderline dimension $d = 3$ between a scaling $(d < 3)$ and
nonscaling $(d \geq 3)$ regime in the mean spherical model as a
consequence of the size dependence of the energy density of the
Gaussian model, for both box and film geometry. More specifically,
we can anticipate nonscaling power laws for $d > 3$ arising from
the nonscaling size dependence of the finite cusp of $U_{surface}
\sim E_{surface}$ according to Eqs. (\ref{gleichung53}) -
(\ref{gleichung81a}).

It turns out that the most natural parameter is not
$\tilde \mu$ but rather the shifted parameter
\be
\label{gleichung178} \Delta \mu \; = \; J_0^{-1}
\left[\tilde \mu \; - \; \tilde \mu_c (L)\right]
\ee
where $\tilde \mu_c (L)$ is given by Eq. (\ref{gleichungB.2}). We
note that $\Delta \mu > 0$ for box geometry for any finite $L$.
In Appendix B we derive the following expression for the
large $L$ behavior of the susceptibility, Eq.
(\ref{gleichung151a}), at fixed $\Delta \mu L^2 - \pi^2 > 0$
\be
\label{gleichung-203} \chi = \frac{4} {J_0 \pi^2} \; L^2
\int\limits_0^\infty dy \; \frac{1 - \exp \left[- (\Delta \mu L^2
- \pi^2) y / \pi^2 \right]} {\Delta \mu L^2 - \pi^2} \; [K (y) - K
(4 y) ]
\ee
where $K (z)$ is defined by Eq. (\ref{gleichung35}). Eq.
(\ref{gleichung-203}) is valid for general $d$.

In Appendix D we analyze Eq. (\ref{gleichung151}) for large
$L/\tilde a$ at fixed shape factor $\tilde s = (L + \tilde a)
/\widetilde L > 0$ near the bulk critical temperature. For small
$t = (T - T_{c,d}) / T_{c,d} \geq 0$ we find for $2 < d < 4$
\be
\label{gleichung185} (\Delta \mu \; L^2)^{(d-2)/2} =
t (L / \xi_0)^{d-2} - \varepsilon (2 A_d)^{-1}\; (L/\tilde a)^{d-3}
E_{surface} (\Delta \mu \; \tilde a^2)
\nonumber\\
\; + \; 2 \varepsilon A_d^{-1}  \; \widetilde{\cal E}_d \left((\Delta
\mu)^{1/2} \; L,\tilde s \right)  \qquad \qquad \qquad
\quad \quad
\ee
where $E_{surface} (z)$ is given by the Gaussian surface function,
Eq. (\ref{gleichung-d71}). The universal finite-size part reads for box
geometry
\be
\label{gleichung191} \widetilde {\cal E}_d (x, \tilde s) \;=\;
\frac{1}{16
\pi^2} \int\limits_0^\infty \; dy \; \Bigg\{
\left(\frac{\pi}{y} \right)^{(d-1) / 2} \; - \; 2
\left(\frac{\pi}{y} \right)^{d/2}
\qquad \qquad \nonumber\\
+ \;\; e^{y/4} \; \left[K \left(\frac{y}{4} \right) -1
\right] \left[\tilde s K (\tilde s^2 y) \right]^{d-1} \Bigg\} e^{- y x^2 / 4
\pi^2}  \; .
\ee
We note that the term $K (y/4) - 1 = 2 \sum_{n = 1}^{\infty} \exp (-
n^2 y / 4)$ comes from the modes with the free boundary
conditions ($z$ direction) whereas the term $\tilde s K (\tilde
s^2 y) = \tilde s \sum_{m = - \infty}^\infty \exp (- \tilde s^2
m^2 y)$ comes from the modes with the periodic
boundary conditions ($d - 1$ horizontal directions). For film
geometry $(\tilde s = 0)$ the latter term is reduced to $(\pi /
y)^{1/2}$, and the universal finite-size part becomes
\begin{eqnarray}
\label{gleichung196} \widetilde {\cal E}_d (x, 0) =
\frac{1}{16 \pi^2} \int\limits_0^\infty dy \;
\left(\frac{\pi}{y} \right)^{(d-1) / 2}
\Bigg\{ 1 \; - \; 2
\left(\frac{\pi}{y} \right)^{1/2}
\qquad \qquad \nonumber\\
+ \; \;e^{y/4} \left[K \left(\frac{y}{4} \right) \; -\; 1
\right] \Bigg\} e^{- y x^2 / 4 \pi^2}.
\end{eqnarray}
Eqs. (\ref{gleichung185}) - (\ref{gleichung196}) determine $\Delta
\mu$ implicitly near $T_{c,d}$ for large $L$ as a function of $t,
L, \tilde L$ and $\tilde a$. In the absence of the $\tilde a$
dependent term $\sim E_{surface}$, Eq. (\ref{gleichung185}) would
yield a universal scaling form for $\Delta \mu L^2$. According to
Eq. (\ref{gleichung-203}), this would then imply a universal
scaling form for the susceptibility. From the analysis of the
Gaussian model in Sect. III we know, however, that different
scaling and nonscaling terms arise from $E_{surface}$ depending on
whether $2 < d < 3, d = 3$, or $3 < d < 4$.

\section*{1. Finite-size scaling in ${\bf 2 < d < 3}$ dimensions}

The asymptotic form of $E_{surface} (\Delta \mu \tilde a^2)$ for
small $\Delta \mu \tilde a^2$ reads for $2 < d < 3$
\be
\label{gleichung195} E_{surface} (\Delta \mu \tilde a^2)
&=& - 4 (d-1) A^+_{surface} (\Delta \mu \tilde
a^2)^{(d-3)/2} \nonumber\\
&+& 8 \tilde b_d \; + \; O\left((\Delta \mu \tilde
a^2)^{(d-2)/2}\right) \; ,
\ee
compare Eq. (\ref{gleichung51}). Substituting Eq.
(\ref{gleichung195}) into Eq. (\ref{gleichung185}) we see that the
dependence on $\tilde a$ is cancelled. This implies, for a given
shape factor $s = L/\widetilde L,$ that $\Delta \mu$ has the
universal scaling form
\be
\label{gleichung157} \Delta \mu \; = \;
L^{-2} \; {\cal M}_d (L/\xi, s)
\ee
where the scaling function ${\cal M}_d (x,s)$ is
determined implicitly by
\be
\label{gleichung158} {\cal M}_d^{(d-2)/2} = x^{d-2} + 2 \epsilon
A_d^{-1} {\cal \widetilde E}_d ({\cal M}_d^{1/2},s) + 2 \epsilon (d-1)
A_d^{-1} A_{surface}^+  {\cal M}_d^{(d-3)/2} \; .
\ee
Substituting Eq. (\ref{gleichung157}) into Eq.
(\ref{gleichung-203}) confirms the finite-size scaling prediction,
Eq. (\ref{gleichung27-a}), with the bulk susceptibility
$\chi_b = \beta J_0^{-1} \xi^2$ and the universal
finite-size scaling function for $2 < d < 3$
\be
\label{gleichung198} f_\chi (x, s) = \frac{4 x^2}{\pi^2}
\int\limits_0^\infty dy \; \frac{1 - \exp \left\{-\left
[M_d (x, s) - \pi^2 \right] y/\pi^2 \right\}} {{\cal M}_d (x, s) - \pi^2} \left[K
(y) - K (4y) \right] \; .
\ee
At $T = T_{c,d}$ this yields the power law Eq.
(\ref{gleichung32a}) with $\gamma/ \nu = 2$ and with the
universal amplitude
\be
\label{gleichung199} B_\chi (s) =
\frac{4}{\pi^2} \int\limits_0^\infty dy \; \frac{1 - \exp \left\{-\left
[M_d (0, s) - \pi^2 \right] y/\pi^2 \right\}} {{\cal M}_d (0, s) - \pi^2} \left[K
(y) - K (4y) \right] \; .
\ee
The scaling results of Eqs. (\ref{gleichung198}) and
(\ref{gleichung199}) are applicable also to film  geometry $(s =
0)$ where $f_\chi (x, 0)$ and $B_\chi (0)$ are finite quantities
for $d < 3$. We note, however, that both $B_\chi(s)$ and $B_\chi
(0)$ diverge for $d \to 3$. Specifically, for $d \to 3$ we find
from Eq. (\ref{gleichung199}) for box geometry at fixed $s > 0$
\be \label{gleichung-212} B_\chi (s) \sim 2 \pi^{-3} s^{-2}
(3-d)^{-1}
\ee
whereas for film geometry
\be \label{gleichung-214} B_\chi (0) \sim 2 \pi^{-4} \;
(3/2)^{2/(3-d)} \; .
\ee
The different types of divergencies are the consequence of a mode
continuum for film geometry and signal different forms of
violations of finite-size scaling at $d = 3$ for box and film
geometry, as will be confirmed in Subsection C2 below.

At fixed $t > 0$ we find from Eqs. (\ref{gleichung157}) -
(\ref{gleichung198}) the leading large-$L$ behavior for both box
and film geometry
\be
\label{gleichung159}
\tilde \mu^{1/2} = J_0^{1/2} \xi
^{-1} \left[1 + 2 \epsilon (d-1) (d-2)^{-1} A_d^{-1} A_{surface}^+
\xi / L + O(\xi^2 / L^2) \right] \; ,
\ee
and
\be
\label{gleichung200} \chi = \chi_b \left\{1 -
[4 \varepsilon (d-1) (d-2)^{-1} A_d^{-1} A_{surface}^+ + 2 ]
\xi/L \; + \; O (\xi^2 / L^2) \right\}
\ee
where $A_{surface}^+ < 0$ is given by Eq. (\ref{gleichung40}).
Eq. (\ref{gleichung200}) yields the surface susceptibility, Eq.
(\ref{gleichung22c0}), with the scaling surface exponent
\be
\label{gleichung203} \gamma_s  \; = \gamma + \nu \; = \; 3 / (d-2)
\ee
and the surface amplitude
\be
\label{gleichung204} A_{\chi, surface}^+ \; = \; - \frac{1}{2} \;
J_0^{-1} \; \xi_0^3 \;  [4 \varepsilon (d-1) (d-2)^{-1} A_d^{-1} A_{surface}^+ + 2
] \; .
\ee

\section*{2. Violation of finite-size scaling in ${\bf d = 3}$ \\dimensions}

We recall that at $d=3$  there exists no sharp transition in box
geometry and in film geometry of finite thickness other than the
bulk transition for $L \to \infty$ at $T_{c,3} > 0$, thus there is
no compelling reason to introduce a shifted reference temperature
or to introduce a physical length scale other than the $d=3$ {\it
bulk} correlation length. At $d = 3$ the Gaussian surface function
reads [compare Eq. (\ref{gleichung63})]
\be
\label{gleichung205} E_{surface} (\Delta \mu \tilde a^2) = - (4
\pi)^{-1} \ln \; (\Delta \mu \tilde a^2) + 8 \left[\tilde b - (16
\pi)^{-1} \right] + O (\Delta \mu^{1/2} \tilde a) \; .
\ee
Substituting Eq. (\ref{gleichung205}) into Eq.
(\ref{gleichung185}) yields
\be
\label{gleichung206} \Delta \mu^{1/2} L &=& t
L/\xi_0 \; + \; \ln \; (\Delta \mu^{1/2} \tilde a)
+ 1/2 - 8 \pi \tilde b  \nonumber\\
&+& 8 \pi {\cal \widetilde E}_3 (\Delta \mu^{1/2} L, s) \; + \; O
(\Delta \mu^{1/2} \tilde a)
\ee
with ${\cal \widetilde E}_3 (x, s)$ given by Eq.
(\ref{gleichung191}). Here we have replaced $\tilde s$ by $s =
L/\widetilde L$ for large $L/\tilde a \gg 1$. Substituting Eq.
(\ref{gleichung206}) into Eq. (\ref{gleichung-203}) yields the
susceptibility. As a consequence of the logarithmic term of
$E_{surface}$ in Eq. (\ref{gleichung205}) we now have a
logarithmic dependence on $\tilde a$ in Eq. (\ref{gleichung206})
that cannot be neglected. One expects that this causes only a {\it
logarithmic} deviation from finite-size scaling. This will be
confirmed for box geometry but not for film geometry where we
shall find a power-law violation of finite-size scaling. The
origin for this unexpected geometry effect at $d = 3$ comes from
the different small $x$ behavior of the finite-size part ${\cal
\widetilde E}_3 (x,s)$ for $s > 0$ and for $s = 0$.

{\large \bf{2.1. Box geometry}}

For fixed $s > 0$ we find from Eq. (\ref{gleichung191}) the small
$x$ behavior
\be
\label{gleichungC.11} {\cal \widetilde E}_3 (x, s) \; = \;
\frac{1}{2} \; s^2 x^{-2} \; - \; \frac{1}{8 \pi} \; \ln x \; + \; O (1) \; .
\ee
Here the first term $\sim x^{-2}$ is the contribution due to the
${\bf k} = {\bf 0}, p = \pi / (L + \tilde a)$
mode which is the lowest mode of the discrete mode spectrum for box geometry.
Eqs. (\ref{gleichung206}) and (\ref{gleichungC.11}) yield the leading
$L$ dependence at $T = T_{c,3}$
\be
\label{gleichung207} \Delta \mu \; = \; 4 \pi s^2 L^{-2} \left[\ln
(L/\tilde a) \right]^{-1} \;
\left[1 \; + \; O \left(\frac{1}{\ln (L/\tilde a)} \right) \right] \; .
\ee
Substituting Eq. (\ref{gleichung207}) into Eq.
(\ref{gleichung-203}) leads to the susceptibility at $T = T_{c,3}$
\be
\label{gleichung208} \chi \; = \; 2 \pi^{-3} J_0^{-1} s^{-2} L^2
\; \ln (L/\tilde a) \; + \; O (L^2) \; .
\ee
At fixed $t > 0$ we find from Eq. (\ref{gleichung206}) the leading
large-$L$ behavior
\be
\label{gleichung165}
\tilde \mu^{1/2} = J_0^{1/2} \xi
^{-1} \left\{1 - \left[\ln (\xi / \tilde a) + 8 \pi \tilde b -
2 \right] \xi / L + O  ([\ln (\xi/\tilde a)]^2 \xi^2 / L^2)
\right\} \; .
\ee
For the susceptibility, Eq. (\ref{gleichung-203}), this implies
\be
\label{gleichung210} \chi = \chi_b \left\{1 + \left
[2 \ln (\xi/\tilde a) + 16 \pi \tilde b - 6 \right]
\xi / L + O ([\ln (\xi/\tilde a)]^2 \xi^2 / L^2) \right\} \;
\ee
which corresponds to the surface susceptibility at $d = 3$
\be
\label{gleichung162} \chi_{surface} (t) \; = \; \frac{1}{2} \;
J_0^{-1} \; \xi^3
\left[2 \ln (\xi/\tilde a) + 16 \pi \tilde b - 6 \right] \; .
\ee
The logarithms in Eqs. (\ref{gleichung208}) - (\ref{gleichung162})
constitute logarithmic deviations from universal finite-size
scaling, with an explicit dependence on the nonuniversal lattice
constant $\tilde a$. Thus there exists no universal scaling form
for box geometry at $d = 3$ in the sense of Eq.
(\ref{gleichung27-a}). This is the consequence of the upper
borderline dimension $d^* = 3$ for the surface energy density of
the Gaussian model.

{\large \bf{2.2. Film geometry}}

A separate analysis is necessary for film geometry in $d = 3$ dimensions
since Eqs. (\ref{gleichung207}) and (\ref{gleichung208}) do not have finite
limits for $s \to 0$. The susceptibility is again given by Eq.
(\ref{gleichung-203}) where now $\Delta \mu L^2$ is determined
implicitly by
\be
\label{gleichung-224} \Delta \mu^{1/2} L = t L/\xi_0 + \ln (\Delta
\mu^{1/2} \tilde a) + 1/2 - 8 \pi \tilde b
+ 8 \pi \widetilde {\cal E}_3 (\Delta \mu^{1/2} L, 0)
\ee
with
\be
\label{gleichung-225} \widetilde {\cal E} (x, 0) = \frac{1}{16
\pi} \int\limits_0^\infty dy \; y^{-1} \left\{1 - 2
\left(\frac{\pi}{y} \right)^{1/2} + e^{y/4} \left[K
\left(\frac{y}{4} \right) - 1 \right] \right\} e^{- y x^2 / 4
\pi^2} \;.
\ee
We find from Eq. (\ref{gleichung-225}) the small $x$
behavior
\be
\label{gleichungC.12} {\cal \widetilde E}_3 (x, 0) \; = \; - \;
\frac{3}{8 \pi} \; \ln x \; + \; \widetilde E_3 + O (x^2 \ln x) \;
\ee
where
\be
\label{gleichung227} \widetilde E_3 &=& \frac{3}{16 \pi} \left[- C
_E + \ln \left(\frac{A}{4 \pi^2} \right) \right] \nonumber\\
&+& \frac{1} {16 \pi} \int\limits_0^A dy \; y^{-1} \left\{1 - 2
\left(\frac{\pi}{y} \right)^{1/2} + e^{y/4} \left[K
\left(\frac{y}{4} \right) - 1 \right] \right\} \nonumber\\
&+& \frac{1} {8 \pi} \int\limits_A^\infty dy \; y^{-1} \left\{ -
\left(\frac{\pi}{y} \right)^{1/2} + \sum_{n = 2}^\infty e^{- (n^2
- 1) y/4} \right\}
\ee
is independent of the arbitrary constant $A > 0$. The absence of a
term $\sim x^{-2}$ in Eq. (\ref{gleichungC.12}) is a consequence
of the fact that for film geometry there exists a mode {\it
continuum}, without a discrete lowest mode. At the bulk critical
temperature $T = T_{c,3}$ Eqs. (\ref{gleichung-224}) and
(\ref{gleichungC.12}) yield the constraint equation in the form
\be
\label{gleichung228} \Delta \mu^{1/2} L \; = \; &&\ln (\Delta \mu^{1/2}
\tilde a) + 1/2 + 8 \pi (\widetilde E_3 - \tilde b)
\nonumber\\
&&-3 \ln (\Delta \mu^{1/2} L) \; + \; O [\Delta \mu L^2 \ln
(\Delta \mu L^2 ) ] \; .
\ee
We recall that the first logarithmic term in Eq.
(\ref{gleichung228}) is the signature of the (upper) borderline
dimension $d^* = 3$ at which the critical exponent $1 - \alpha -
\nu$ of the Gaussian surface energy density vanishes whereas the
second logarithmic term in Eq. (\ref{gleichung228}) is the
signature of the (lower) borderline dimension $d_l = 3$ at which
the film critical temperature $T_{c,3} (L)$ vanishes, in accord
with the Mermin - Wagner theorem \cite{mermin}. Both logarithmic
terms can be combined as
\be
\label{gleichung229} \ln (\Delta \mu^{1/2} \tilde a) - 3 \ln
(\Delta \mu^{1/2} L) \; = \; \ln (\tilde a / L) \; - \; \ln
(\Delta \mu L^2)
\ee
and, after substituting into Eq. (\ref{gleichung228}), yield the
solution with a power-law (rather than logarithmic) $L$ dependence
\be
\label{gleichung212} \Delta \mu \; = \; \widetilde A_\mu
\tilde a L^{-3} \; [1 \; + \; O (\tilde a^{1/2} L^{- 1/2}) ]
\ee
with the nonuniversal amplitude
\be
\label{gleichung213} \widetilde A_\mu \; = \; \exp \left\{
1/2 + 8 \pi (\widetilde E_3 - \tilde b) \right\} \; .
\ee
Thus $\Delta \mu$ has a nonscaling size dependence at $T =
T_{c,3}$. Substituting Eq. (\ref{gleichung212}) into Eq.
(\ref{gleichung-203}) leads to the susceptibility at $T_{c,3}$
\be
\label{gleichung214} \chi \; = \; (J_0 \widetilde A_\mu \tilde
a)^{-1}  L^3 \left[1 \; + \; O (\tilde a^{1/2} L^{- 1/2}) \right]
\ee
which constitutes a strong power-law violation of the scaling
prediction $\chi \sim L^2$, Eq. (\ref{gleichung32a}), in contrast
to the logarithmic deviation in Eq. (\ref{gleichung208}) for box
geometry. The same violation persists in the critical region $\xi
L \gtrsim 1$. The unexpected difference between Eqs.
(\ref{gleichung214}) and (\ref{gleichung208}) results from the
difference between the discrete mode spectrum for box geometry and
the mode continuum for film geometry at the lower borderline
dimension $d_l = 3$ above which the film critical temperature
becomes finite. We emphasize that the existence of this borderline
dimension $d_l = 3$ which causes the second logarithmic term in
the constraint equation (\ref{gleichung-224}) is not restricted to
the spherical model. It remains to be seen whether similar
geometry-dependent effects exist at $d = 3$ also in other models
of $O (n)$ symmetric systems with $n \geq 2$ and with free (or
Dirichlet) boundary conditions.

At fixed $T > T_{c,3}$ for film geometry, we find the same leading
large-$L$ behavior as already given in Eqs. (\ref{gleichung165}) -
(\ref{gleichung162}) for box geometry, with a logarithmic (rather
than power-law) violation of finite-size scaling. In summary, it
is not possible to write $\chi$ in a universal finite-size scaling
form, in the sense of Eqs. (\ref{gleichung0}) and
(\ref{gleichung-2}),
in the region $T \geq T_{c,3}$ for film geometry at $d = 3$.

{\large \bf{2.3. Comparison with Barber and Fisher}}

The susceptibility of the mean spherical model in film geometry
with free boundary conditions was calculated by Barber and Fisher
\cite{barber-1973} by a different mathematical technique.
For $d = 3$ they introduced a "quasicritical temperature shift"
and found that, for large $n \equiv L/\tilde a$, there exists a
scaling representation in terms of a scaled temperature variable
$n \Delta \dot K$
\be
\label{gleichung233} \chi_{BF} \; = \; \frac{n^2} {2 J} \; X (n
\Delta \dot K)
\ee
with a shifted inverse temperature deviation from the $d = 3$ bulk
critical temperature $T_{c,3}$
\be
\label{gleichung234} \Delta \dot K &=& \frac{J} {k_B T_{c,3}} \;\; -
\; \; \frac{J} {k_B T} \;\; - \; \;\frac{1} {8 \pi n} \ln \; n \; \; +
\;\; \widetilde C_{BF}/n \; ,\\
\label{gleichung235}
\widetilde C_{BF} &=& \frac{1}{2} \left[W_3 (0) \;\; - \; \; \frac{1}
{2} W_2 (4) \;\; - \;\; \frac{7 \ln 2} {16 \pi} \right] \; .
\ee
The scaling function was represented parametrically via
\be
\label{gleichung236} X (z) &=&  y^{-2} \left[1 - (2/y) \tanh
(y/2) \right] \; , \\
\label{gleichung237} 8 \pi z &=& \ln [(\sinh \; y) / y ]
\ee
and was plotted for $z > 0$ in Fig. 4 of Ref.
\cite{barber-1973}. For finite $L / \tilde a \gg 1$ and at $\Delta
\dot K = 0$, Eq. (\ref{gleichung234}) defines a "quasicritical"
\cite{barber-1973} temperature $\widetilde T (L) > T_{c,3}$ where
\be
\label{gleichung238} \frac{J} {k_B \widetilde T (L)} \;\; = \; \;
\frac{J} {k_B T_{c,3}} \; \; - \; \; \frac{\tilde a} {8 \pi L} \;
\ln \left(\frac{L} {\tilde a}\right) \; \; + \; \; \widetilde C_{BF}
\;\; \tilde a / L \; .
\ee
We note that all thermodynamic quantities are smooth functions of
$T$ near $\widetilde T (L)$ and that there exists no {\it
physical} criterion for defining $\widetilde T (L)$. The
analysis of the susceptibility in terms of $\Delta \dot K$ in
Ref. \cite{barber-1973} was restricted to $\Delta \dot K
\geq 0$ and did not include the region $T_{c,3} \leq T <
\widetilde T (L)$, $\xi / L \gg 1$ which is of principal
interest for testing the scaling predictions of Eqs.
(\ref{gleichung0}) and (\ref{gleichung-2}).

Our dimensionless susceptibility $\chi / \tilde a^2$ corresponds
to $\chi_{BF}$. Our explicit result for $\chi / \tilde a^2$,
however, is at variance with that of Barber and Fisher since Eqs.
(\ref{gleichung-203}), (\ref{gleichung-224}) and
(\ref{gleichung-225}) cannot be reduced to the simple form of Eqs.
(\ref{gleichung233}) - (\ref{gleichung237}). Our result differs
from that of Ref. \cite{barber-1973} both in the region $T_{c,3}
\leq T < \widetilde T (L)$ and in the region $T \geq \widetilde T
(L)$. A unique analytic comparison between $\chi / \tilde a^2$ and
$\chi_{BF}$ can be made in the region $L / \xi \gg 1$ at fixed $T$
above $\widetilde T (L)$. Substituting into $\chi_{BF}$ the
large-$z$ representation according to Eq. (9.17) of Ref.
\cite{barber-1973}
\be
\label{gleichung239} X (z) \; = \; (8 \pi z)^{-2} \; - \; 2 (8 \pi
z)^{-3} \; \ln (16 \pi e z ) \; + \; O (z^{-4} \; \ln z)
\ee
and rewriting the resulting expression in terms of the bulk
susceptibility $\chi_{BF}^{bulk}$ and the asymptotic bulk
correlation length [compare our Eqs. (\ref{gleichung138}),
(\ref{gleichung142})]
\be
\label{gleichung240} \xi &=& \frac{\tilde a} {8 \pi} \; \; \;
\frac{k_B T_{c,3}} {J} \; t^{-1}  \; ,
\ee
we find
\be
\label{gleichung241} \chi_{BF} &=& \chi_{BF}^{bulk} \left\{ 1 +
\left[2 \ln \left(\frac{\xi}{\tilde a}\right) \; + \; C_{BF}
\right] \frac{\xi} {L} \; + O \left(\frac{\xi^2} {L^2} \right)
\right\}
\ee
with
\be
\label{gleichung242} C_{BF} &=& - \; 2 \; + \; \frac{3}{2} \; \ln 2
\; + \; 16 \pi \left[ \frac{1}{4} \; W_2 (4) \; - \; \frac{1}{2} \; W_3
(0) \right] \; .
\ee
While the leading logarithmic term $\sim [\ln (\xi / \tilde a)]
\xi / L$ of Eq. (\ref{gleichung241}) agrees with ours in Eq.
(\ref{gleichung210}), the leading correction term $\sim \xi/L$
differs from ours. We note that our constant $\tilde b$, Eq.
(\ref{gleichung-82}), does contain the last two terms of $C_{BF}$
but the additional integral expressions in Eq.
(\ref{gleichung-82}) are missing in $C_{BF}$, Eq.
(\ref{gleichung242}). Our integral expressions come from the
finite part of $W_{d-1} (0)$ for $d \to 3$, after subtracting the
pole term $\sim (d-3)^{-1}$, see Eq. (\ref{gleichung91}). We
believe that our result for $\tilde b$ is correct since it has
been obtained both by a calculation directly at $d = 3$ and by a
calculation at $d \neq 3$, taking the limits $d \to 3 +$ and $d
\to 3 -$. A further analytic comparison between our $\chi / \tilde
a^2$ and $\chi_{BF}$ can be made directly at $T = \widetilde T
(L)$ where $\chi_{BF}$ is simply given by $\chi_{BF} = (n^2 / 2 J)
X (0)$ with $X (0) = 1/12$ according to Eq. (9.18) of Ref.
\cite{barber-1973}. Our result at $T = \widetilde T (L)$ depends
on $\tilde b$ and differs from the simple result for $\chi_{BF}$.
Thus we doubt the correctness of the previous result
\cite{barber-1973} for $\chi_{BF}$ for free boundary conditions at
$d = 3$.

\section*{3. Violation of finite-size scaling in ${\bf 3 < d < 4}$\\
dimensions}

From Eqs. (\ref{gleichung-d70}), (\ref{gleichung53}) and
(\ref{gleichung86}) we have the Gaussian surface function for $d > 3$
\be
\label{gleichung215} E_{surface} (\Delta \mu \tilde a^2) &=& 8
\tilde a^{3-d} \widetilde B_d \; - \; 4 (d-1) A_{surface}^+
(\Delta \mu \tilde a^2)^{(d-3)/2} \nonumber\\
&+& O \left((\Delta \mu \tilde a^2)^{(d-2) / 2} \right)
\; ,
\ee
where $A_{surface}^+ > 0$ and $\widetilde B_d$ are given by Eqs.
(\ref{gleichung49b}) and (\ref{gleichung56}). Substituting Eq.
(\ref{gleichung215}) into Eq. (\ref{gleichung185}) yields
\be
\label{gleichung216} (\Delta \mu L^2)^{(d-2)/2} = t
(L/\xi_0)^{d-2}
+ 2 \varepsilon (d-1) A_d^{-1} A_{surface}^+ (\Delta \mu
L^2)^{(d-3)/2} \nonumber\\
- 4 \varepsilon A_d^{-1} \widetilde B_d (L / \tilde a)^{d-3}
+ 2 \varepsilon A_d^{-1} {\cal \widetilde E}_d (\Delta \mu^{1/2} L,
s)\; .\qquad  \qquad
\ee
We see that the finite value of $E_{surface} (0) \sim \widetilde B_d >
0$ causes a nonnegligible term $\sim (L / \tilde a)^{d-3}$ in Eq.
(\ref{gleichung216}) that depends on the lattice spacing $\tilde
a$. This will imply power-law violations of finite-size scaling for
box geometry.

At fixed $s = L/\widetilde L > 0$, Eq. (\ref{gleichung216})
yields the $L$ dependence at $T = T_{c,d}$ for $3 < d < 4$

\be
\label{gleichung217} \Delta \mu \; = \; \frac{1}{2} \; \tilde a^{d-3}
\widetilde B_d^{-1} \; s^{d-1} L^{1-d} \left[1 \; + \; O \left(\tilde
a^{d-3} L^{3-d} \right) \right] \; .
\ee
For the susceptibility at $T = T_{c,d}$ this implies
\be
\label{gleichung218} \chi \; = \; 4 \; J_0^{-1} \; \tilde a^{3-d}
\; \widetilde B_d \; s^{1-d} L^{d-1} \left[1 \; + O \;
\left(\tilde a^{d-3} L^{3-d} \right) \right] \; ,
\ee
in contrast to the scaling prediction $\chi \sim L^2$, Eq.
(\ref{gleichung32a}).

At fixed $T > T_{c,d}$ we find from Eq.
(\ref{gleichung216}) the leading large-$L$ behavior
\be
\label{gleichung219} \tilde \mu^{1/2} \; = \; J_0^{1/2} \;
\xi^{-1} \left[1 - \tilde m_d (\xi / \tilde a) \xi/L \; + \; O
((\xi/\tilde a)^{d-3} \xi^2 / L^2) \right]
\ee
with the nonuniversal function for $3 < d < 4$
\be
\label{gleichung220} \tilde m_d (\xi / \tilde a) \; = \; 2
\varepsilon (d - 2)^{-1} \; A_d^{-1} \; \left[2 \widetilde B_d
(\xi/\tilde a)^{d-3} \; - \; (d - 1) A^+_{surface} \right] \; .
\ee
Substituting Eqs. (\ref{gleichung219}) and (\ref{gleichung220})
into Eq. (\ref{gleichung-203}) yields the susceptibility
\be
\label{gleichung221} \chi \; = \; \chi_b \left\{1 \; + \; \left[2
\tilde m_d (\xi / \tilde a) - 2\right] \xi/L \; + \; O
((\xi/\tilde a)^{d-3} \xi^2 / L^2) \right\} \; .
\ee
The corresponding surface susceptibility is for $\xi \gg \tilde a$
and for $3 < d < 4$
\be
\label{gleichung222} \chi_{surface} \; = \; \frac{1}{2} \;
J_0^{-1}\; \left[2 \tilde m_d (\xi/\tilde a) - 2 \right]
\xi^3 \; \sim \; \xi^d \; \sim \;  t^{- d / (d-2)} \; .
\ee
This is in contrast to the scaling prediction $\chi_{surface} \sim
\chi_b \; \xi \sim t^{-3/ (d-2)}$, Eqs. (\ref{gleichung22c}) and
(\ref{gleichung22c0}).

We see that the amplitudes of the leading nonscaling terms are
proportional to $\widetilde B_d$, both for the size dependence at
$T_{c,d}$\;, Eq. (\ref{gleichung218}), and for the temperature
dependence above $T_{c,d}$\;, Eqs. (\ref{gleichung221}) and
(\ref{gleichung222}). Thus it is the cusp of the Gaussian surface
energy density that is the origin of the nonscaling effects in the
mean spherical model for $3 < d < 4$ rather than an enhanced
transition temperature which does not exist for box geometry.

For film geometry in $3 < d < 4$ dimensions a new analysis of our
solution would be necessary since there exists a sharp critical temperature
$T_{c,d} (L) > T_{c,d} (\infty)$. Here we only note that the leading size
dependence at fixed $T > T_{c,d} (L)$ for film geometry is the same
as given in Eqs. (\ref{gleichung219}) - (\ref{gleichung222}) for box geometry.
It is expected that a full description of the crossover from the
$L$-dependent film critical behavior near $T_{c,d} (L)$ to the $d$
dimensional bulk critical behavior near $T_{c,d} (\infty)$ would
involve {\it two} different correlation lengths. Our solution for
$\chi$ does provide the basis for such an analysis of a dimensional
crossover which, however, is beyond the scope of our present paper.

{\bf Acknowledgment}\\
Support by NASA under contract number 1226553 and by DLR under
contract number 50WM9911 is acknowledged. One of the authors (V.D.)
is grateful to Max-Planck-Gesellschaft for supporting his visit at the
Institute of Theoretical Physics of the Chinese
Academy of Sciences at Beijing where part of this work was completed. Hospitality at
the Chinese Academy of Sciences is also gratefully acknowledged.

\newpage

\section* {Appendix A. Gaussian free energy density}

\renewcommand{\theequation}{A\arabic{equation}}
\setcounter{equation}{0}%

In this Appendix we derive the asymptotic form, Eq.
(\ref{gleichung31}), for the free energy density of the Gaussian lattice
model for box geometry. We start from Eq. (\ref{gleichung24}). Using the
representation
\be
\label{gleichungA1} \ln \; z = \int\limits_0^\infty dy \; y^{-1}
(e^{-y} - e^{- z y} )
\ee
we rewrite the $L$ dependent part
\be
\label{gleichungA2} \Delta f (t, L, \widetilde L, \tilde a)
\equiv f (t, L, \widetilde L, \tilde a) - f_b (t) + \frac{1}{2}
\; \tilde a^{-d} (\widetilde L / \tilde a)^{1-d} \ln 2
\ee
of the free energy density in the form
\be
\label{gleichungA3} \Delta f (t, L, \widetilde L, \tilde a) =
\frac{1}{2} \; \tilde a^{-d} \int\limits^\infty_0 dy \; y^{-1} \; e^{-
\tilde r_0 y} \; \Phi (L/\tilde a, \widetilde L/\tilde a, y)
\ee
with $\tilde r_0 \equiv r_0 \tilde a^2 / (2 J)$ and
\be
\label{gleichungA4} \Phi (L / \tilde a, \widetilde L / \tilde a,
y) = \Big[S (\infty, y) \Big]^d - \left[S (\widetilde L /
\tilde a, y) \right]^{d-1} S_D (\widetilde L / \tilde a, y)
\ee
where
\be
\label{gleichungA5} S (\widetilde L / \tilde a, y) = (\tilde a /
\widetilde L) \sum_k \exp \left[- 2 y (1 - \cos k) \right] \; ,
\ee
\be
\label{gleichungA6} S (\infty, y) = (2 \pi)^{-1} \int\limits_{-
\pi}^\pi d x \exp \left[- 2 y (1 - \cos x) \right] \; ,
\ee
\be
\label{gleichungA7} S_D (L / \tilde a, y) = (\tilde a /L) \sum_p
\exp \left[- 2 y (1 - \cos p) \right] \; .
\ee
The sums $\sum_k$ and $\sum_p$ run over dimensionless wave numbers
in the range $- \pi \leq k = 2 \pi \tilde a m / \widetilde L <
\pi$ and $0 < p = \pi \tilde a n / (L + \tilde a) < \pi$ with
integers $m = 0, \pm 1, \pm 2, \ldots$ and $n = 1, 2, \ldots,
L/\tilde a$, as is appropriate for periodic and Dirichlet boundary
conditions, respectively. In determining the large $L / \tilde a$ and large
$\widetilde L / \tilde a$ behavior of $\Delta f$ at fixed finite
ratio $L / \widetilde L$ it is important to distinguish the
regimes $0 \leq y \lesssim y_0$ and $y \gtrsim y_0$ with
\be
\label{gleichungA8} y_0 \; = \; \frac{L + \tilde a} {\tilde a}
\ee
in the integral representation (\ref{gleichungA3}). Accordingly we split
\be
\label{gleichungA8a} \Delta f = \frac{1}{2} \; \tilde a^{-d}
\left(\Delta f_1 + \Delta f_2 \right) \qquad \qquad \quad
\ee
where
\be
\label{gleichungA9} \Delta f_1 = \int\limits_0^{y_0} dy \; y^{-1}
\; e^{- \tilde r_0 y} \; \Phi (L / \tilde a, \widetilde L / \tilde a,
y) \; ,
\ee
\be
\label{gleichungA10} \Delta f_2 = \int\limits_{y_0}^\infty dy \; y^{-1}
\; e^{- \tilde r_0 y} \; \Phi (L / \tilde a, \widetilde L / \tilde a,
y) \; .
\ee
First we derive the leading $L / \tilde a$ and $\widetilde L /
\tilde a$ dependence of $\Delta f_1$. Since $\cos k$ is a periodic
function the sum $S (\widetilde L / \tilde a, y)$ satisfies the
Poisson identity \cite{morse-feshbach}
\be
\label{gleichungA11} S (\widetilde L / \tilde a, y) = \sum_{N = -
\infty}^\infty \; (2 \pi)^{-1} \int\limits_{- \pi}^\pi dk \; e^{i
k N \widetilde L / \tilde a} \exp [- 2 y (1 - \cos k) ]
\ee
\be
\label{gleichungA12} = S (\infty, y) + 2 e^{- 2 y} \;\sum_{N =
1}^\infty \; F (N \widetilde L / \tilde a, y) \quad
\ee
with
\be
\label{gleichungA13} F (M, y) = (2 \pi)^{-1} \;
\int\limits_{- \pi}^\pi \; dk \; e^{i k M} \; \exp (2 y \cos k)
= I_M (2 y)
\ee
\be
\label{gleichungA14} S (\infty, y) = e^{- 2 y} \; I_0 (2 y) \; ,
\qquad \qquad \qquad \qquad \qquad \qquad \quad
\ee
where
\be
\label{gleichungA15} I_M (z) = \frac{1} {\pi} \;
\int\limits_0^\pi \; d \theta \; e^{z \cos \theta} \; \cos (M
\theta)
\ee
are the Bessel functions of integer order $M$ (see, e.g., 9.6.19
of Ref. \cite{handbook}, see also Eqs. (3.6) - (3.10) of Ref.
\cite{chen-dohm-2000}). For $y < y_0$ at fixed $L / \widetilde
L$, the large $\widetilde L / \tilde a$ behavior of $F (N
\widetilde L / \tilde a, y)$ for $N \geq 1$ is $F (N \widetilde L
/ \tilde a, y) \sim O (e^{- \widetilde L / \tilde a})$, thus
\be
\label{gleichungA16} S (\widetilde L / \tilde a, y) = S (\infty,
y) + O (e^{- \widetilde L / \tilde a}) \; . \qquad \qquad
\ee
In order to determine the leading $\L / \tilde a$ dependence
of the sum $S_D (L / \tilde a, y)$ for $y < y_0$ we first derive
a representation of the one-dimensional integral
\be
\label{gleichungA17}
I (a,b) \; = \; \int\limits^b_a f(x) dx
\ee
in terms of summations. The derivation is similar to that in Eqs.
(A.21) - (A.30) of Ref. \cite{chen-dohm-1998-a}.
We assume the arbitrary real function $f(x)$
of the real variable $x$ to be well behaved in the interval
$a \leq x \leq b$, in particular we assume that $f(x)$ has a convergent
Taylor expansion around any $x$ in this interval. We split the interval
$a \leq x \leq b$ into $N$ subintervals of length $\Delta x = (b-a)/N > 0$
between the points $x_i = a + i \Delta x$, $i = 0, 1, \cdots, N$, with
$x_0 = a, x_N = b$. The integral $I$ can be represented as
\be
\label{gleichungA18}
I (a,b) \; = \;  \sum\limits^{N-1}_{i=0} \; \int\limits^{x_{i+1}}_{x_i}
f(x) dx   \;.
\ee
For each interval we expand $f(x)$ into a Taylor series around
$x_{i + 1}$ (rather than around $x_i$ as in Ref. \cite{chen-dohm-1998-a})
\begin{eqnarray}
\label{gleichungA19}
\int\limits^{x_{i+1}}_{x_i} f(x)dx & = &  \int\limits^{x_{i+1}}_{x_i}
\left [ f(x_{i+1}) \; + \; \sum\limits_{n=1}^\infty \; \frac{1}{n!} \; f^{(n)}
(x_{i+1}) (x-x_{i+1})^n \right ] dx  \qquad \\
& = & f(x_{i+1}) \Delta x \; + \;
\sum\limits_{n=1}^\infty \; \frac{(-1)^n}{(n+1)!} \; f^{(n)}
(x_{i+1}) (\Delta x)^{n+1}
\end{eqnarray}
where $f^{(n)} (x) \;  \equiv \;  d^n f(x) /dx^n$. Thus we obtain
\be
\label{gleichungA21}
\int\limits_a^b \; f(x) d x \; = \; \sum\limits^{N-1}_{i=0}
f(x_{i+1}) \Delta x \; + \; \sum\limits^{\infty}_{n=1}
\; \frac{(-1)^n (\Delta x)^n}{(n+1)!}
K_N^{(n)} (a,b)
\ee
where
\be
\label{gleichungA22}
K_N^{(n)}(a,b) \; = \;
\sum\limits^{N-1}_{i=0} f^{(n)} (x_{i+1}) \Delta x \; .
\ee
Since $f(x)$ is an arbitrary function we may also apply Eq.
(\ref{gleichungA21}) to the function $f'(x)$ instead of $f(x)$.
This yields an expression for $K^{(1)}_N (a,b)$ in terms of higher
derivatives,
\be
\label{gleichungA23}
K_N^{(1)}(a,b) \; = \;
f(b) \; - \; f(a) \; - \; \sum\limits_{n=1}^{\infty}
\frac{(-1)^n (\Delta x)^n}{(n+1)!}
\; K_N^{(n+1)} (a,b)\; ,
\ee
which can be substituted into the $n=1$ term of Eq.
(\ref{gleichungA21}). Successive application of this procedure
permits one to express the difference
\be
\label{gleichungA24}
\int\limits^b_a f(x) dx \; - \;
\sum\limits^{N-1}_{i=0} f(x_{i+1}) \Delta x \;  \equiv \;
\widetilde R_N (a,b)
\ee
in terms of the differences of the derivatives at $a$ and $b$,
\be
\label{gleichungA25}
\Delta f^{(k)} \; = \; f^{(k)} (b) \; - \; f^{(k)} (a) \quad .
\ee
Note that $\widetilde R_N (a,b)$ differs from $R_N (a,b)$ of Ref.
\cite{chen-dohm-1998-a}. The result is
\begin{eqnarray}
\label{gleichungA26}
\widetilde R_N(a,b) \; = \; - \; \frac{\Delta x}{2} \big [ f(b) - f (a) \big ]
\; - \; \frac{(\Delta x)^2}{12} \Delta f^{(1)}
\; + \; {\cal{O}} \left( (\Delta x)^4 \right ) \; .
\end{eqnarray}
The coefficient of the ${\cal{O}} \left ((\Delta x)^3 \right )$
term vanishes. Since $\Delta x \sim {\cal{O}}(N^{-1})$ this representation is
expected to converge rapidly for large $N$ if $\Delta f^{(k)}$
remains sufficiently well-behaved for large $k$. Eq.
(\ref{gleichungA26}) differs from Eq. (A.30) of Ref.
\cite{chen-dohm-1998-a} by a minus sign in the first term.

We apply Eqs. (\ref{gleichungA24}) - (\ref{gleichungA26}) to the
integral
\be
\label{gleichungA27} S(\infty, y) = \frac{1} {\pi}
\int\limits_0^\pi dp \; \exp [- 2 y \; (1 - \cos p) ]
\ee
where the integration variable $p$ plays the role of $x$ in
the integral of Eq. (\ref{gleichungA24}). The sum on the
left-hand side of Eq. (\ref{gleichungA24}) now corresponds to
\be
\label{gleichungA28} \frac{1}{\pi} \sum_{n = 1}^N \Delta p \;
\exp \Big\{- 2 y \left[1 - \cos (n \Delta p) \right] \Big\}
\equiv \widetilde I_N (y)
\ee
with $\Delta p = \pi / N$. Setting $N = L / \tilde a$ we see
that
\be
\label{gleichungA29} \widetilde I_{L/\tilde a} (y) \; = \; \frac{L +
\tilde a} {L} \; S_D (L / \tilde a, y) \; .
\ee
We note that the derivative of the integrand of Eq.
(\ref{gleichungA27}) with respect to $p$ vanishes at $p = 0$ and $p = \pi$. Eqs.
(\ref{gleichungA24}) - (\ref{gleichungA29}) yield
the leading large $L / \tilde a$ behavior, for $y < (L
+ \tilde a) / \tilde a$,
\be
\label{gleichungA30} S_D (L / \tilde a, y) = S (\infty, y) -
\frac{\tilde a} {2 L} \left[1 + e^{- 4 y} - 2 e^{- 2 y} I_0 (2 y)
\right] + O (\tilde a^4 / L^4) \; .
\ee
In order to ensure that the discretization
intervals $\Delta p$ become sufficiently small for large $L/\tilde a$
the restriction $y \lesssim O (L /\tilde a)$ was necesssary. For
this reason, Eq. (\ref{gleichungA30}) is applicable only to
$\Delta f_1$, Eq. (\ref{gleichungA9}), but not to $\Delta f_2$,
Eq. (\ref{gleichungA10}). Using Eq. (\ref{gleichungA16}) and
substituting Eq. (\ref{gleichungA30}) into Eqs. (\ref{gleichungA4})
and (\ref{gleichungA9}) we arrive at
\begin{eqnarray}
\label{gleichungA31} \Delta f_1 = \frac{\tilde a}{2 L}
\int\limits_0^{y_0} &dy& \left\{y^{-1} \; e^{- \tilde r_0 y}
\left[S (\infty, y) \right]^{d-1} \left[1 + e^{- 4 y} - 2 S
(\infty, y) \right] \right\} \qquad \nonumber\\ &+& O \left(\tilde a^4 / L^4, \; e^{-
\widetilde L / \tilde a} \right) \; .
\end{eqnarray}
This can be combined with $\Delta f_2$ in the form
\begin{eqnarray}
\label{gleichungA32} \Delta f_1 + \Delta f_2 &=& \frac{\tilde
a}{2 L} \int\limits_0^\infty dy \left\{y^{-1} e^{- \tilde r_0 y}
[S (\infty, y)]^{d-1} \left[1 + e^{- 4 y} - 2 S (\infty, y)
\right] \right\} \qquad \nonumber\\
&+& \Delta f_3 (L / \tilde a, \widetilde L / \tilde a) \; + \; O
(\tilde a^4 / L^4, \; e^{- \widetilde L / \tilde a}) \; ,
\end{eqnarray}
\begin{eqnarray}
\label{gleichungA33} \Delta f_3 &=& \int\limits_{y_0}^\infty dy \;
y^{-1} \; e^{- \tilde r_0 y} \Big\{ [S (\infty, y)]^d \; (1 + \tilde a
/ L) \nonumber\\ &-& [S (\widetilde L / \tilde a, y)]^{d-1} \; S_D (L
/ \tilde a, y) \; - \; \frac{\tilde a}{2 L} \; [S (\infty, y)]^{d-1}
\Big\} \; .
\end{eqnarray}
The integral term in Eq. (\ref{gleichungA32}) represents the
surface contribution of $O (L^{-1})$ to $\Delta f$ whereas $\Delta
f_3$ will yield the finite-size part of $O (L^{-d})$. Since $y >
y_0$ in Eq. (\ref{gleichungA33}) is sufficiently large it
suffices to use the small $k$ approximation $- 2 y (1 - \cos k)
\approx - k^2 y$ in Eq. (\ref{gleichungA5}), and similarly in
Eqs. (\ref{gleichungA6}) and (\ref{gleichungA7}),
\begin{eqnarray}
\label{gleichungA34} S (\widetilde L / \tilde a, y) \approx
(\tilde a / \widetilde L) \sum_k \; e^{- k^2 y} \; = \; (\tilde a /
\widetilde L) K (4 \pi^2 \tilde a^2 L^{-2} y) \; + \;  O (e^{- \pi^2
y}) \; , \\
\label{gleichungA35} S (\infty, y) \approx \pi^{-1}
\int\limits_0^\pi dk \; e^{- k^2 y} \; = \; (2 \pi)^{-1}
(\pi/y)^{1/2}
\; + \; O (e^{- \pi^2 y}) \; ,\qquad \quad \\
\label{gleichungA36} S_D (L / \tilde a, y) \approx (\tilde a / L)
\sum_p \; e^{- p^2 y} \qquad \qquad \qquad \qquad \qquad \qquad
\qquad \qquad \quad \\
\label{gleichungA37} = \frac{1} {2} \; (\tilde a / L) \; \Big[K
(\pi^2 \tilde a^2 (L + \tilde a)^{-2} y) -1 \Big] \; + \; O (e^{-
\pi^2 y}) \; , \qquad \quad
\end{eqnarray}
where $K(y)$ is given by Eq. (\ref{gleichung35}). Furthermore it
is useful to turn to the integration variable
\be
\label{gleichungA38} z = 4 \pi^2 \tilde a^2 y / (L + \tilde a)^2
\; .
\ee
Instead of $y_0$ we then have the lower integration limit $z_0 = 4
\pi^2 \tilde a / (L + \tilde a) \to 0$ for large $L / \tilde a$.
This leads to
\begin{eqnarray}
\label{gleichungA39} \Delta f_3 &=& \frac{\tilde a^d L^{-1}} {(L +
\tilde a)^{d-1}}
\int\limits_0^\infty dz \Bigg(z^{-1} \Bigg\{ \Big(\frac{\pi}{z}\Big)^{d/2} - \;
\frac{1}{2} \left[\tilde s K (\tilde s^2 z)]^{d-1} [K (z/4) -
1 \right] \nonumber\\
&-& \frac{1}{2} \Big(\frac{\pi}{z} \Big)^{(d-1)/ 2} \Bigg\} \exp
\left[- \frac{r_0 (L + \tilde a)^2} {8 J \pi^2} \; z \right]
\Bigg)
\; + \; \left[1 + O (\tilde a^2 / L^2) \right]
\end{eqnarray}
with the shape factor
\be
\label{gleichungA40} \tilde s \; = \; \frac{L + \tilde a}
{\widetilde L} \; .
\ee
For $L \gg \tilde a$ we finally obtain Eqs. (\ref{gleichung31}) -
(\ref{gleichung58}).

The surface free energy, Eq. (\ref{gleichung32}), can be expressed
in terms of the generalized Watson function, Eqs.
(\ref{gleichung73}) and (\ref{gleichung-74}), as follows
\be
\label{gleichungA42} f_{surface} (t) = \frac{\tilde a^{1-d}}{8}
\int\limits_{r_0 \tilde a^2 J_0^{-1}}^\infty dz \left[W_{d-1} (z)
+ W_{d-1} (z + 4) - 2 W_d (z) \right] \; .
\ee
It can be shown that for $d \neq 3$ there exists the following
common representation of the coefficients $\tilde b_d$, Eq.
(\ref{gleichung36b}), and $\widetilde B_d$, Eq. (\ref{gleichung56}),
of the regular term of $f_{surface}$ linear in $r_0$
\be
\label{gleichungA43} \tilde b_d \; = \; \widetilde B_d &=&
\frac{1}{8}\; \left[W_{d-1} (4) \; - \; 2 \; W_d (0) \right]
\nonumber\\
&+& \frac{1}{8} \int\limits_0^A dy \left[e^{- 2 y} I_0 (2 y)
\right]^{d-1} \nonumber\\
&+& \frac{1}{8} \int\limits_A^\infty \left\{ \left[e^{- 2 y} I_0 (2 y)
\right]^{d-1} \; - \; (4 \pi y)^{(1-d)/2} \right\} \nonumber\\
&+& 2^{-d-1} \pi^{(1-d)/2} (d-3)^{-1} \; A^{(3-d) / 2} \; .
\ee
This expression is independent of the arbitrary constant $A > 0$.
For $d \to 3 +$ and $d \to 3 -$, the first two terms have a finite
limit $[W_2 (4) - 2 W_3 (0) ] / 8$ whereas the last term exhibits
a divergence $\sim (d-3)^{-1}$ that originates from $W_{d-1} (0) /
8$ for $d \to 3 +$ according to Eq. (\ref{gleichung91}). The
same divergence is contained in $A^+_{surface}$, see Eqs.
(\ref{gleichung-79}) and (\ref{gleichung92}).

\newpage

\section* {Appendix B. Susceptibility}

\renewcommand{\theequation}{B\arabic{equation}}
\setcounter{equation}{0}

We rewrite Eq. (\ref{gleichung96}) as
\be \label{gleichung-B1} \chi &=& \frac{\tilde a^2} {L (L + \tilde
a)} \; \sum_p \; \left\{2 - \left[1 + (-1)^n \right] \right\} \;
\frac{\cot^2 (p \tilde a / 2)} {r_0 + J_p} \ee
\be
\label{gleichung-B2} &=& \frac{2 \tilde a^4} {J_0 L (L + \tilde
a)} \; \sum_p \; \frac{1 + \cos p} {(1 - \cos p) \;\;  [\tilde r_0 + 2 (1 -
\cos p) ]} \nonumber\\
&-& \frac{2 \tilde a^4} {J_0 L (L + \tilde
a)} \; \sum_q \; \frac{1 + \cos q} {(1 - \cos q) \;\;  [\tilde r_0 + 2 (1 -
\cos q) ]}
\ee
with $\tilde r_0 = r_0 \tilde a^2 / J_0$. The sums $\sum_p$ and
$\sum_q$ run over dimensionless wave numbers $p = \pi \tilde a n /
(L + \tilde a)$ with integers $n = 1,2, \ldots, L/\tilde a$ and $q
= 2 \pi \tilde a m / (L + \tilde a)$ with integers $m = 1,2,
\ldots, L/(2 \tilde a)$ where we have assumed that $L / \tilde a$
is an even integer. Using the decomposition
\be
\label{gleichung-B3} \frac{1 + \cos x} {(1 - \cos x) \; \; [\tilde
r_0 + 2 (1 - \cos x) ]} \; \; =   &-&  \frac{1} {\tilde r_0
+ 2 (1 - \cos x)} \nonumber\\
+ \; \; \frac{4} {\tilde r_0} \; \Big[ \frac{1} {2 (1 - \cos x) }
&-&  \frac{1} {\tilde r_0 + 2 (1 - \cos x)} \Big]
\ee
and applying the representation Eq. (\ref{gleichungA1}) we obtain
\be \label{gleichung-B4} \chi &=& \frac{2 \tilde a^5} {J_0 L^2 (L
+ \tilde a)} \int\limits_0^\infty dy \left[ \frac{4} {\tilde r_0}
(1 - e^{- \tilde r_0 y}) - e^{\tilde r_0 y} \right] \Psi (L/\tilde
a, y) \ee
with
\be \label{gleichung-B5} \Psi (L / \tilde a, y) &=& S_D (L /
\tilde a, y) \; - \; \frac{1}{2} \; \overline S_D (L / \tilde a,
y) \ee
where $S_D (L / \tilde a, y)$ is given by Eq. (\ref{gleichungA7})
and
\be \label{gleichung-B6} \overline S_D (L / \tilde a, y) &=&
\frac{2 \tilde a} {L} \; \sum_q \; \exp [- 2 y (1 - \cos q) ] \; .
\ee
We distinguish the regions $0 \leq y \lesssim y_0 = (L + \tilde a)
/ \tilde a$ and $y \gtrsim y_0$. The large $L$ behavior of $S_D (L
/ \tilde a, y)$ in the former region is given by Eq.
(\ref{gleichungA30}), the corresponding behavior of $\overline
S_D$ is
\be \label{gleichung-B7} \overline S_D (L / \tilde a, y) &=& (1 +
\tilde a/L) \; S (\infty, y) \; - \; \pi \tilde a / (L + \tilde a)
\; + \; O (\tilde a^3 / L^3) \ee
which follows from the Poisson identity (see, e.g., Eq. (3.6) of
Ref. \cite{chen-dohm-2000}). In the region $y \gtrsim y_0$ we may
use the approximation (\ref{gleichungA36}) and
\be \label{gleichung-B8} \overline S_D (L / \tilde a, y) &=& (2
\tilde a/L) \sum_{n = 1}^\infty \exp \left[- 4 \pi^2 \tilde a^2 y
n^2 / (L + \tilde a)^2 \right] \; + \; O (e^{- \pi^2 y} ) \; . \ee
While the contributions of the region $y \lesssim y_0$ are
important for the large $L$ behavior of $\chi$ at fixed $T > T_c$,
the contributions of the region $y \gtrsim y_0$ yield the
finite-size scaling behavior of $\chi$ in the critical region $L /
\tilde a \gg 1$, $\xi / \tilde a \gg 1$ at fixed ratio $x = L/\xi
\geq 0$, including the leading terms of the scaling function for
large $x$, as given by Eqs. (\ref{gleichung115}) -
(\ref{gleichung100}) for the Gaussian model.

The derivation of $\chi$ from Eq. (\ref{gleichung151a}) is
parallel to that given above, except that $\tilde r_0$ is to be
replaced by
\be
\label{gleichung-B9} \tilde \mu / J_0 &=& \Delta \mu \; - \; \pi^2
/ (L + \tilde a)^2 \; + \; O (\tilde a^2 L^{-4}) \; ,
\ee
where $\Delta \mu$ is defined by Eq. (\ref{gleichung178}). At
fixed $\widetilde M \equiv \Delta \mu (L + \tilde a)^2 \geq 0$
this leads to the large $L$ behavior
\be \label{gleichung-B10} \chi &=& \frac{4 \beta (L + \tilde a)^3}
{J_0 \pi^2 L} \; \int\limits_0^\infty \; dy \;  \Bigg\{ \frac{1 -
\exp [- (\widetilde M - \pi^2) y / \pi^2 ]} {\widetilde M - \pi^2}
\nonumber\\
&-& \frac{\tilde a^2} {(L + \tilde a)^2} \; \exp \; [- (\widetilde M -
\pi^2) y / \pi^2]  \Bigg\} \left[K(y) - K(4 y) \right]
\ee
where $K(z)$ is given by Eq. (\ref{gleichung35}). For $L \gg
\tilde a$ this yields Eq. (\ref{gleichung-203}).

\newpage

\section* {Appendix C. Film critical temperature for $d > 3$}

\renewcommand{\theequation}{C\arabic{equation}}
\setcounter{equation}{0}

In the following we consider Eq. (\ref{gleichungB.3}) for
$d > 3$. Subtracting
\be
\label{gleichungB6} \beta_{c,d} (\infty) &=& \tilde a^{d-2}
\int\limits_p \int\limits_{\bf k} \left(J_{{\bf k},d-1} \; + \;
J_p \right)^{-1}
\ee
and using the representation
\be
\label{gleichungB7} \frac{1}{z} \; = \;  \int\limits_0^\infty dy
\; e^{-zy}
\ee
for $z > 0$ we obtain
\be
\label{gleichungB8} 2J \left[\beta_{c,d} (\infty) \; - \;
\beta_{c,d} (L) \right] \; = \; \int\limits_0^\infty dy \;
\widetilde \Phi (L/\tilde a, y) \; ,
\ee
\be
\label{gleichungB9} \widetilde \Phi (x, y) = \left[S (\infty, y)
\right]^{d-1} \left\{ S (\infty, y) - S_D (x, y) \exp
\left[2 y \left(1 - \cos \frac{\pi} {1 + x}\right) \right] \right\}
\ee
where $S (\infty, y)$ and $S_D (L/\tilde a, y)$ are defined by Eqs.
(\ref{gleichungA6}) and (\ref{gleichungA7}) [see also Eq.
(\ref{gleichungA14})]. It is important to distinguish the regimes
$0 \leq y \lesssim y_0$ and $y \gtrsim y_0$ with $y_0$ given by Eq.
(\ref{gleichungA8}). Accordingly we split
\be
\label{gleichungB10} \int\limits_0^\infty dy \; \widetilde \Phi (x, y)
\; = \;\int\limits_0^{y_0} dy \; \widetilde \Phi (x,y) \; + \;
\int\limits_{y_0}^\infty dy \; \widetilde \Phi (x, y)
\; \equiv \; \widetilde \Delta_1 \; + \; \widetilde \Delta_2 \; .
\ee
In $\widetilde \Delta_1$ we use Eq. (\ref{gleichungA30}). A
treatment similar to that in Eqs. (\ref{gleichungA31}) -
(\ref{gleichungA38}) leads to
\be
\label{gleichungB11} \widetilde \Delta_1 + \widetilde \Delta_2 =
4 \widetilde B_d \; \frac{\tilde a}{L}
- \widetilde C_d \left(\frac{\tilde a}{L + \tilde a} \right)^{d-2}
\left(1 + \frac{\tilde a}{L} \right) \; + \; O (\tilde a^{d/2} \; L^{- d/2})
\ee
with the nonuniversal amplitude $\widetilde B_d$, Eq.
(\ref{gleichung56}), and the universal amplitude
\be
\label{gleichungB12} \widetilde C_d = \frac{1}{8 \pi^2}
\int\limits_0^\infty dz \left\{1 - 2
\left(\frac{\pi}{z}\right)^{1/2} + e^{z/4} \left[K
\left(\frac{z}{4} \right) - 1 \right] \right\} \left(\frac{\pi}{z}
\right)^{(d-1)/2}
\ee
with $\widetilde C_d > 0$. The first term $\sim L^{-1}$ in Eq.
(\ref{gleichungB11}) has a nonscaling $L$ dependence whereas the
second term $\sim L^{2-d}$ has the scaling $L$ dependence $\sim
L^{1 / \nu}$. Eqs. (\ref{gleichungB8}) - (\ref{gleichungB12}) lead
to Eq. (\ref{gleichung187}). Rewriting
\be
\label{gleichungB16} 1 + e^{-4y} - 2 e^{- 2 y} I_0 (2y) \; = \; 2
e^{- 2 y} \left[\cosh (2y) \; - \; I_0 (2 y) \right]
\ee
and using (see 9.6.39 of Ref. \cite{handbook})
\be
\label{gleichungB17} \cosh (z) \; - \; I_0 (z) \; = \; 2 I_2 (z)
\; + \; 2 I_4 (z) \; + \; \ldots \geq 0
\ee
we see that $\widetilde B_d$ is {\it positive} and finite
for $d > 3$, thus $T_{c,d} (L) > T_{c,d} (\infty)$ for $L \gg
\tilde a$. Using the representation
\be
\label{gleichungB13} 2J \beta_{c,d} (\infty) \; = \;
\int\limits_0^\infty dy \; \left[S (\infty, y) \right]^d
\ee
we obtain the fractional shift of the film critical temperature
as given in Eqs. (\ref{gleichung189}) - (\ref{gleichung190}).
We have verified that $a_4$, Eq. (\ref{gleichung189a}), agrees
with the corresponding amplitude of Barber and Fisher
\cite{barber-1973} at $d = 4$ which was expressed in terms
of the generalized Watson function, Eqs. (\ref{gleichung73})
and (\ref{gleichung-74}). The amplitude $a_d = 4 \widetilde
B_d / W_d (0)$, Eq. (\ref{gleichung189a}), diverges for $d \to 3$.
This divergence is cancelled by the next term of $O (L^{2-d})$ in
Eqs. (\ref{gleichungB11}) and (\ref{gleichung189}).

\newpage

\section* {Appendix D. Constraint equation}

\renewcommand{\theequation}{D\arabic{equation}}
\setcounter{equation}{0}

We start from the constraint equation (\ref{gleichung148}) for box
geometry where we decompose $\tilde \mu = J_0 \Delta \mu + \tilde
\mu_c (L)$ and subtract $\beta_{c,d}$ in the form of Eq.
(\ref{gleichungB6}). Furthermore we add and subtract
\be
\label{gleichungC1} \tilde \beta (\Delta \mu) \; \equiv \; \tilde a^{d-2}
\int\limits_{\bf k} \int\limits_p \left(J_0 \Delta \mu \; + \; J_{{\bf
k},d-1} \; + \; J_p \right)^{-1} \; .
\ee
This yields
\be
\label{gleichungC2}  \beta_{c,d} \; - \; \beta
\; = \; M_1 \; + \;
M_2 \; ,
\ee
\be
\label{gleichungC3} M_1 = \tilde \beta (\Delta \mu) \; - \;
\tilde a^{d-2} \; \widetilde L^{1-d} L^{-1} \; \sum_{{\bf k},p}
\left(J_0 \Delta \mu + \tilde \mu_c + J_{{\bf k}, d-1} \; + \; J_p \right)^{-1}
\; ,
\ee
\be
\label{gleichungC4} M_2 = \tilde a^{d-2} \; \Delta \mu
\int\limits_{\bf k} \int\limits_p
\left(J_0 \Delta \mu + J_{{\bf k}, d-1} \; + \; J_p \right)^{-1}
\left(J_{{\bf k}, d-1} \; + \; J_p \right)^{-1} \; .
\ee
Using the representation Eq. (\ref{gleichungB7}) we obtain
\be
\label{gleichungC5} J_0 \; M_1 \; = \; \int\limits_0^\infty dy \; \Phi
(L/\tilde a, \widetilde L/\tilde a, y) \exp \left(- \Delta \mu
\tilde a^2 y \right)
\ee
where $\Phi (L/\tilde a, \widetilde L/\tilde a, y)$ is given by
Eq. (\ref{gleichungA4}). Again we split the integral in Eq.
(\ref{gleichungC5}) as $\int_0^\infty = \int_0^{y_0} +
\int_{y_0}^\infty \equiv I_1 + I_2$ with $y_0 = (L + \tilde a) /
\tilde a$. For large $L/\tilde a$ and $\widetilde L/\tilde a$ we find
\be
\label{gleichungC6} I_1 \;=\; \frac{\tilde a}{2 L}
\int\limits_0^{y_0} dy \; \left[1 + e^{-4y} - 2 S (\infty, y)
\right] \left[S (\infty, y) \right]^{d-1} \exp \left(- \Delta \mu
\tilde a^2 y \right) \nonumber\\
+ \;O \left(e^{- \widetilde L / \tilde a}, \tilde a^{d/2} L^{-
d/2} \right) \; , \qquad \qquad \qquad \qquad \qquad
\ee
\be
\label{gleichungC7} I_2 = \frac{\tilde a}{L} \left(\frac{L +
\tilde a}{\tilde a} \right)^{3-d} \frac{1}{8 \pi^2}
\int\limits_{z_0}^\infty dz \Big\{e^{z/4} \left[K \left(\frac{z}{4}
\right) - 1
\right] \left[\tilde s K (\tilde s^2 z) \right]^{d-1}
+ \left(\frac{\pi}{z} \right)^{(d-1) / 2} \nonumber\\
- 2 \left(\frac{\pi}{z} \right)^{d/2}\Big\} + O \left(\tilde a^{d/2}
L^{-d/2} \right) \qquad \qquad \qquad
\ee
with $z_0 = 4 \pi^2 \tilde a / (L + \tilde a)$ and
$\tilde s = (L + \tilde a) / \widetilde L$ where $K (z)$
is given by Eq. (\ref{gleichung35}). For large $L/\tilde a$ we can
let $z_0 \to 0$ in Eq. (\ref{gleichungC7}). Evaluating the integral in
Eq. (\ref{gleichungC4}) for small $\Delta \mu$ yields for $2 < d < 4$
\be
\label{gleichungC8} J_0 M_2 = \varepsilon^{-1} A_d (\Delta \mu
\tilde a^2)^{(d-2) / 2} \; + \; O (\Delta \mu
\tilde a^2) \; .
\ee
Eqs. (\ref{gleichungC2}) - (\ref{gleichungC8}) lead to
\be
\label{gleichungC9} J_0 (\beta_{c,d} - \beta) = \varepsilon^{-1}
A_d (\Delta \mu \tilde a^2)^{(d-2)/2}
- \; E_{surface} (\Delta \mu \tilde a^2) (\tilde a/2L)
\nonumber\\
+\; 2 \widetilde{\cal E}_d \left((\Delta
\mu)^{1/2} \; (L + \tilde a), \tilde s \right)
\left[(L + \tilde a) / \tilde a \right]^{2-d} \qquad \qquad
\ee
where $E_{surface} (z)$ and $\widetilde{\cal E}_d (x, s)$ are
given by Eqs. (\ref{gleichung-d71}) and (\ref{gleichung191}).
Multiplying Eq. (\ref{gleichungC9}) by $\varepsilon A_d^{-1}
(L / \tilde a)^{d-2}$ and using
\be
\label{gleichungC10} \varepsilon A_d^{-1} J_0 (\beta_c - \beta) =
t (\xi_0 / \tilde a)^{2-d} \; + \; O (t^2)
\ee
we obtain Eq. (\ref{gleichung185}) for $L
\gg \tilde a$.



\newpage


\begin{thebibliography}{99}
%
\bibitem{fisher}
M.  E. Fisher, in {\it Critical Phenomena, Proceedings of the 1970
International School of Physics ``Enrico Fermi'',}  Course 51,
edited by M. S. Green (Academic, New York, 1971), p. 1.
%
\bibitem{barber}
M. N. Barber, in {\it Phase Transitions and Critical Phenomena}, edited
by C. Domb, J.L. Lebowitz (Academic, New York, 1983),
Vol.~8, p. 145.
%
\bibitem{finite}
{\it Finite Size Scaling and Numerical Simulation of Statistical
Systems}, edited by V. Privman (World Scientific, Singapore, 1990).
%
\bibitem{privman}
V. Privman, A. Aharony, and P.C. Hohenberg, in {\it Phase
Transitions and Critical Phenomena}, edited by C. Domb and J.L.
Lebowitz (Academic, New York, 1991), Vol. 14, p. 1.
%
\bibitem{chen-dohm-1999}
X.S. Chen and V. Dohm, Eur. Phys. J. B {\bf 7}, 183 (1999).
%
\bibitem{chen-dohm-99}
X.S. Chen and V. Dohm, Eur. Phys. J. B {\bf 10}, 687 (1999).
%
\bibitem{chen-dohm-2000}
X.S. Chen and V. Dohm, Eur. Phys. J. B {\bf 15}, 283 (2000). On
the right-hand side of Eq. (1.1) the factor $(2d)^{-1}$ is
missing.
%
\bibitem{chen-dohm-cond}
X.S. Chen and V. Dohm, Phys. Rev. E {\bf 66}, 016102 (2002); E
{\bf 66}, 059901 (2002) (E); cond-mat/0108202; cond-mat/0112310;
Physica B \ldots (in press); cond-mat/0209321.
%
\bibitem{dohm-1993}
V. Dohm, Physica Scripta T{\bf{49}}, 46 (1993).
%
\bibitem{gasparini}
F.M. Gasparini and I. Rhee, in {\it{Prog. Low Temp. Phys.}}XIII,
ed. D.F. Breuer (North-Holland, Amsterdam, 1992), p. 1; I. Rhee,
F.M. Gasparini and D.J. Bishop, Phys. Rev. Lett. {\bf 63}, 410
(1989); M.O. Kimball, S. Mehta, and F.M. Gasparini, J. Low Temp.
Phys. {\bf 121}, 29 (2000); F.M. Gasparini, M.O. Kimball, and K.P.
Moonay, J. Phys.: Condens. Matter {\bf 13}, 4871 (2001); M.O. Kimball,
M. Diaz-Avila, and F.M. Gasparini, LT 23 Proceedings, to be
published in Physica B.
%
\bibitem{lipa}
J.A. Lipa, D.R. Swanson, J.A. Nissen, Z.K. Geng, P.R. Williamson,
D.A. Stricker, T.C.P. Chui, U.E. Israelsson, and M. Larson,
Phys. Rev. Lett. {\bf 84}, 4894 (2000); J.A. Lipa, M.
Coleman, and D.A. Stricker, J. Low Temp. Phys. {\bf 124}, 443
(2001).
%
\bibitem{fig-2}
See also Fig. 2 in V. Dohm, in {\it Forschung unter
Weltraumbedingungen}, Proceedings DLR-Bilanzsymposium Norderney,
edited by M.H. Keller and P.R. Sahm (WPF RWTH Aachen, 2000), p. 150.
%
\bibitem{kuehn}
K. Kuehn, S. Mehta, H. Fu, E. Genio, D. Murphy, F. Liu, Y. Liu,
and G. Ahlers, Phys. Rev. Lett. {\bf 88}, 095702 (2002).
%
\bibitem{schultka}
N. Schultka and E. Manousakis, J. Low Temp. Phys. {\bf 109}, 733
(1997).
%
\bibitem{barber-1973}
M. N. Barber and M. E. Fisher, Ann. Phys. (N.Y.) {\bf 77}, 1 (1973).
%
\bibitem{barber-1974}
M.N. Barber, J. Stat. Phys. {\bf 10}, 59 (1974).
%
\bibitem{dantchev1}
D. Danchev , J. Stat. Phys. {\bf 73}, 267 (1993).
%
\bibitem{barber-2}
M.N. Barber, Aust. J. Phys. {\bf 26}, 483 (1973).
%
\bibitem{barber-3}
M.N. Barber and M.E. Fisher, Phys. Rev. {\bf A8}, 1124 (1973).
%
\bibitem{singh-1985}
S. Singh and R.K. Pathria, Can. J. Phys. {\bf 63}, 358 (1985).
%
\bibitem{gelfand}
M.P. Gelfand and M.E. Fisher, Physica A {\bf 166}, 1 (1990).
%
\bibitem{cardy-1990}
J.L. Cardy, Phys. Rev. Lett. {\bf 65}, 1443 (1990).
%
\bibitem{eisenriegler}
E. Eisenriegler, Z. Phys. {\bf B 61}, 299 (1985).
%
\bibitem{symanzik}
K. Symanzik, Nucl. Phys. {\bf B 190} [FS {\bf 3}], 1 (1981).
%
\bibitem{krech-1992}
M. Krech and S. Dietrich, Phys. Rev. Lett. {\bf 66}, 345 (1991);
Phys. Rev. A {\bf 46}, 1886 (1992); {\bf 46}, 1922 (1992).
%
\bibitem{sutter}
P. Sutter and V. Dohm, Physica B {\bf 194 - 196}, 613 (1994).
%
\bibitem{schmolke}
R. Schmolke, A. Wacker, V. Dohm, and D. Frank, Physica B {\bf 165}
\& {\bf 166}, 575 (1990).
%
\bibitem{frank-dohm}
D. Frank and V. Dohm, Phys. Rev. Lett. {\bf 62}, 1864 (1989);
Z. Phys. B {\bf 84}, 443 (1991).
%
\bibitem{mohr-dohm-2000}
U. Mohr and V. Dohm, Physica B {\bf 284} - {\bf 288}, 43 (2000).
%
\bibitem{chen-dohm-1998}
X.S. Chen and V. Dohm, Physica {\bf A 251}, 439 (1998).
%
\bibitem{chen-dohm-1998-a}
X.S. Chen and V. Dohm, Eur. Phys. J. {\bf B 5}, 529 (1998).
%
\bibitem{chen-dohm-1998-b}
X.S. Chen and V. Dohm, Int. J. Mod. Phys. {\bf C 9}, 1007 (1998).
%
\bibitem{chen-dohm-1998-c}
X.S. Chen and V. Dohm, Int. J. Mod. Phys. {\bf C 9}, 1073 (1998).
%
\bibitem{chen-dohm-2001}
X.S. Chen and V. Dohm, Phys. Rev. E {\bf 63}, 016113 (2001).
%
\bibitem{dohm-1989}
V. Dohm, Z. Phys. B {\bf 75}, 109 (1989).
%
\bibitem{mermin}
N.D. Mermin and H. Wagner, Phys. Rev. Lett. {\bf 17}, 1133 (1966).
%
\bibitem{privman-fisher}
V. Privman and M.E. Fisher, Phys. Rev. B {\bf 30}, 322 (1984).
%
\bibitem{privman1}
V. Privman, Phys. Rev. {\bf B 38}, 9261 (1988).
%
\bibitem{wegner}
F. Wegner, Phys. Rev. {\bf B 5}, 4529 (1972).
%
\bibitem{fisher-burford}
M.E. Fisher and R.J. Burford, Phys. Rev. {\bf 156}, 583 (1967).
%
\bibitem{goldschmidt}
Y.Y. Goldschmidt and D. Jasnow, Phys. Rev. {\bf B 29}, 3990
(1984).
%
\bibitem{mikheev}
L.V. Mikheev and M.E. Fisher, J. Low Temp. Phys. {\bf 90}, 119
(1993).
%
\bibitem{privman-rudnick}
V. Privman and J. Rudnick, J. Phys. {\bf A 19}, L 1215 (1986).
%
\bibitem{dantchev-2001}
D. Dantchev and J. Rudnick, Eur. Phys. J. {\bf B 21}, 251 (2001).
%
\bibitem{stanley-1968}
H.E. Stanley, Phys. Rev. {\bf 176}, 718 (1968).
%
\bibitem{fisher-1968}
M.E. Fisher, Phys. Rev. {\bf 176}, 257 (1968).
%
\bibitem{shapiro}
J. Shapiro and J. Rudnick, J. Stat. Phys. {\bf 43}, 51 (1986).
%
\bibitem{morse-feshbach}
P.M. Morse and H. Feshbach, {\it Methods of Theoretical Physics}
(Mc Graw-Hill, New York, 1953).
%
\bibitem{handbook}
{\it Handbook of Mathematical Functions}, edited by M. Abramowitz
and I.A. Stegun (Dover Publ., New York, 1972).


\end{thebibliography}
\end{document}